\documentclass{ws-ijbc}
\usepackage{ws-rotating}     % used only when sideways tables/figures are used
\usepackage{graphicx}
\usepackage{epstopdf}

\usepackage[colorlinks=true, linkcolor=black, citecolor=black, urlcolor=black]{hyperref}
\usepackage{amsmath}
\usepackage{amssymb}
\usepackage{hyperref}
 \usepackage{multicol}
 \usepackage{multirow}
 \usepackage{array}
 \usepackage[abs]{overpic}
 \usepackage{mathptm,rotate,color}

\begin{document}

\catchline{}{}{}{}{} % Publisher's Area please ignore

\markboth{Morgan R. Frank}{Standing Swells Surveyed Showing Surprisingly Stable Solutions for the Lorenz '96 Model}

%\title{\uppercase{Standing Swells Surveyed Showing Surprisingly Stable Solutions for the Lorenz '96 Model}}
%
%\author{\uppercase{Morgan R. Frank}}
%
%\address{Department of Mathematics, University of Vermont\\
%Burlington, Vermont 05401, USA\\
%fauthor@university.com\footnote{State completely without
%abbreviations the affiliation and mailing address, including
%country. Typeset in 11~pt Times Italic.}}
%
%\author{SECOND AUTHOR}
%\address{Group, Company, Address\\
%City, State ZIP/Zone, Country\\
%sauthor@company.com}

 \title{\fontsize{11}{12}{\uppercase{Standing Swells Surveyed Showing Surprisingly Stable Solutions for the Lorenz '96 Model}}}
 \author{\fontsize{11}{12}{Morgan R. Frank$^{1}$, Lewis Mitchell$^{2}$, Peter Sheridan Dodds$^3$, Christopher M. Danforth$^{4}$\\
 \emph{Computational Story Lab, Department of Mathematics and Statistics,\\ Vermont Complex Systems Center, Vermont Advanced Computing Core,\\
University of Vermont, Burlington, Vermont \\
$^{1}$mrfrank@uvm.edu, $^{2}$lmitchel@uvm.edu, $^{3}$pdodds@uvm.edu, $^4$cdanfort@uvm.edu}}}

\maketitle

\begin{history}
\received{(to be inserted by publisher)}
\end{history}

\begin{abstract}
The Lorenz '96 model is an adjustable dimension system of ODEs exhibiting chaotic behavior representative of dynamics observed in the Earth's atmosphere. In the present study, we characterize statistical properties of the chaotic dynamics while varying the degrees of freedom and the forcing. Tuning the dimensionality of the system, we find regions of parameter space with surprising stability in the form of standing waves traveling amongst the slow oscillators. The boundaries of these stable regions fluctuate regularly with the number of slow oscillators. These results demonstrate hidden order in the Lorenz '96 system, strengthening the evidence for its role as a hallmark representative of nonlinear dynamical behavior.
%The abstract should summarize the context, content and conclusions
%of the paper. It should not contain any references or displayed
%equations. Typeset the abstract in 10~pt Times Roman with
%baselineskip of 12 pt, making an indentation of 1.6~cm on the left
%and right margins.
\end{abstract}


%\begin{multicols}{2}
\section{Introduction}
\begin{figure}[!hHb]
	$\begin{array}{cc}
	\fbox{\begin{overpic}[width=.47\columnwidth,trim=8.5cm -1cm 5cm -2.5cm,clip]{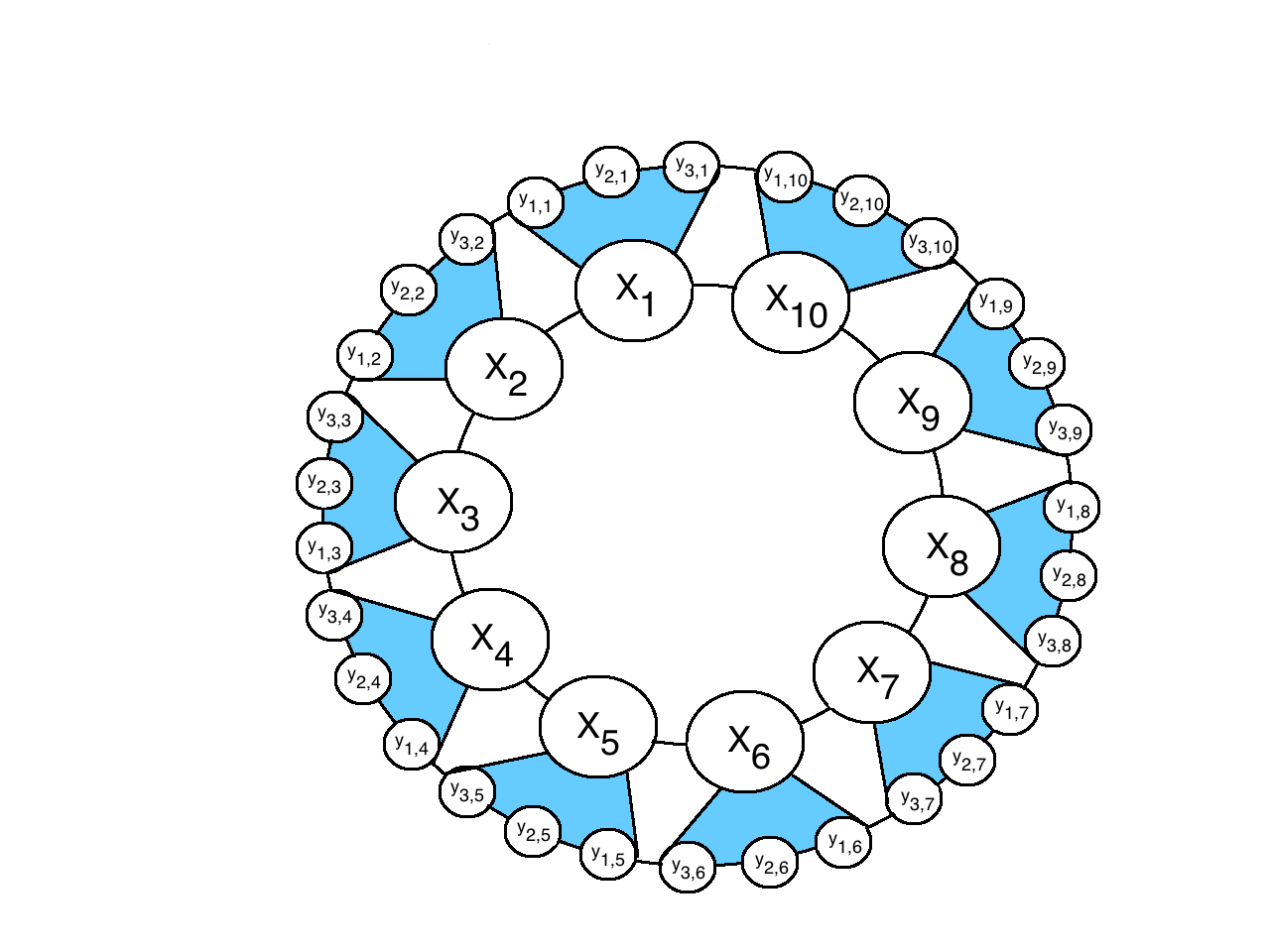}\put(2,70){\fbox{A}}\end{overpic}}&
	\hspace{-.12cm}\fbox{\hspace{.5cm}\begin{overpic}[width=.42\columnwidth,trim=6cm 8cm 5cm 6cm]{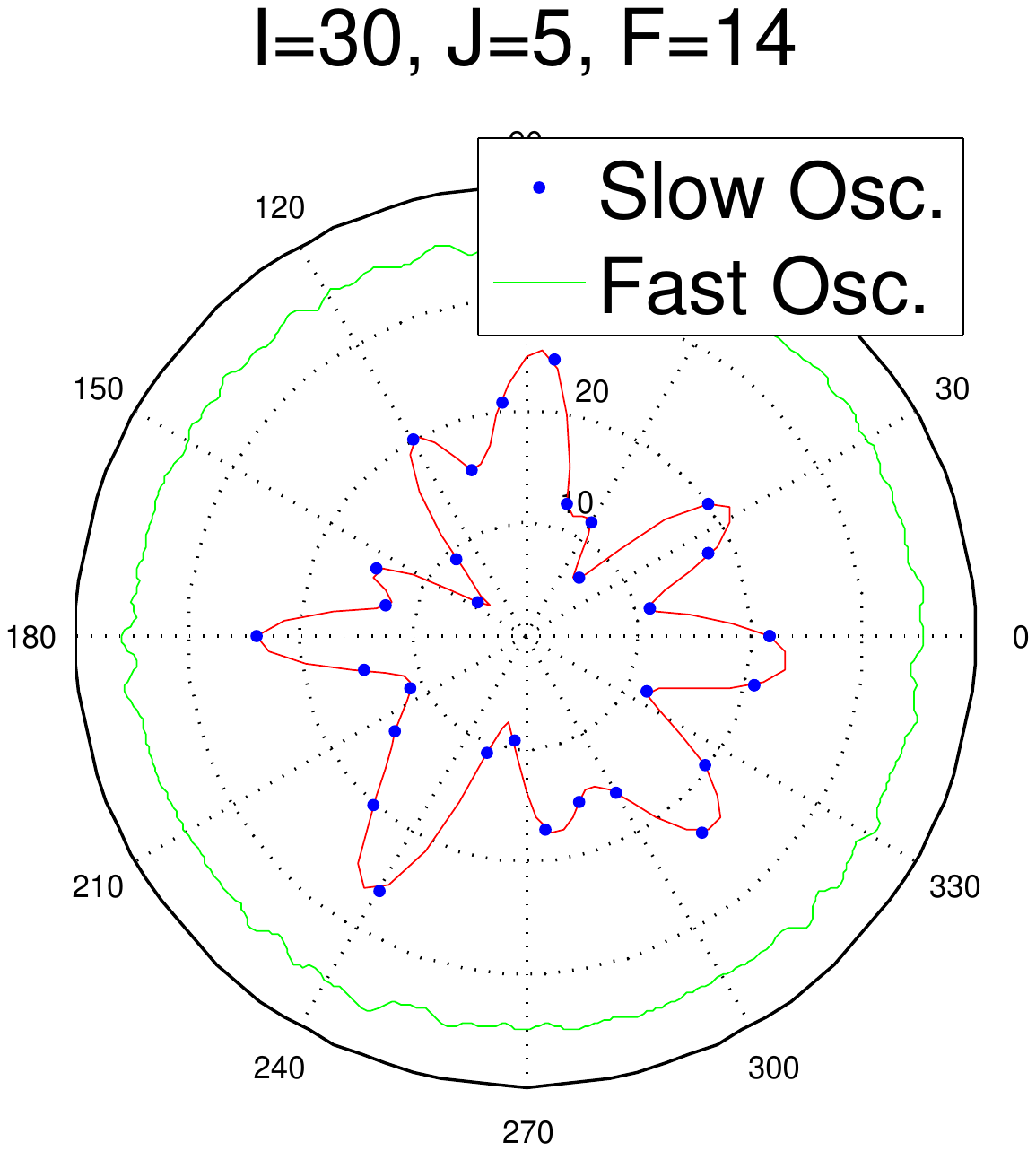}\put(2,70){\fbox{B}}\end{overpic}\hspace{.5cm}}
	\end{array}$
	\caption{ (A) A visual representation for the coupling of the fast and slow time systems in the Lorenz '96 model. There are $I=10$ slow  large amplitude oscillators, each of which are coupled to $J=3$ fast small amplitude oscillators. The slow oscillators are arranged in a circle representing a given latitude. (B) An example snapshot from an actual trajectory with $I=30$, $J=5$, and $F=14$. The blue dots represent the slow oscillators, and the green represents the flow of information among the fast oscillators. See Fig. (\ref{fig5}) for further examples and a more detailed explanation.}
	\label{fig1}
\end{figure}

\indent Modern society often depends on accurate weather forecasting for daily planning, efficient air-travel, and disaster preparation \cite{kerr00}. Predicting the future state of physical systems, such as the atmosphere, proves to be difficult; chaotic systems exhibit sensitive dependence on initial conditions, meaning that small errors in any state approximation will lead to exponential error growth \cite{yorke00}. Furthermore, weather prediction requires the use of computationally expensive numerical models for representing the atmosphere. Most scientists trying to advance current predictive techniques cannot afford to run experiments using these real-world weather models. To this end, computationally manageable ``simple models" are used instead to represent interesting atmospheric characteristics while reducing the overall computational cost. \\ 
\indent Scientists have long wrestled with chaotic behavior limiting the predictability of weather in the Earth's atmosphere \cite{lorenz00,lorenz01,farmer00,lorenz02,DY00}. In the case of atmospheric forecasting, simple models exhibiting exponential error growth provide an ideal environment for basic research in predictability. Edward Lorenz, one of the great pioneers in predictability research, introduced the following $I$-dimensional model which exhibits chaotic behavior when subject to sufficient forcing
 \begin{equation}
 	\frac{dx_{i}}{dt}=x_{i-1}(x_{i+1}-x_{i-2})-x_{i}+F,
	\label{eq1}
 \end{equation}
 where $i=1,2,\dots,I$ and $F$ is the forcing parameter \cite{lorenz03}. Each $x_{i}$ can be thought of as some atmospheric quantity, e.g. temperature, evenly distributed about a given latitude of the globe, and hence there is a modularity in the indexing that is described by $x_{i+I}=x_{i-I}=x_{i}$. \\
 \indent In an effort to produce a more realistic growth rate of the large-scale errors, Lorenz went on to introduce a multi scale model by coupling two systems similar to the model in Eq.(\ref{eq1}), but differing in time scales. The equations for the Lorenz '96 model \cite{lorenz03} are given as
 \begin{equation}
 	\frac{dx_{i}}{dt}=x_{i-1}(x_{i+1}-x_{i-2})-x_{i}+F-\frac{hc}{b}\displaystyle\sum_{j=1}^{J}y_{(j,i)},
	\label{eq2}
 \end{equation}
 \begin{equation}
 	\frac{dy_{(j,i)}}{dt}=cby_{(j+1,i)}(y_{(j-1,i)}-y_{(j+2,i)})-cy_{(j,i)}+\frac{hc}{b}x_{i},
	\label{eq3}
 \end{equation}
 where $i=1,2,\dots,I$ and $j=1,2,\dots,J$. The parameters $b$ and $c$ indicate the time scale of solutions to Eq. (\ref{eq3}) relative to solutions of Eq. (\ref{eq2}), and $h$ is the coupling parameter. The coupling term can be thought of as a parameterization of dynamics occurring at a  spatial and temporal scale unresolved by the $x$ variables. Again, each $x_{i}$ can be thought of as an atmospheric quantity about a latitude that oscillates in slow time, and the set of $y_{(j,i)}$ are a set of $J$ fast time oscillators that act as a damping force on $x_{i}$. The $y$'s exhibit a similar modularity described by $y_{(j+IJ,i)}=y_{(j-IJ,i)}=y_{(j,i)}$. A snapshot of a solution state is shown as an example in Fig. (\ref{fig1}).\\
  \indent This system has been used to represent weather related dynamics in several previous studies as a low-dimensional model of atmospheric dynamics \cite{orrell00,wilks00,danforth00,lieb00}. There are many advantages to using the Lorenz '96 model. Primarily, the model allows for flexibility in parameter tuning to achieve varying relative levels of nonlinearity, coupling of timescales, and spatial degrees of freedom. Unless otherwise noted, we fix the time scaling parameters $b=c=10$ and the coupling parameter $h=1$ for the remainder of this study. These parameter choices are consistent with the literature in terms of producing chaotic dynamics quantitatively similar to those observed in the atmosphere \cite{karimi00}. We vary $I$, $J$, and $F$ to explore different spatial degrees of freedom and different levels of nonlinearity in the system dynamics.\\
%%%%%%%%%%%%%%
%%%%%%%%%%%%%%
 \indent In this study, we characterize the parameter space of the Lorenz '96 system revealing patterns of order and chaos in the system. We discuss our methods in Section 2. In Section 3 and 4, we provide our results along with evidence for stability in the Lorenz '96 model in the form of standing waves traveling around the slow oscillators. We discuss the implications of our findings in Section 4.\\
%%%%%%%%%%%%%%%%%%%%%%%%%%%%%%%%%%%%%%%%%%%
 \section{Methods}
\begin{figure}[!h]
	\centering
	$\begin{array}{cc}
		\hspace{-.5cm}\begin{overpic}[width=.4\columnwidth,trim=3cm 8cm 3cm 8cm,clip]{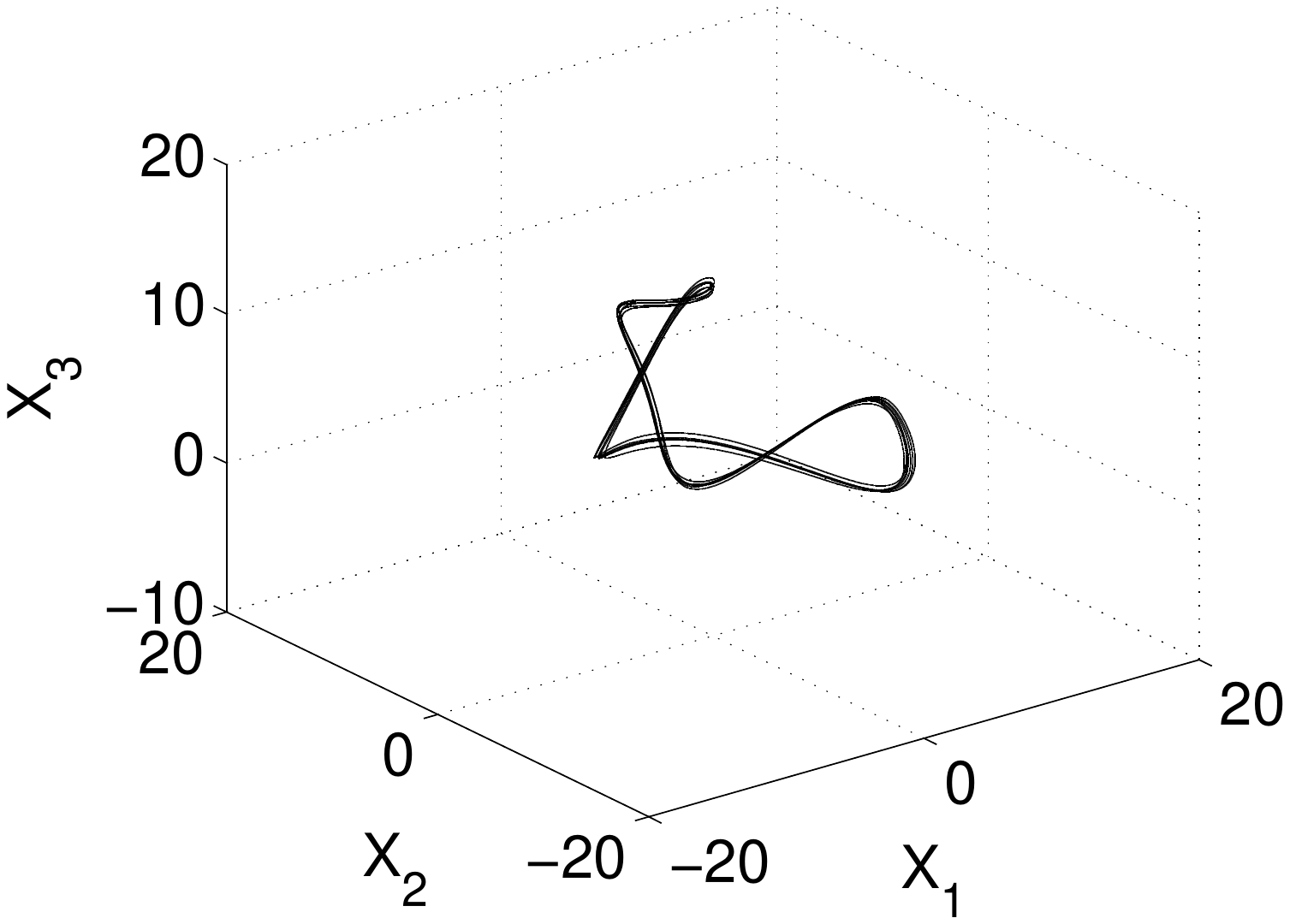}\put(15,35){\colorbox{white}{\fbox{\text{A}}}}\end{overpic}&\begin{overpic}[width=.4\columnwidth,trim=3cm 8cm 3cm 8cm,clip]{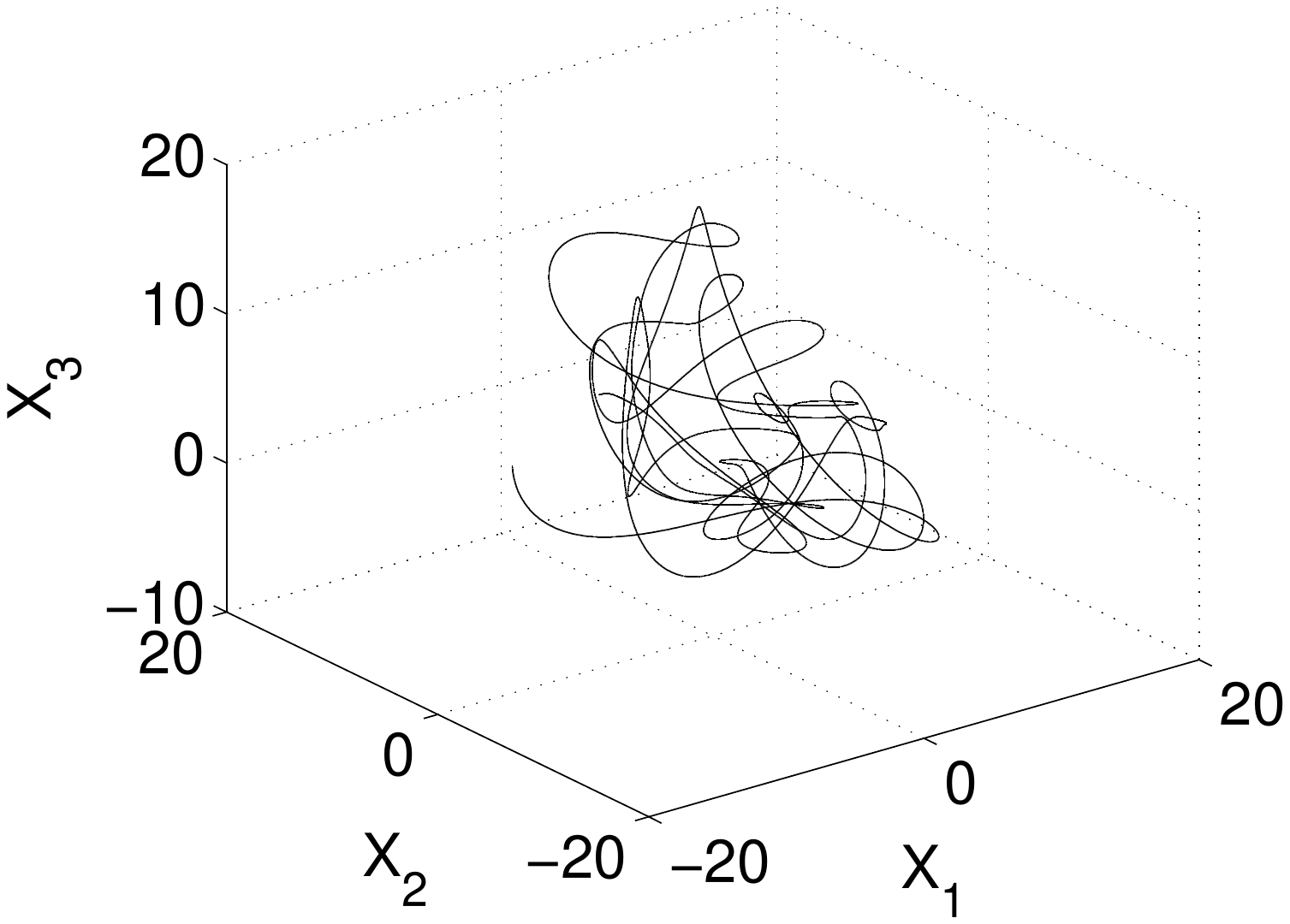}\put(15,35){\colorbox{white}{\fbox{\text{B}}}}\end{overpic}
	\end{array}$
	$\begin{array}{cc}
	\hspace{-.5cm}\begin{overpic}[scale=.24,trim=3cm 8cm 4cm 8cm,clip]{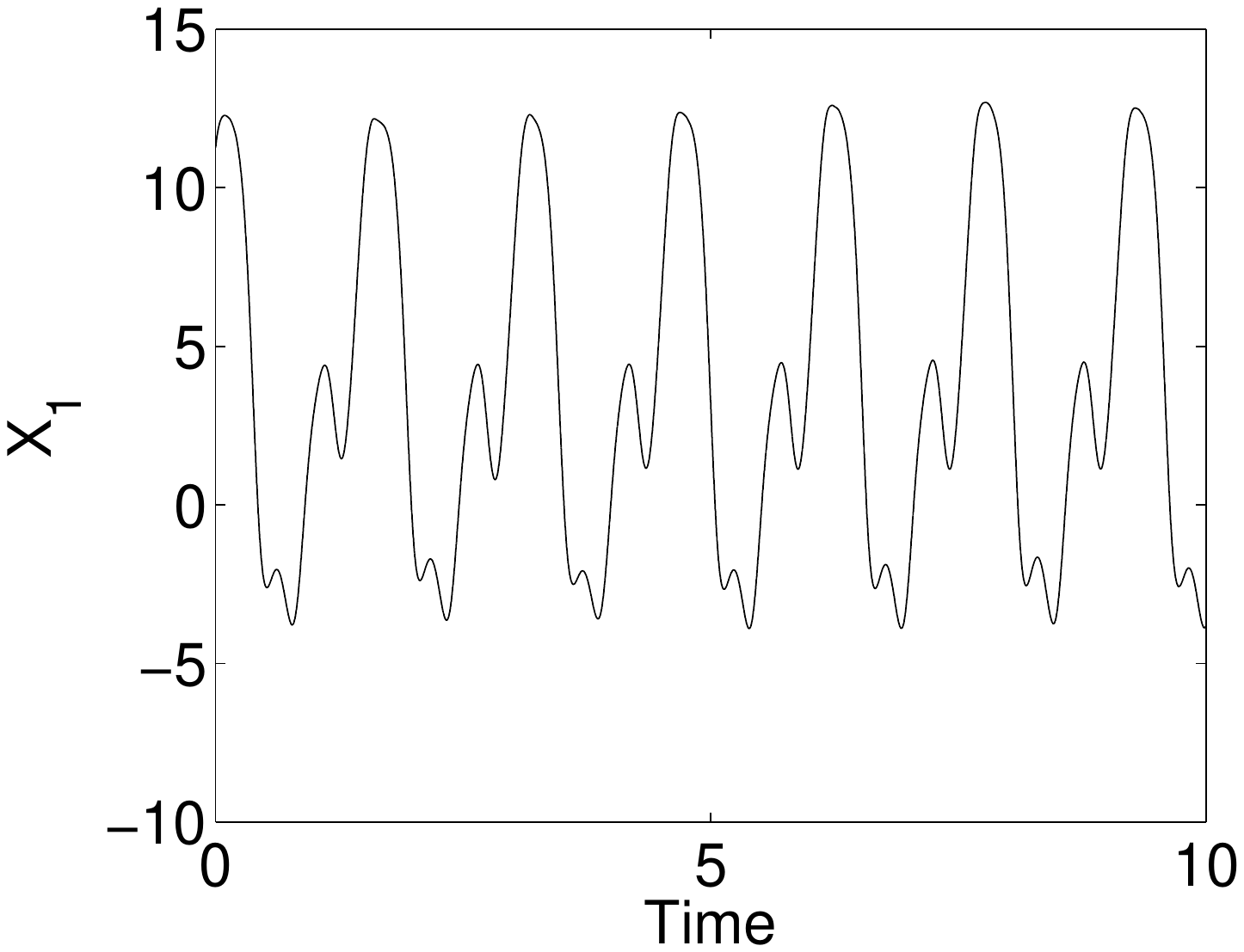}\put(7,25){\colorbox{white}{\fbox{C}}}\end{overpic}&
	\begin{overpic}[scale=.24,trim=3cm 8cm 4cm 8cm,clip]{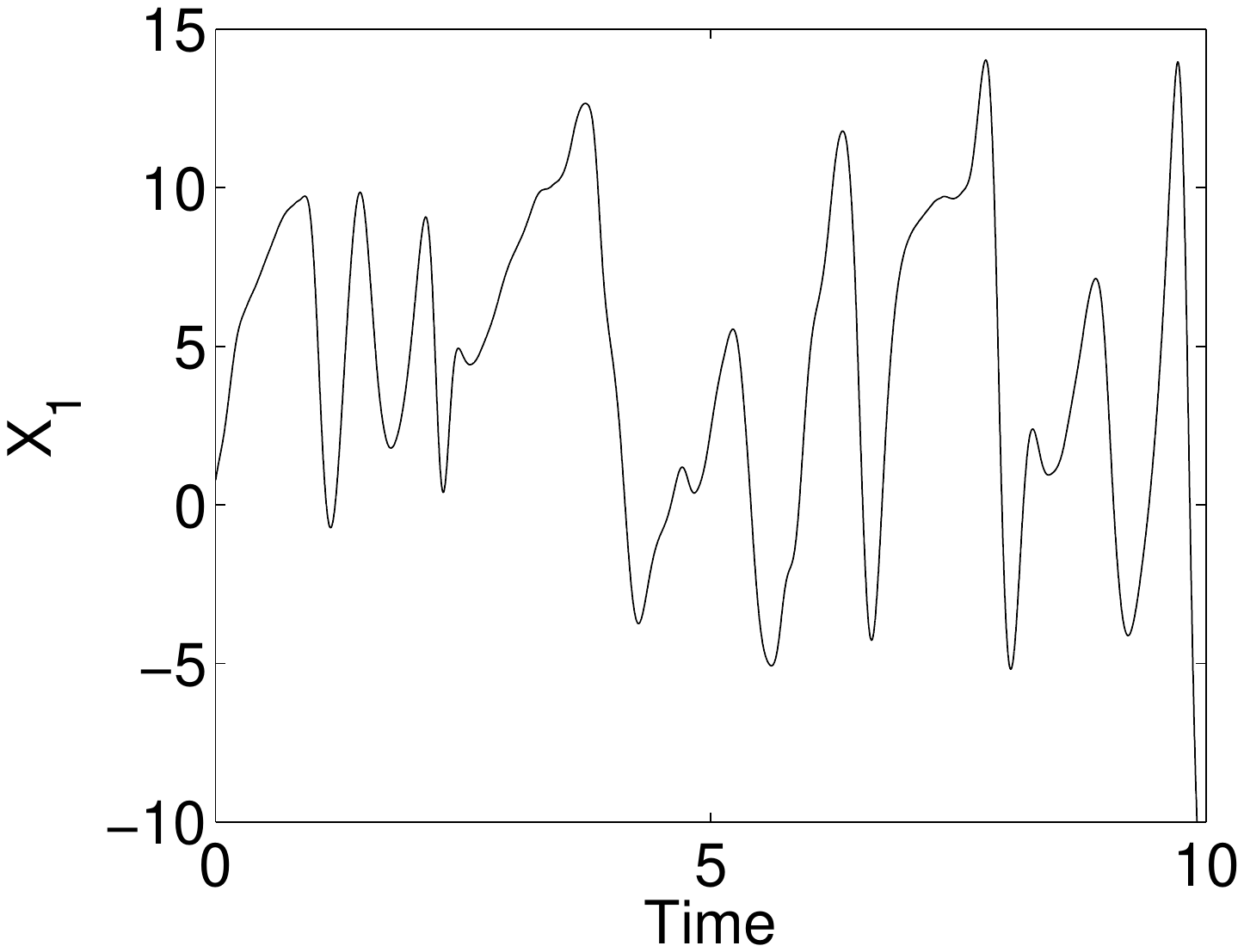}\put(7,25){\colorbox{white}{\fbox{D}}}\end{overpic}\\
	\hspace{-.5cm}\begin{overpic}[scale=.24,trim=3cm 8cm 4cm 8cm,clip]{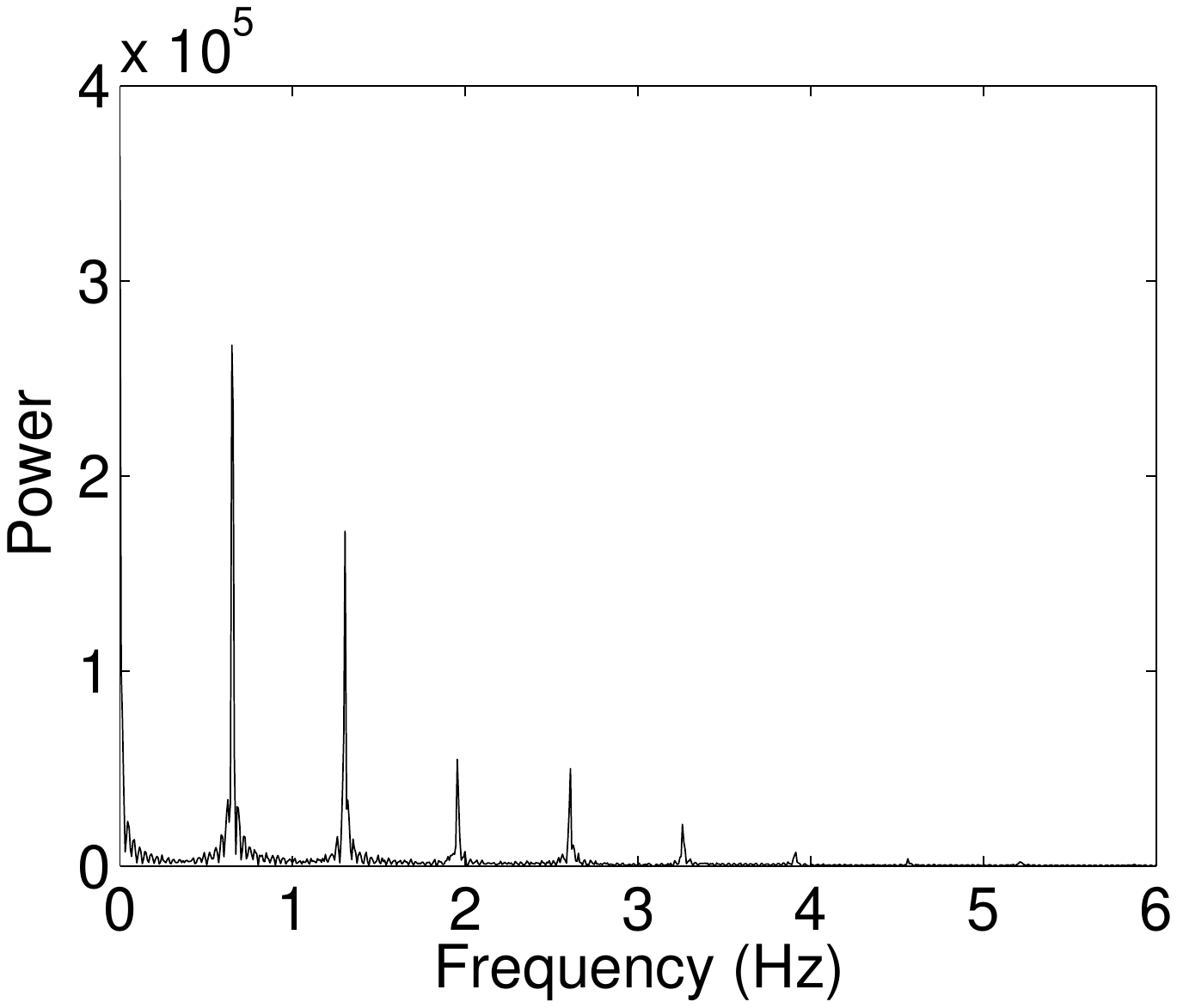}\put(5,21){\colorbox{white}{\fbox{E}}}\end{overpic}&
	\begin{overpic}[scale=.24,trim=3cm 8cm 4cm 8cm,clip]{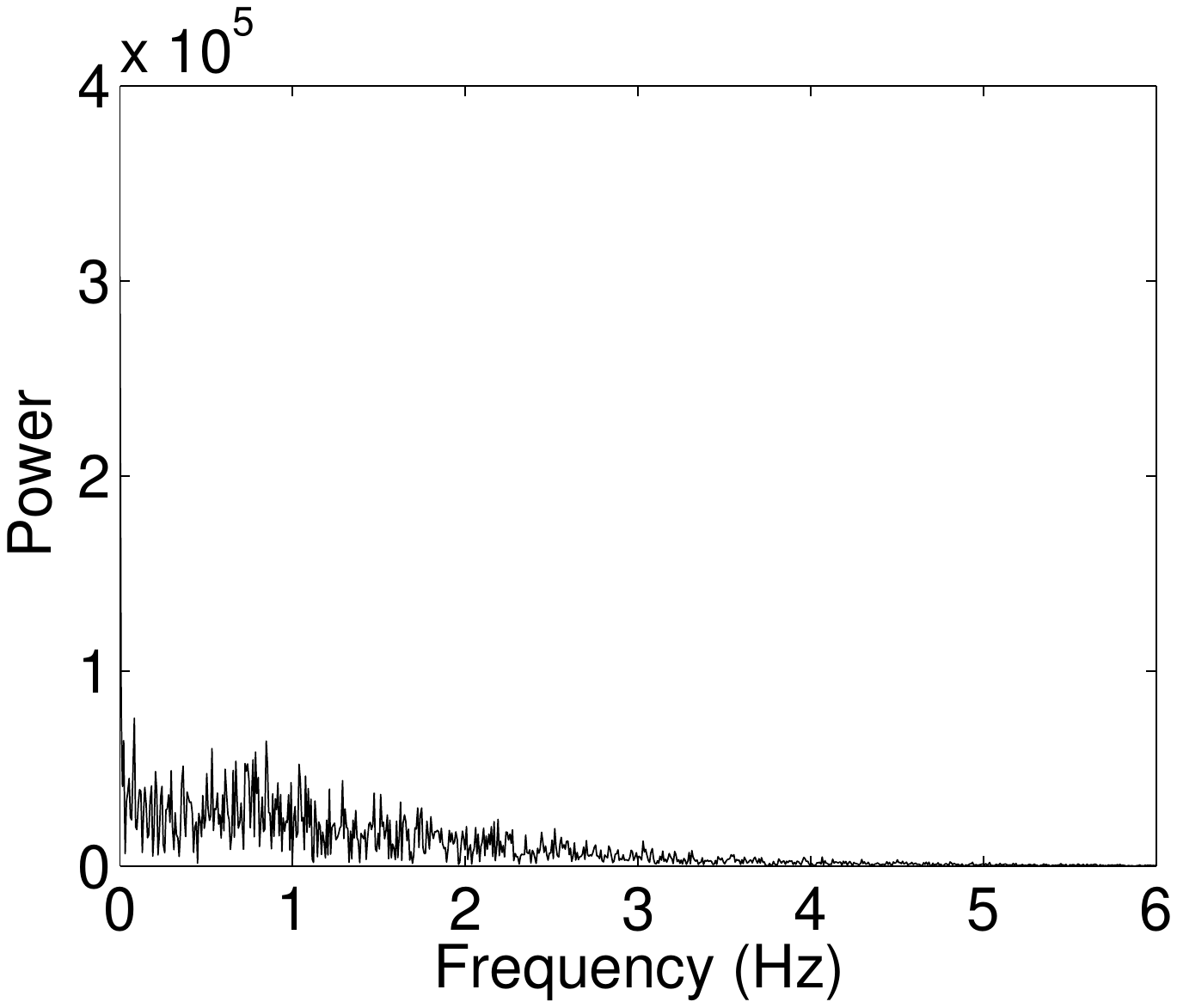}\put(5,21){\colorbox{white}{\fbox{F}}}\end{overpic}\\
        \begin{overpic}[scale=.2,trim=1cm 6cm 1cm 7cm,clip]{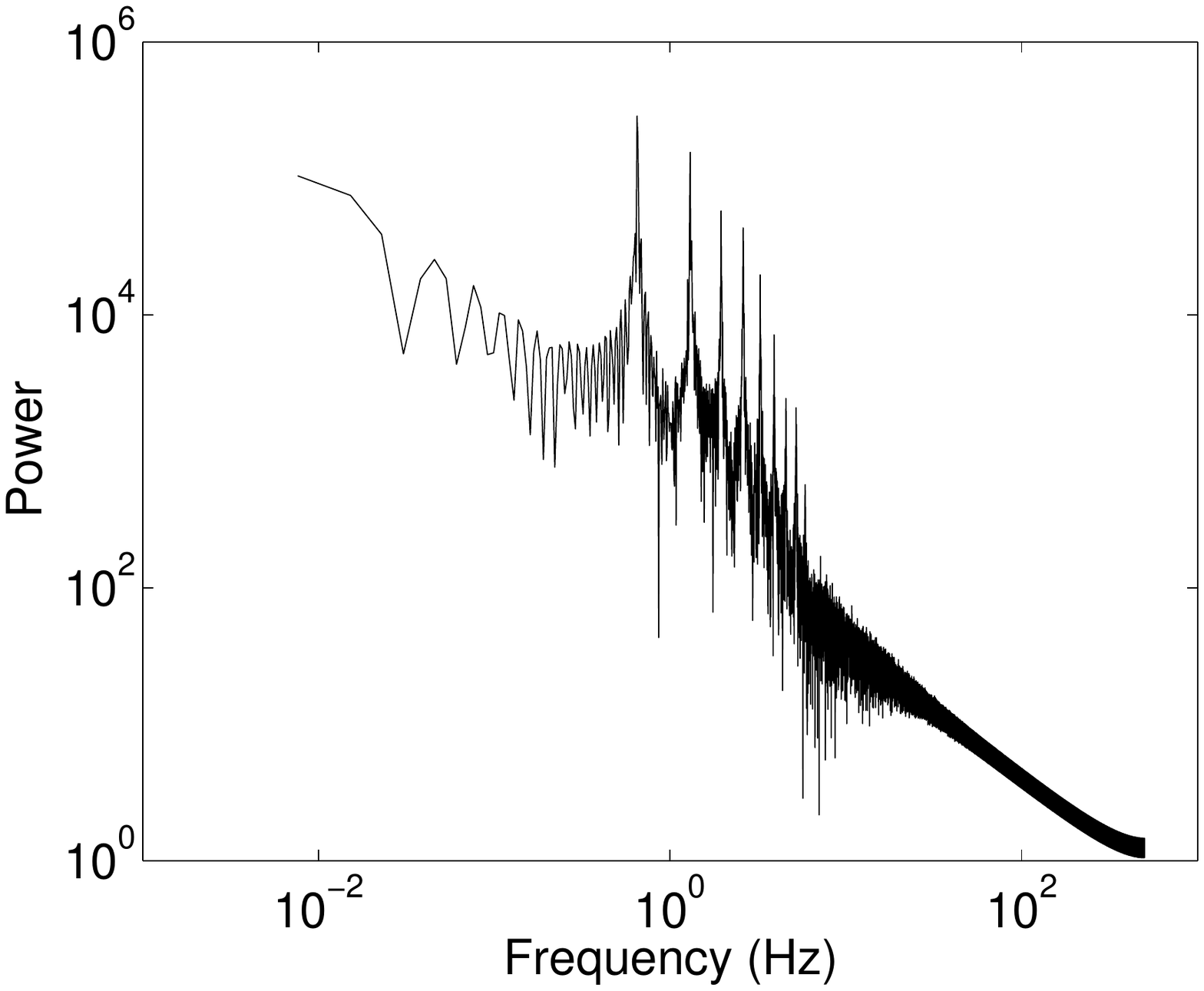}\put(5,15){\colorbox{white}{\fbox{\small G}}}\end{overpic}&
        \begin{overpic}[scale=.2,trim=1cm 6cm 1cm 7cm,clip]{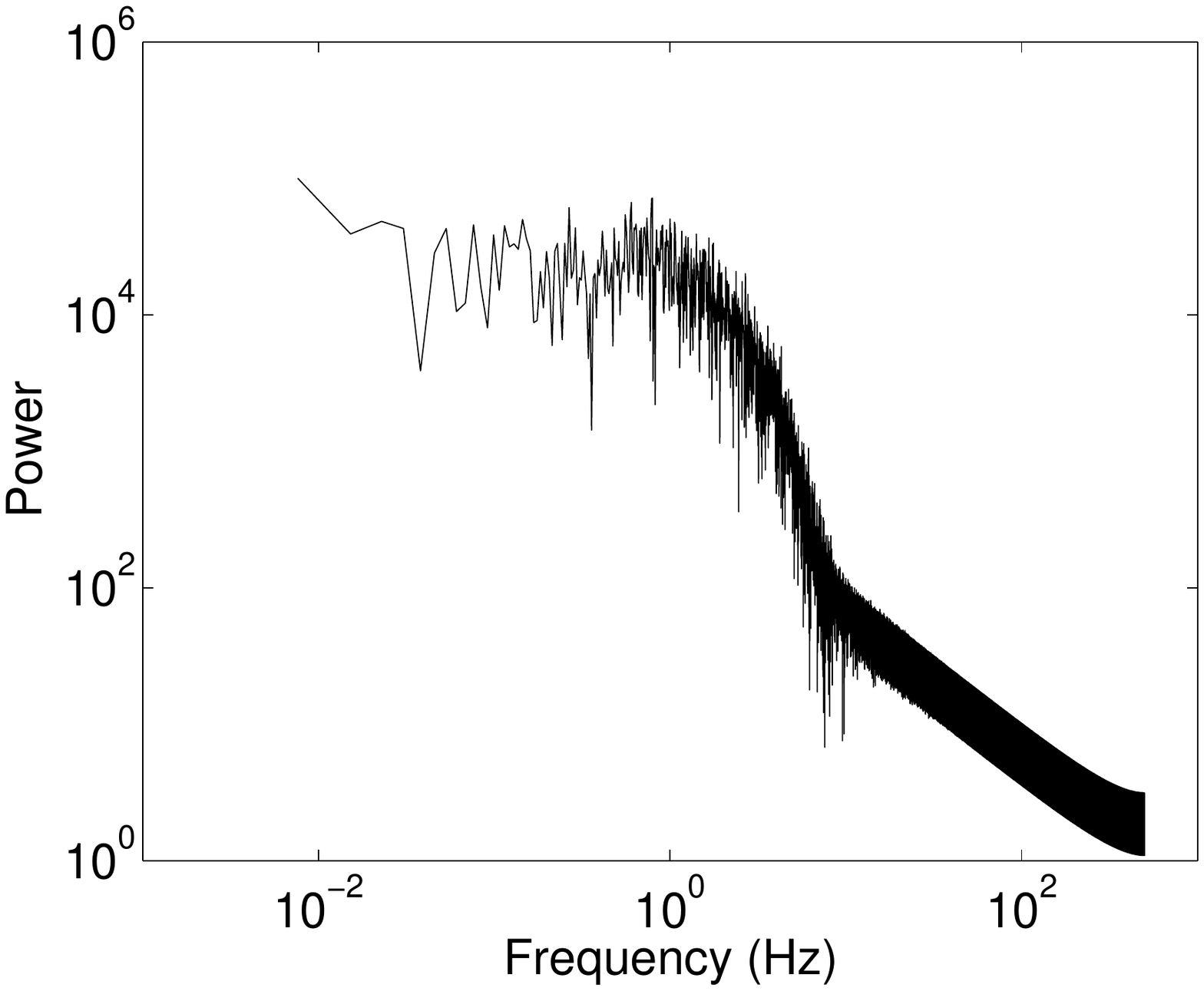}\put(5,15){\colorbox{white}{\fbox{\small H}}}\end{overpic}
	\end{array}$
	\caption{Two example trajectories of the Lorenz '96 model, along with periodograms for the corresponding trajectories of the slow oscillators. (A, C, E, \& G) $I=4$, $J=8$, and $F=14$. We observe a fairly regular trajectory. The periodogram for this system supports this by showing that only a few isolated frequencies have significant power. (B, D, F, \& H) $I=10$, $J=5$, and $F=14$. We observe an irregular trajectory. The periodogram exhibits some power at many frequencies.}
	\label{fig2}
\end{figure}

 \indent We examine the Lorenz '96 model for forcings $F\in[1,18]$, and integer spatial dimensions $I\in[4,50]$ and $J\in[0,50]$. For each choice of $F$, $I$, and $J$, we integrate the Lorenz '96 model with a randomly selected initial condition in the basin of attraction for the system attractor. We use the Runge-Kutta method of order-4 \cite{england00} with a time step of .001 to integrate the initial point along its trajectory. Initially, we iterate the point 500 time units without performing any analysis so that the trajectory is allowed to approach the attractor; thus transient activity is ignored. From here, we integrate an additional 500 time units for analysis. Results were insensitive to increases in integration time, specific choices of initial condition, and decreases in time step size. We show examples of stable and chaotic trajectories in Fig. (\ref{fig2}).\\
 \indent We use the largest Lyapunov exponent, the percentage of positive Lyapunov exponents, and the normalized Lyapunov dimension to characterize the nonlinearity of the system.  We approximate the Lyapunov exponent for the
$i$th \\
 % It is common to think of the Lorenz '96 model as an $I$-dimensional model where the trajectory of each slow 
%%%%%%%%%%%%
%%%%%%%%%%%%
\noindent dimension of the slow modes $X$ along the trajectory $\vec{v}$ as
 \begin{equation}
 	L_{i}(\vec{v})\approx\frac{1}{\Delta \text{time}_{total}}\displaystyle\sum_{n=1}^{N} \ln(|f(\vec{v}_{i}^{(n)})|)
 \end{equation}
where $N$ is the number of iterations, $\vec{v}_{i}^{(n)}$ is the $i$th coordinate of the trajectory at the $n$th iterate, $\Delta \text{time}_{total}$ is the total model integration time, and $f$ is the stretch factor measured from the trajectories of an $I$-dimensional ensemble near a point on the trajectory over a unit time step. This calculation can be thought of as an average of the natural-log of the stretching/shrinking dynamics of the system acting on an ensemble of points very near to the trajectory over time. The Lyapunov dimension is given by
\begin{equation}
	L=D+\frac{1}{|L_{D+1}(\vec{v})|}\displaystyle\sum_{d=1}^{D}L_{d}(\vec{v})
\end{equation}
where $D$ is the largest whole number such that $\sum_{d=1}^{D}L_{d}(\vec{v})\geq0$. This calculation yields an approximation of the slow mode attractor fractal dimension. In general, the fractal dimension compared to the number of slow mode dimensions in the model (namely $I$) provides a reasonable measure of the nonlinearity in the system which we can subsequently compare to the dynamics resulting from different parameter choices \cite{kaplan00}.\\
\indent From the one-dimensional time series in Fig. (\ref{fig2}), we see examples of the dynamics exhibited by the slow variables $X$ ($x_1$ is a representative example of $X$). We also measure nonlinearity by looking at the frequency spectrum for the trajectories of individual slow oscillators. Given a time series, the frequency spectrum can be approximated using the fourier transform \cite{orrell00,orrell01}. Chaotic systems typically exhibit power at a large number of frequencies, while stable systems will exhibit power at only a small number of frequencies. Furthermore, the frequency spectrum illuminates which frequencies the $X$ variables will tend to exhibit.\\
\indent Lorenz suggested that the slow oscillators represent measurements of some atmospheric quantity about a given latitude \cite{lorenz03}. With this in mind, it is meaningful to visualize the system accordingly. Different from the images provided in Fig. (\ref{fig2}), we will visualize states for all of the slow oscillators during a given trajectory as points evenly spaced around a circle centered at the origin, where the origin represents the lowest value ($x_{\min}$) obtained by any of the slow oscillators along their respective trajectories. Each point's distance from the origin is given by $x_i$'s current value minus $x_{\min}$. Treating the points in polar-coordinates $(r,\theta)$, where $r$ is the oscillator's distance from $x_{\min}$ and $\theta$ indicates the subscript of the oscillator, we fit a cubic spline to the shifted slow oscillator values to obtain approximations for the flow of the atmospheric quantity between the slow oscillators. For clarity, the slow oscillators' radial positions ($\theta$) remain fixed, while their distance from the origin varies over the course of the trajectory (see Fig. 1B). Note that this method of visualization allows us to observe all of the slow oscillators at once for any state on a trajectory. A similar method is performed to represent the activity of the fast oscillators in the same plot (the outer ring).
\section{Results}
\begin{figure}[!t]
	\centering
	$\begin{array}{cccc}
		\hspace{-.7cm}\includegraphics[width=.28\columnwidth,trim=1cm 5cm .5cm 6cm,clip]{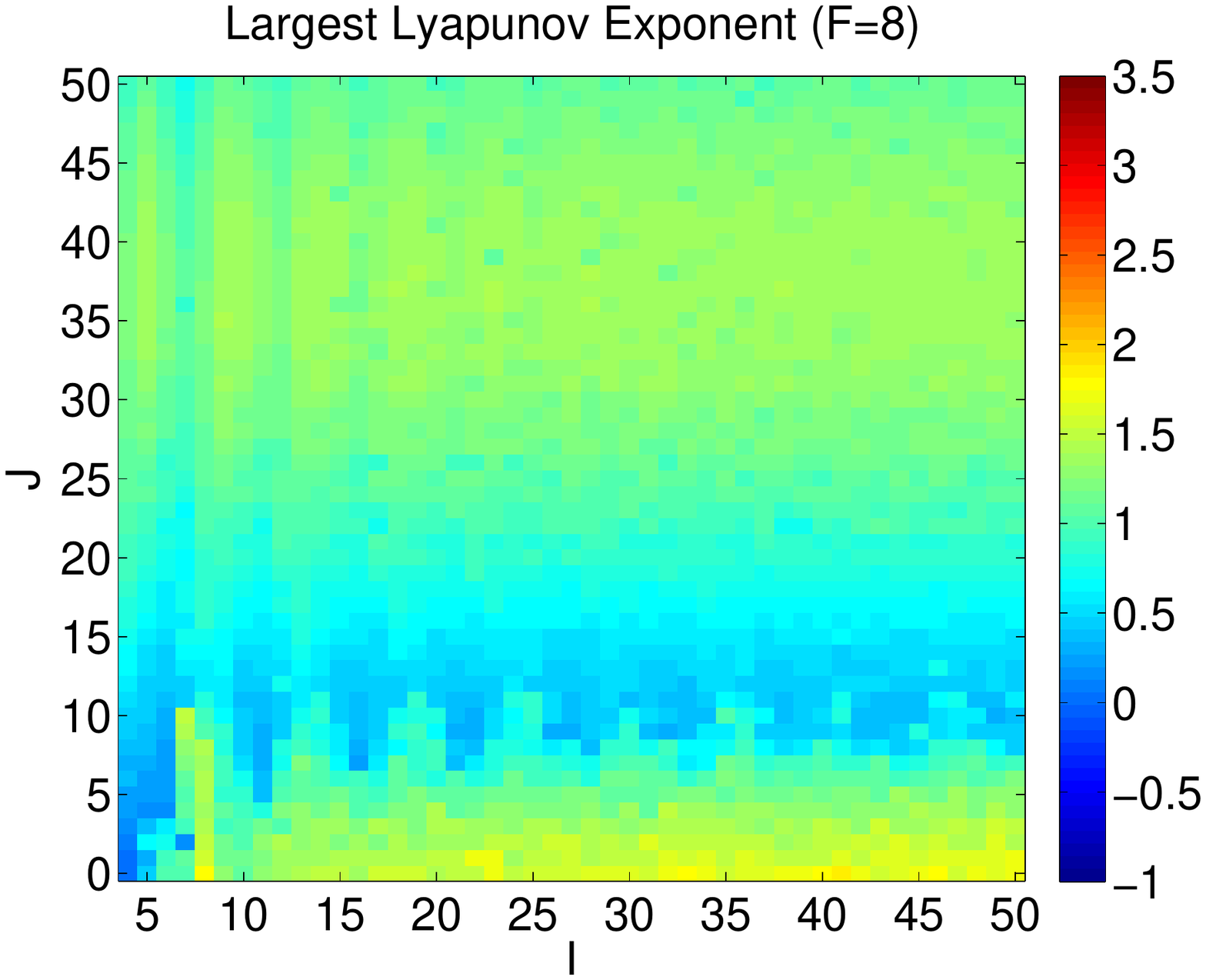}&
		\hspace{-.7cm}\includegraphics[width=.28\columnwidth,trim=1cm 5cm .5cm 6cm,clip]{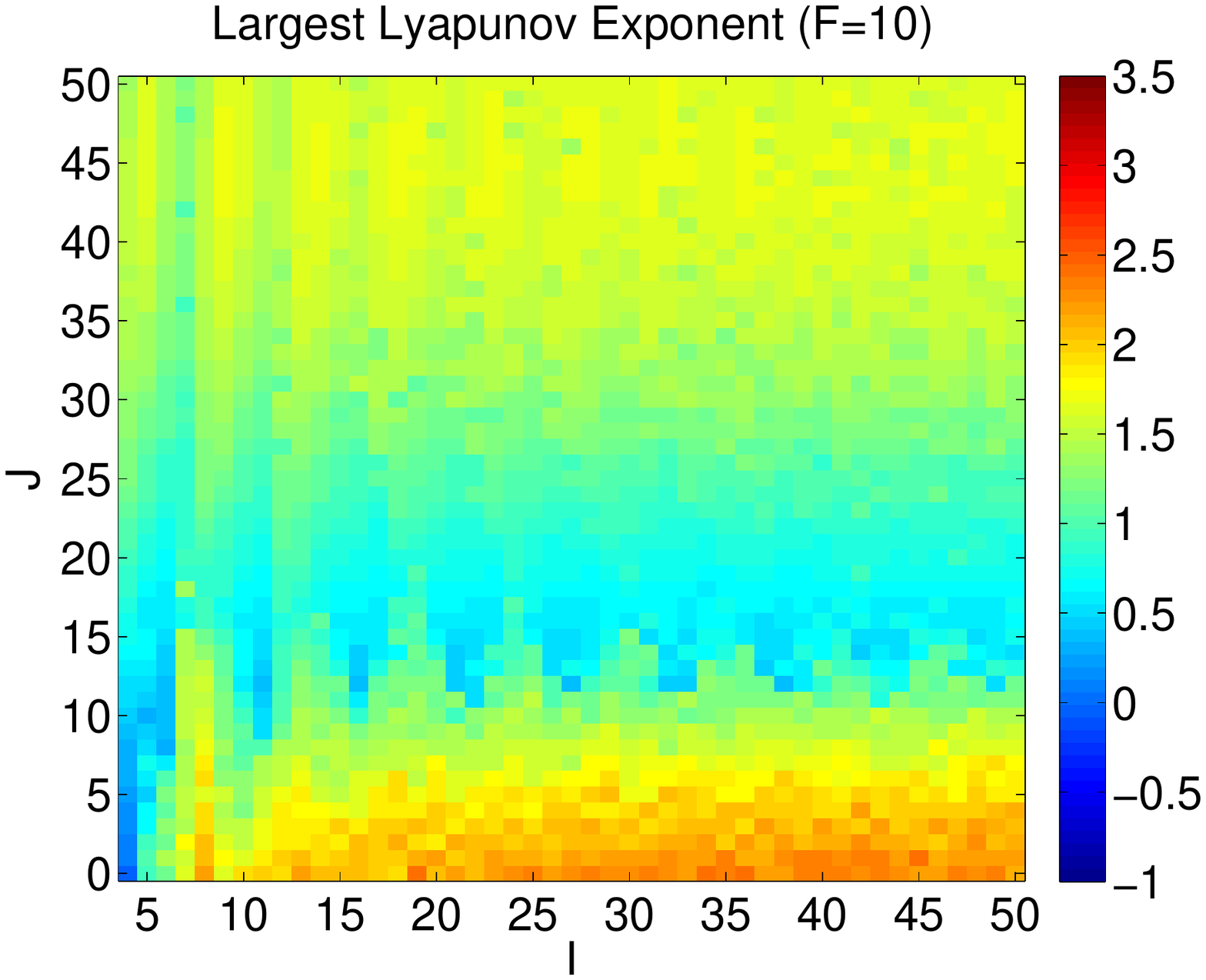}&
		\hspace{-.7cm}\includegraphics[width=.28\columnwidth,trim=1cm 5cm .5cm 6cm,clip]{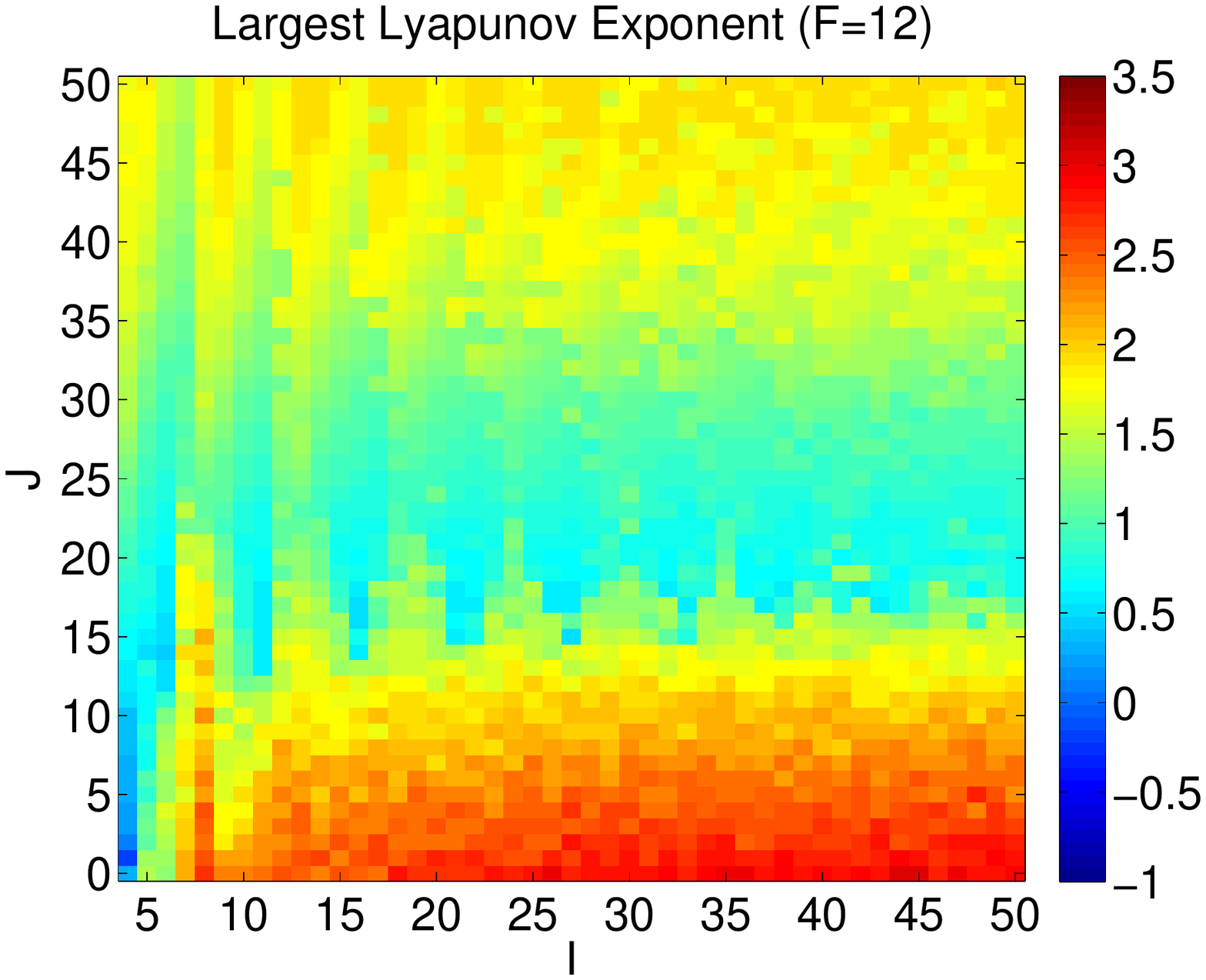}&
		\hspace{-.7cm}\includegraphics[width=.28\columnwidth,trim=1cm 5cm .5cm 6cm,clip]{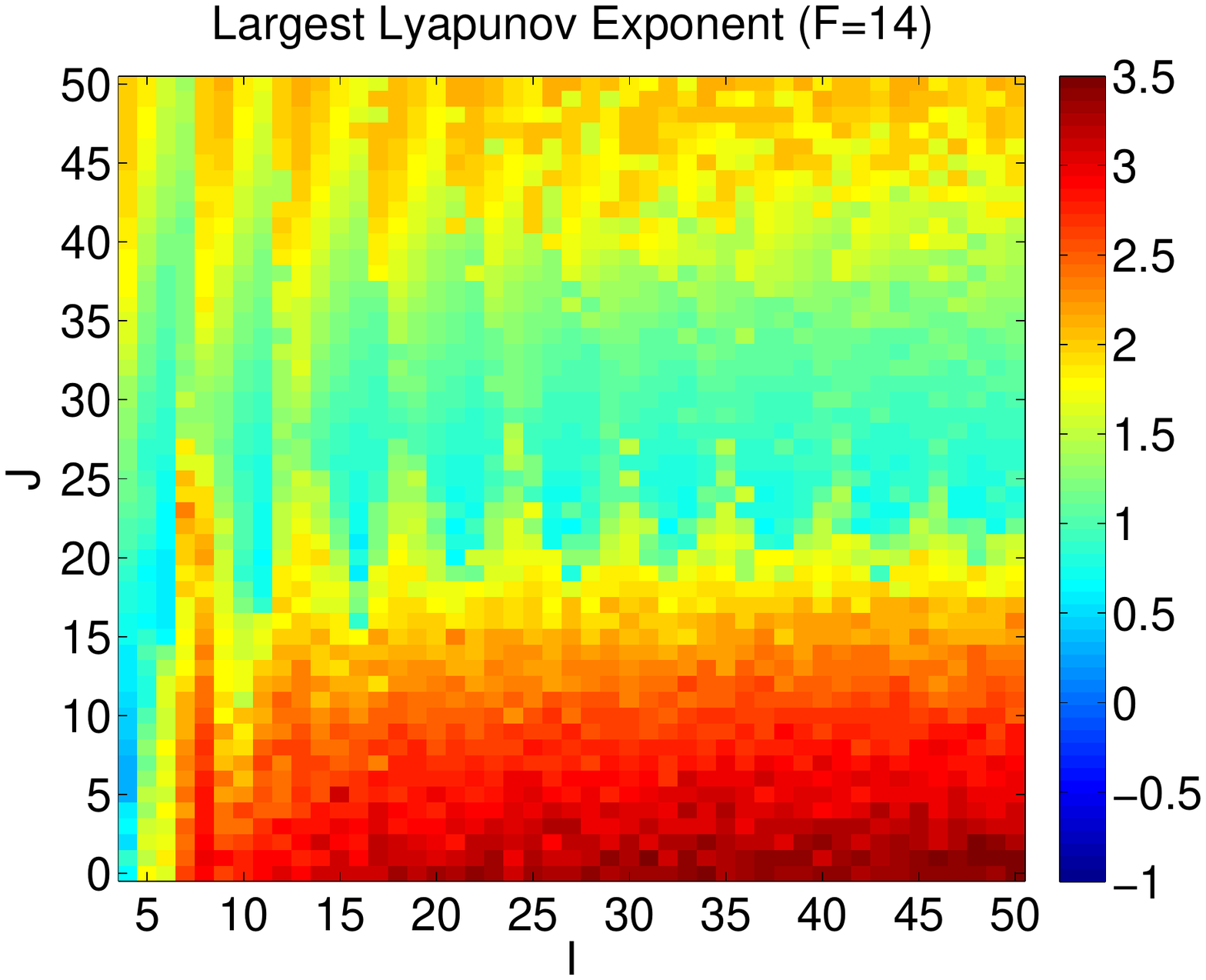}\\
		%%%%%%%%%%%%%%%
		\hspace{-.7cm}\includegraphics[width=.28\columnwidth,trim=1cm 5cm .5cm 6cm,clip]{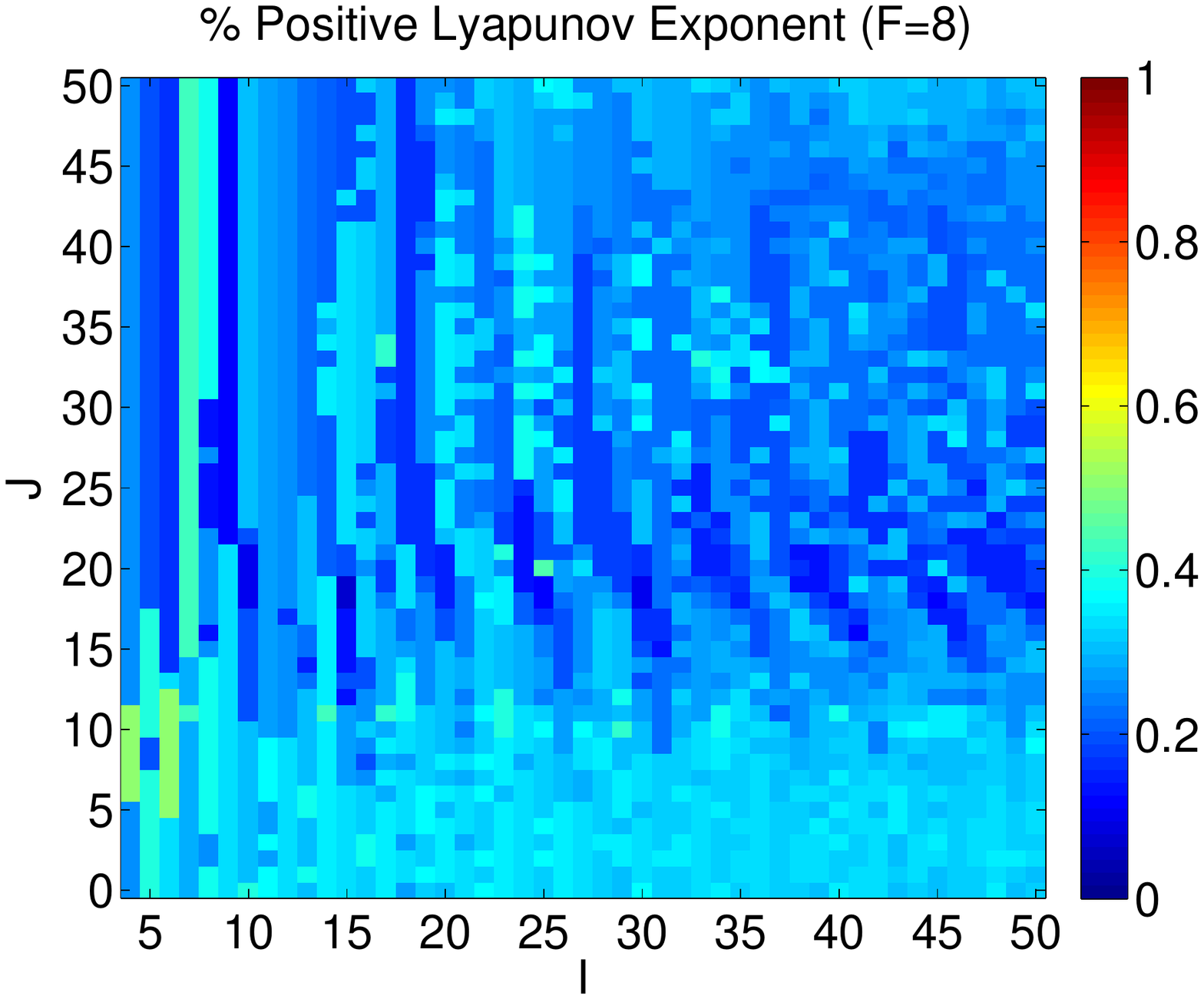}&
		\hspace{-.7cm}\includegraphics[width=.28\columnwidth,trim=1cm 5cm .5cm 6cm,clip]{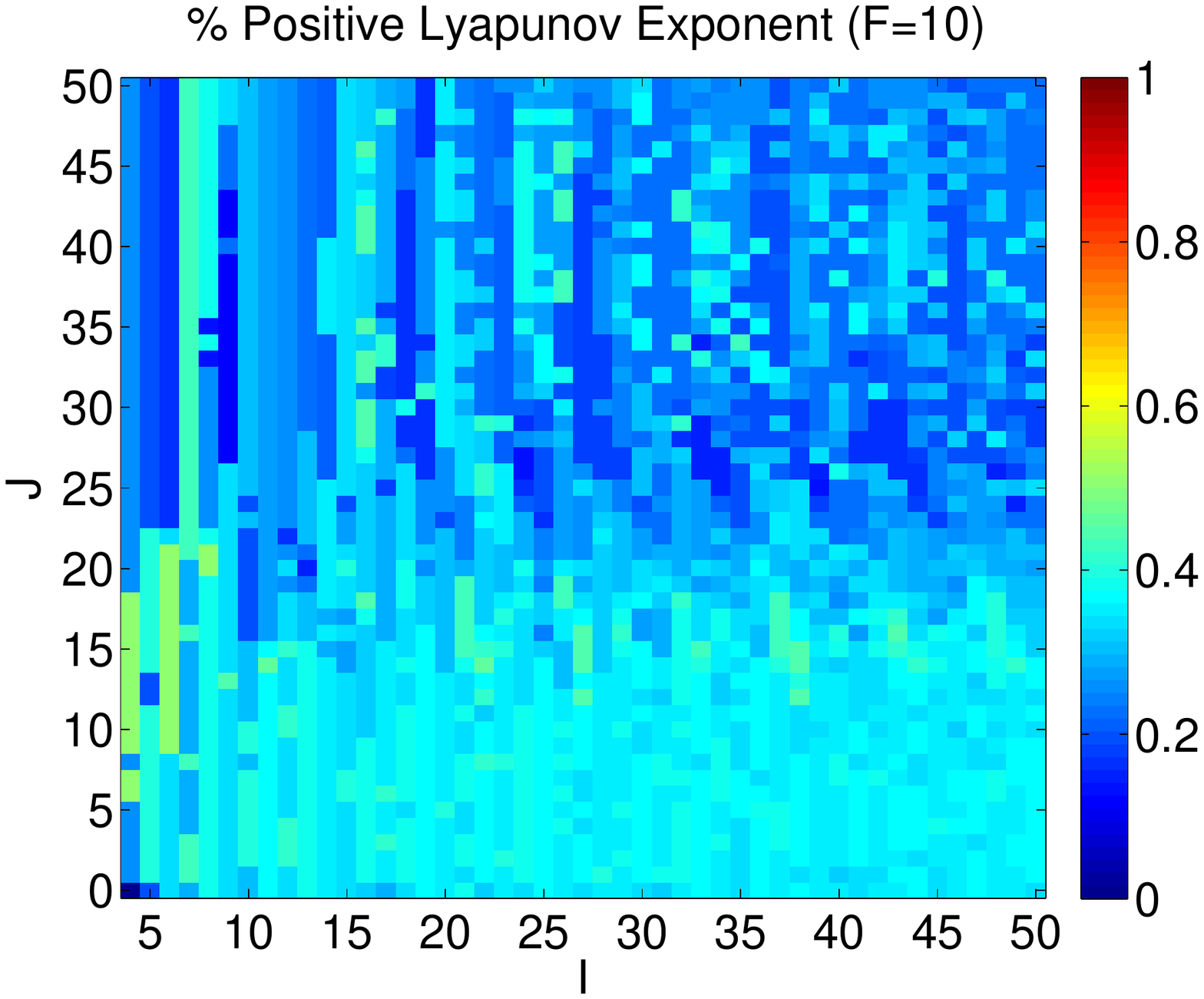}&
		\hspace{-.7cm}\includegraphics[width=.28\columnwidth,trim=1cm 5cm .5cm 6cm,clip]{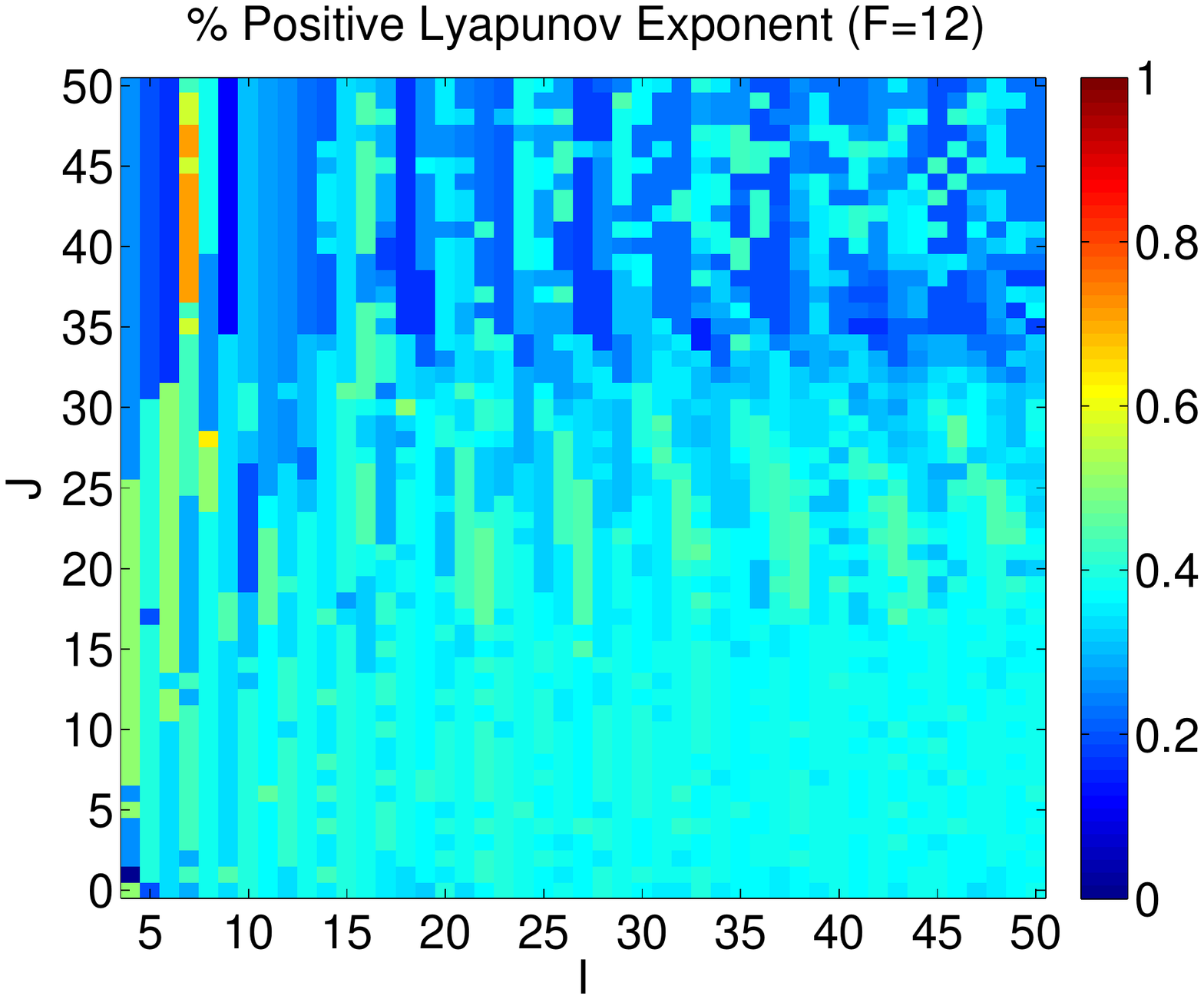}&
		\hspace{-.7cm}\includegraphics[width=.28\columnwidth,trim=1cm 5cm .5cm 6cm,clip]{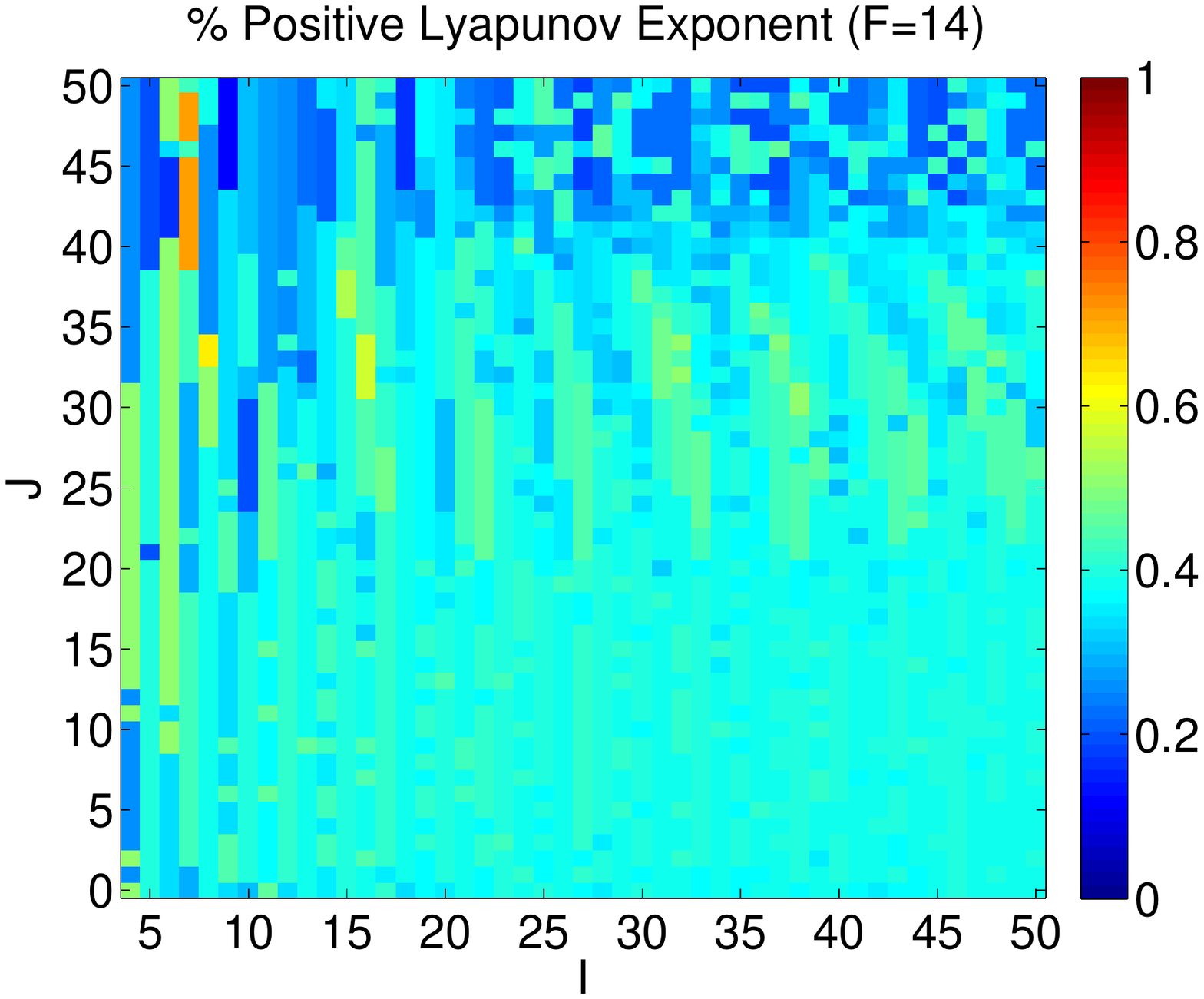}\\
		%%%%%%%%%%%%%%%
		\hspace{-.7cm}\includegraphics[width=.28\columnwidth,trim=1cm 5cm .5cm 6cm,clip]{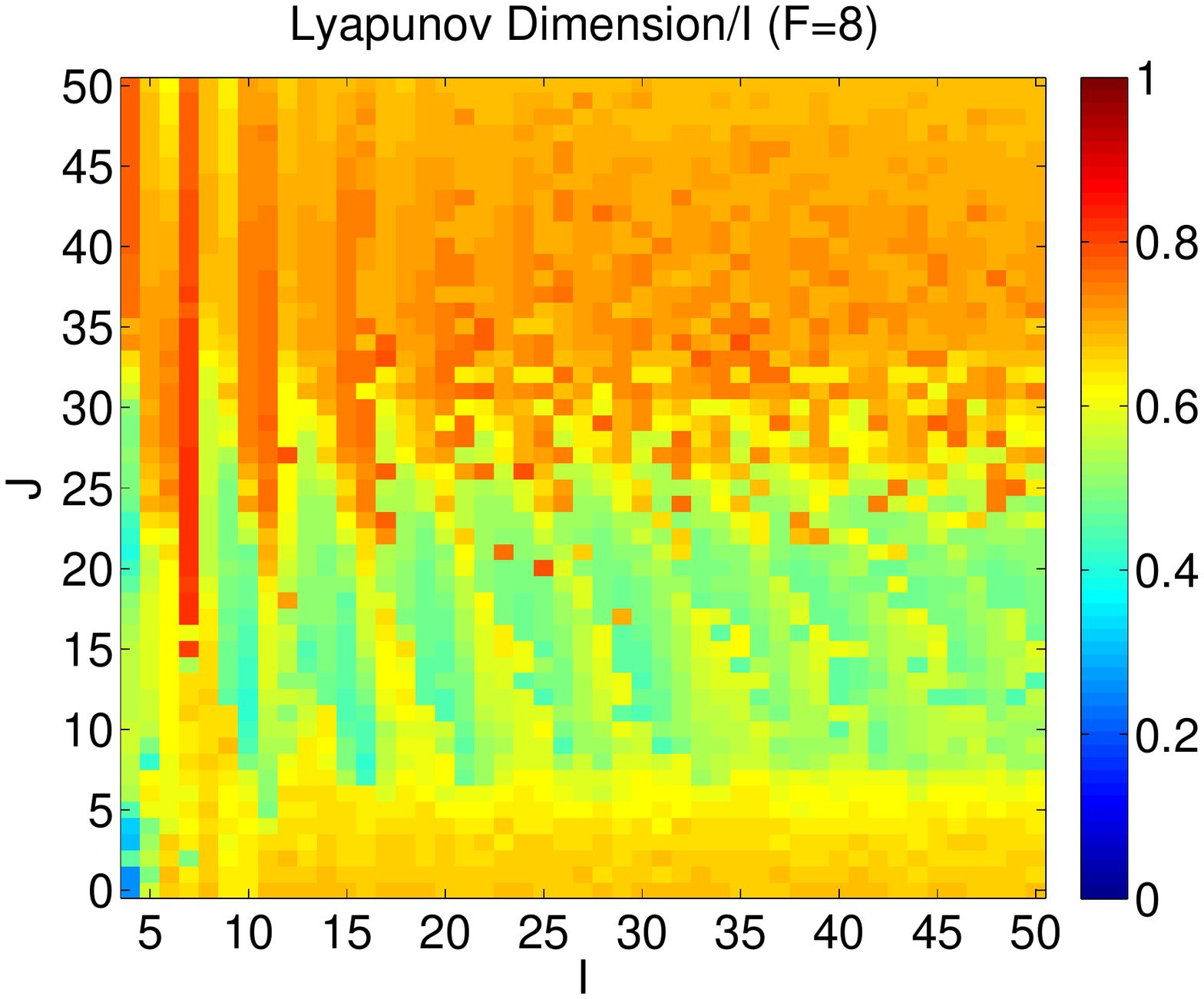}&
		\hspace{-.7cm}\includegraphics[width=.28\columnwidth,trim=1cm 5cm .5cm 6cm,clip]{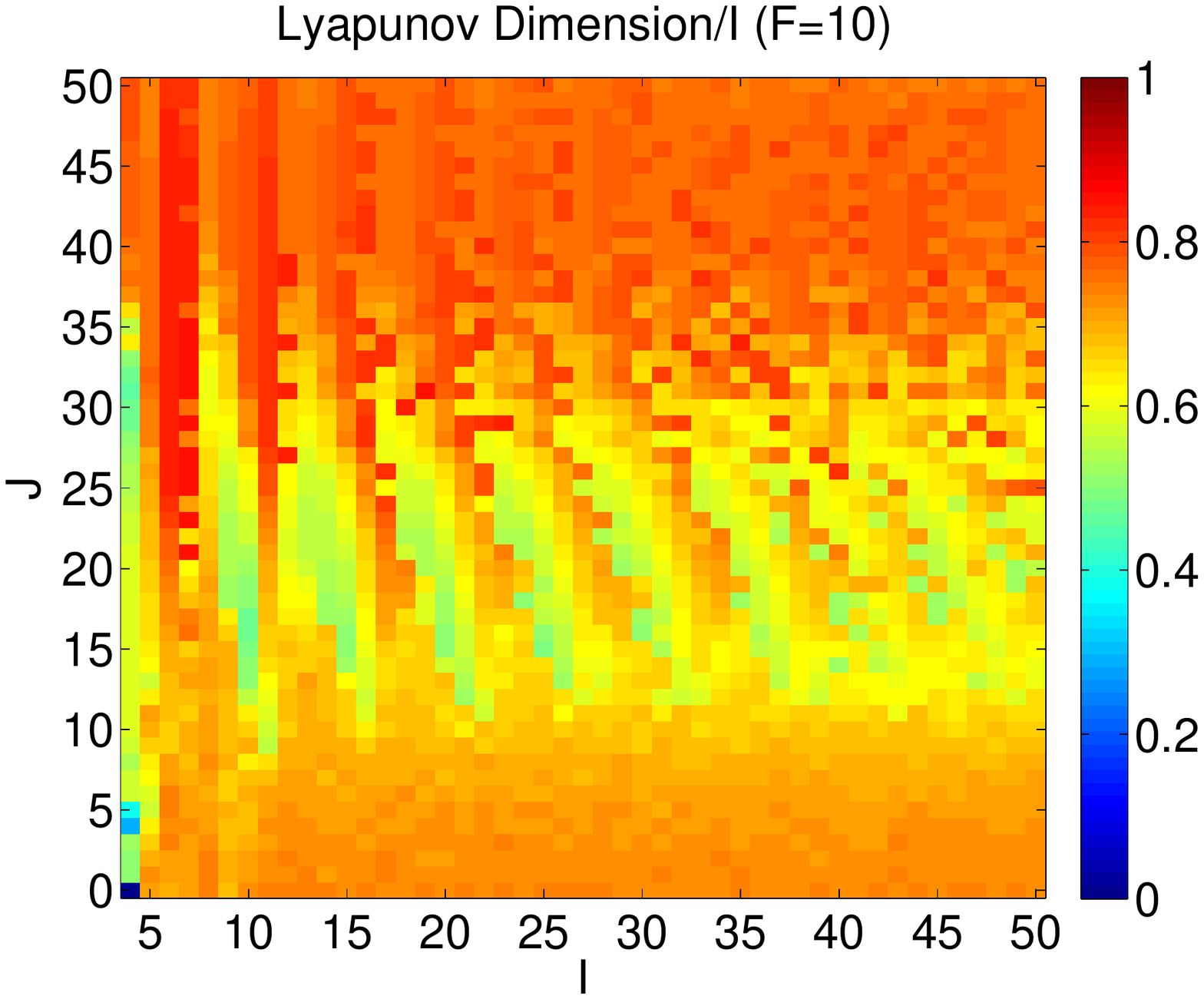}&
		\hspace{-.7cm}\includegraphics[width=.28\columnwidth,trim=1cm 5cm .5cm 6cm,clip]{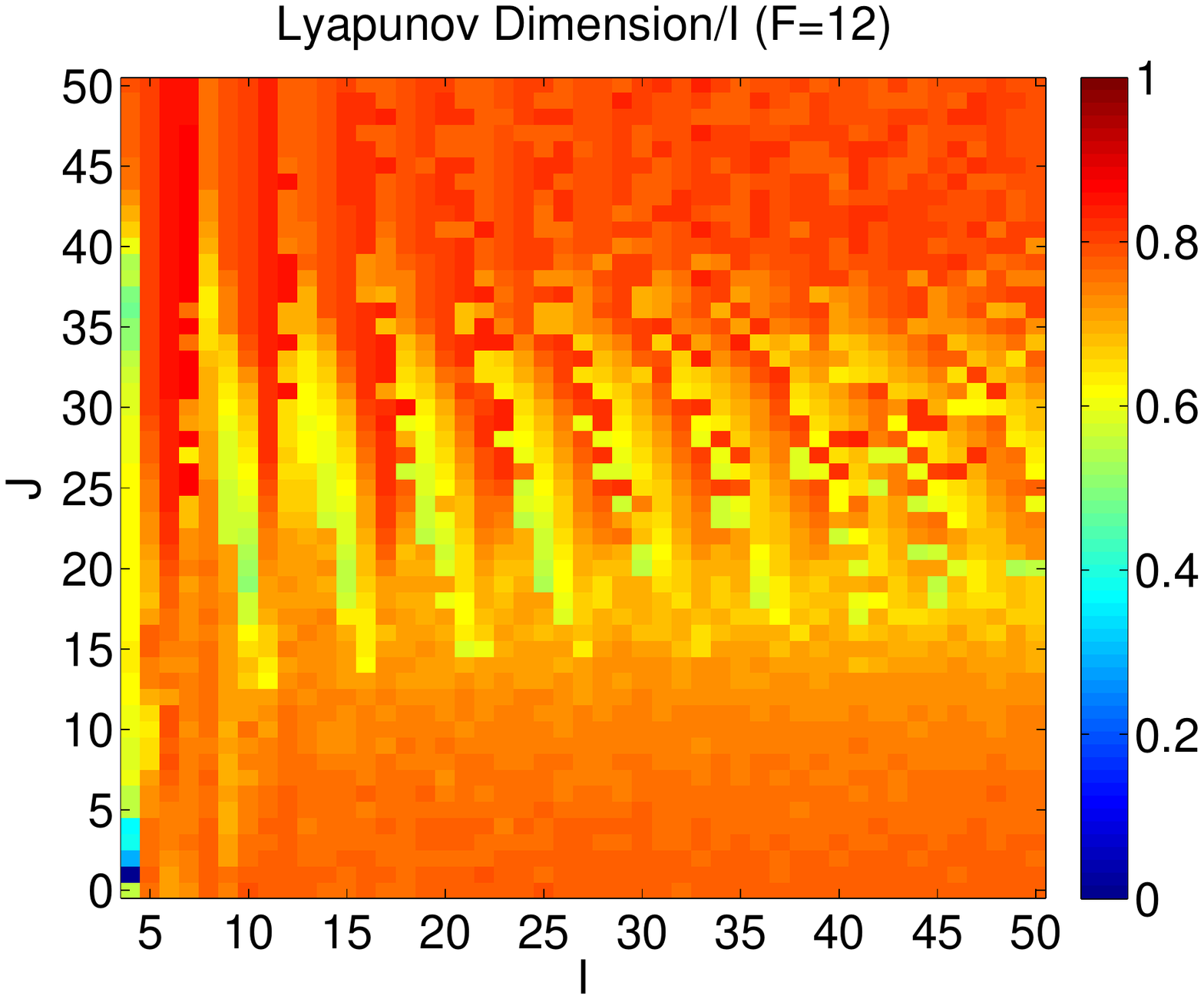}&
		\hspace{-.7cm}\includegraphics[width=.28\columnwidth,trim=1cm 5cm .5cm 6cm,clip]{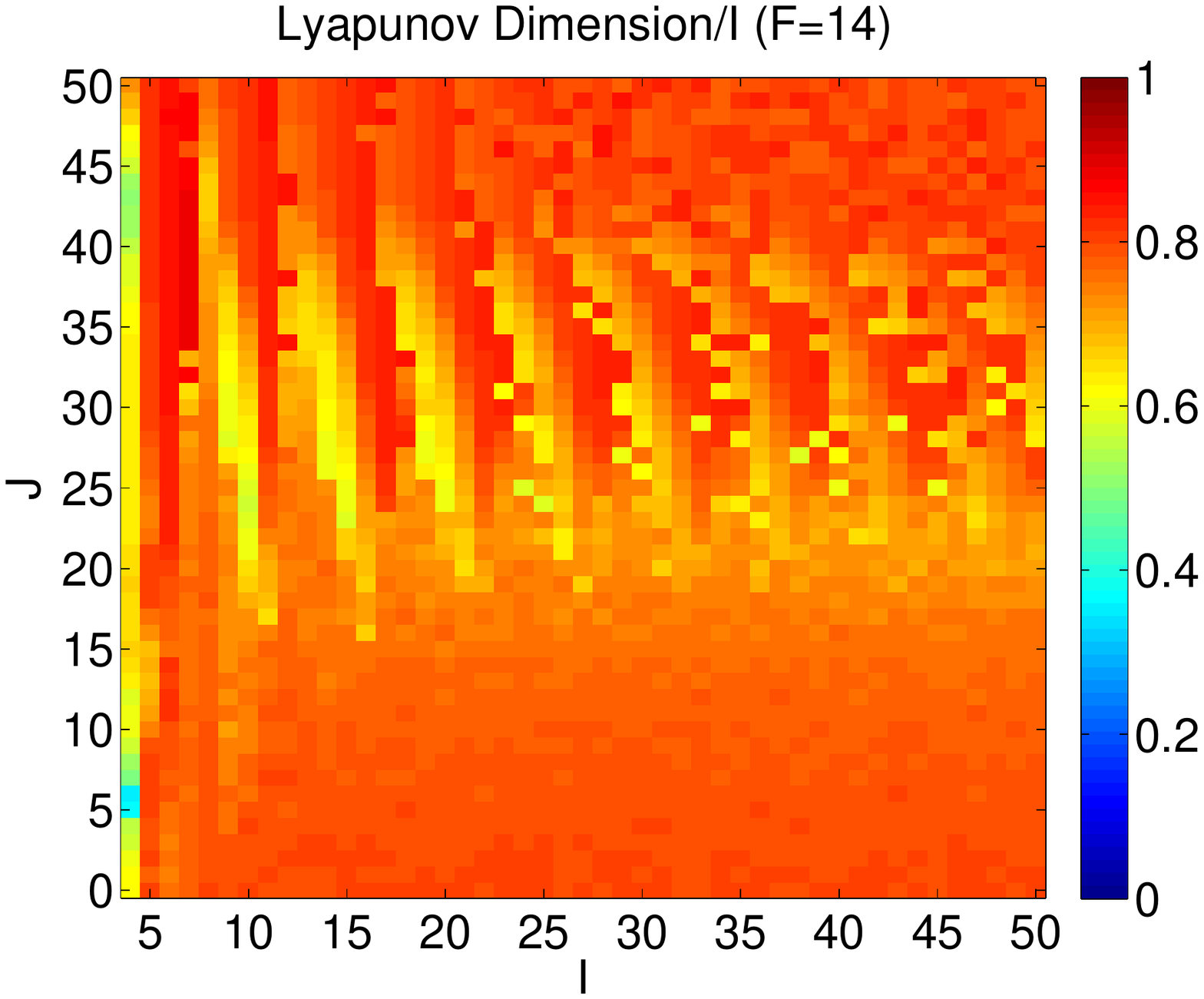}\\
	\end{array}$
	\caption{For these plots, the axes represent integer values of the model dimensions $I$ (slow) and $J$ (fast). Each cell in the resulting plot represents a single integration with 500 unit time steps (or $10^{6}$ iterations) of the Lorenz '96 model. Note that these images are insensitive to changes in initial condition. (Far Left Column) $F=8$. (Center Left Column) $F=10$. (Center Right Column) $F=12$. (Far Right Column) $F=14$. (Top Row) The largest Lyapunov exponent. (Middle Row) The percent of positive Lyapunov exponents. (Bottom Row) The normalized Lyapunov dimension. }
	\label{fig3}
\end{figure}

\begin{figure}[!t]
	\centering
	$\begin{array}{cccc}
		\hspace{-.7cm}\includegraphics[width=.29\columnwidth,trim=1cm 5cm .5cm 6cm,clip]{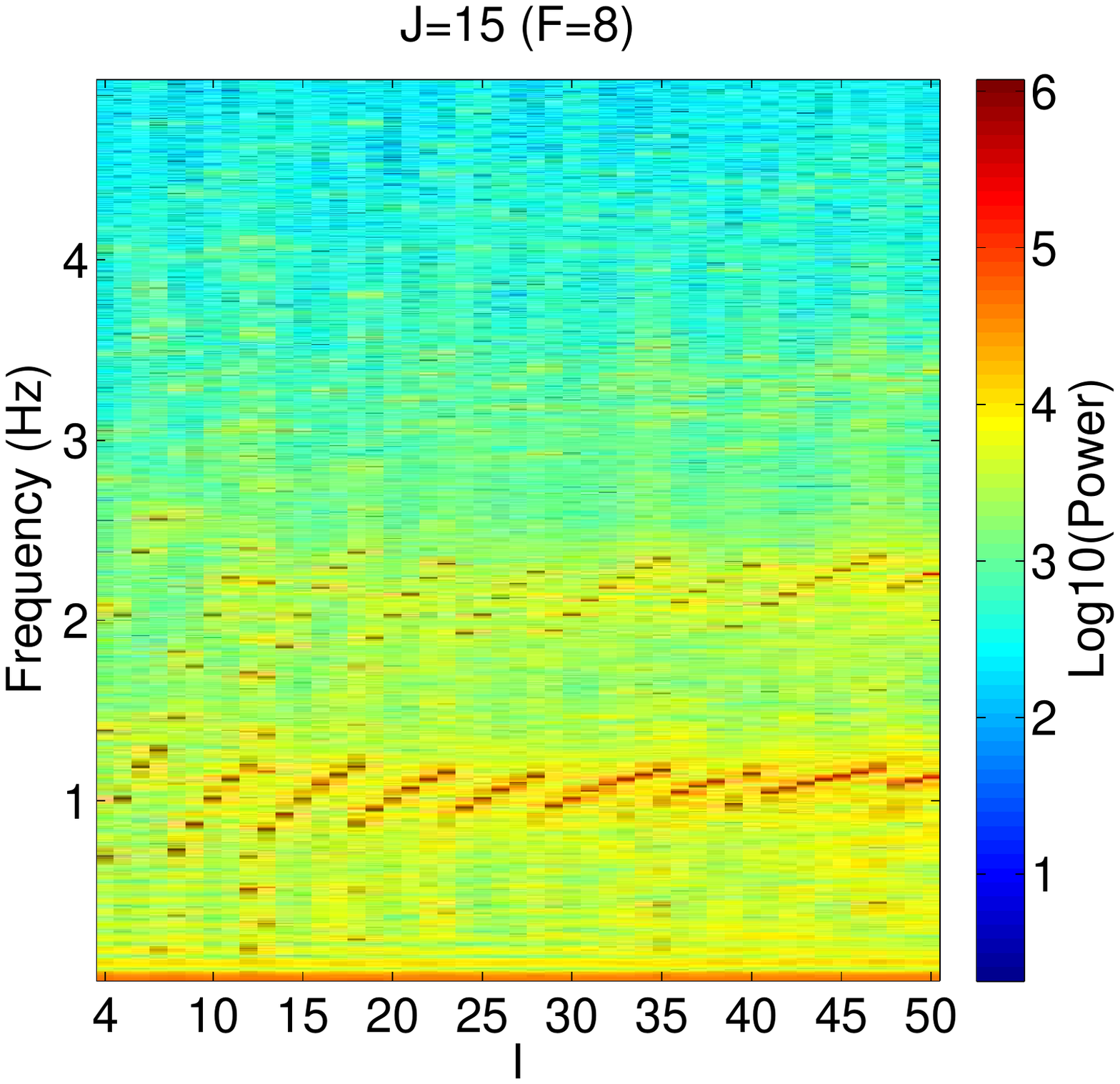}&
		\hspace{-.7cm}\includegraphics[width=.29\columnwidth,trim=1cm 5cm .5cm 6cm,clip]{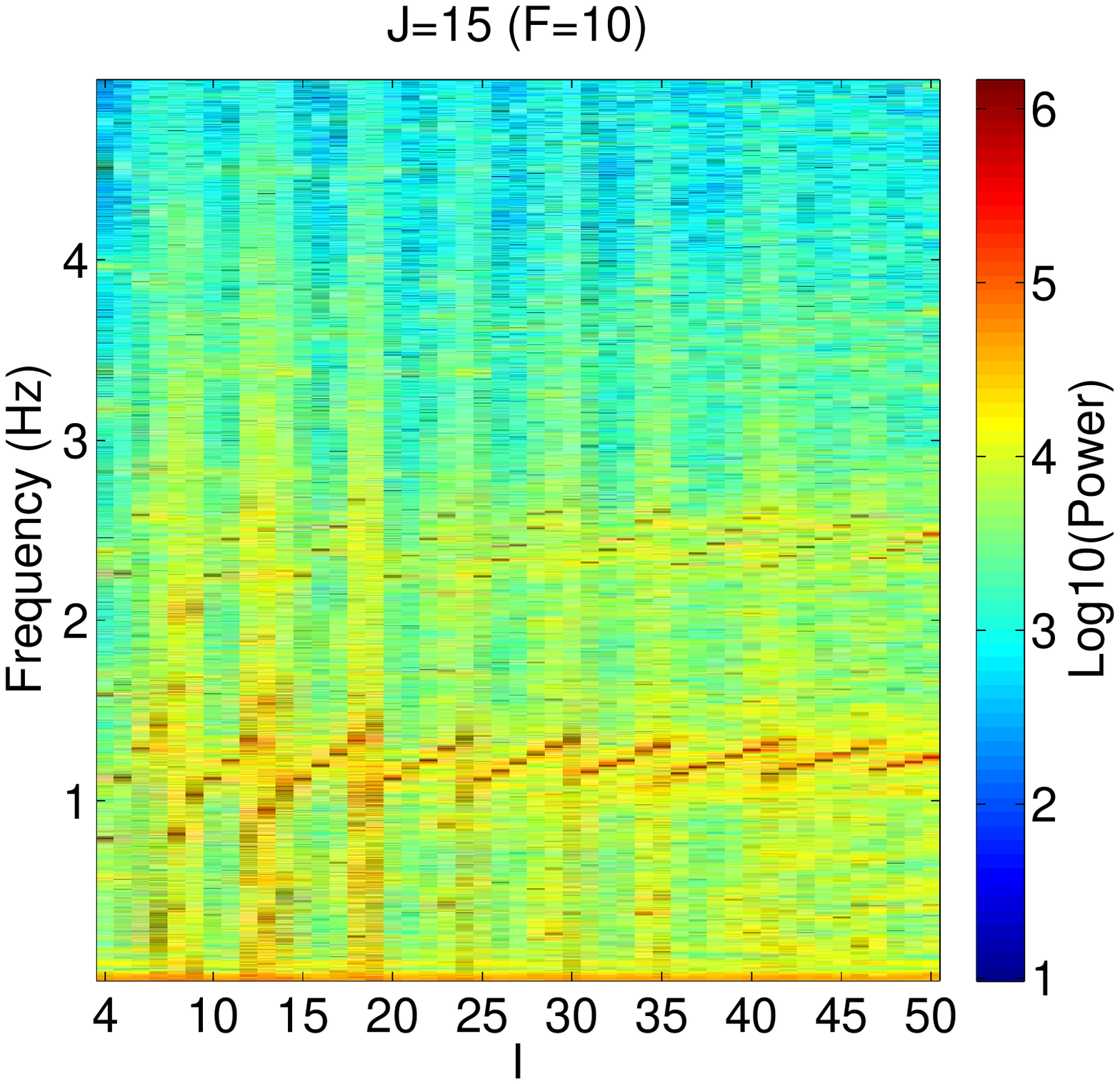}&
		\hspace{-.7cm}\includegraphics[width=.29\columnwidth,trim=1cm 5cm .5cm 6cm,clip]{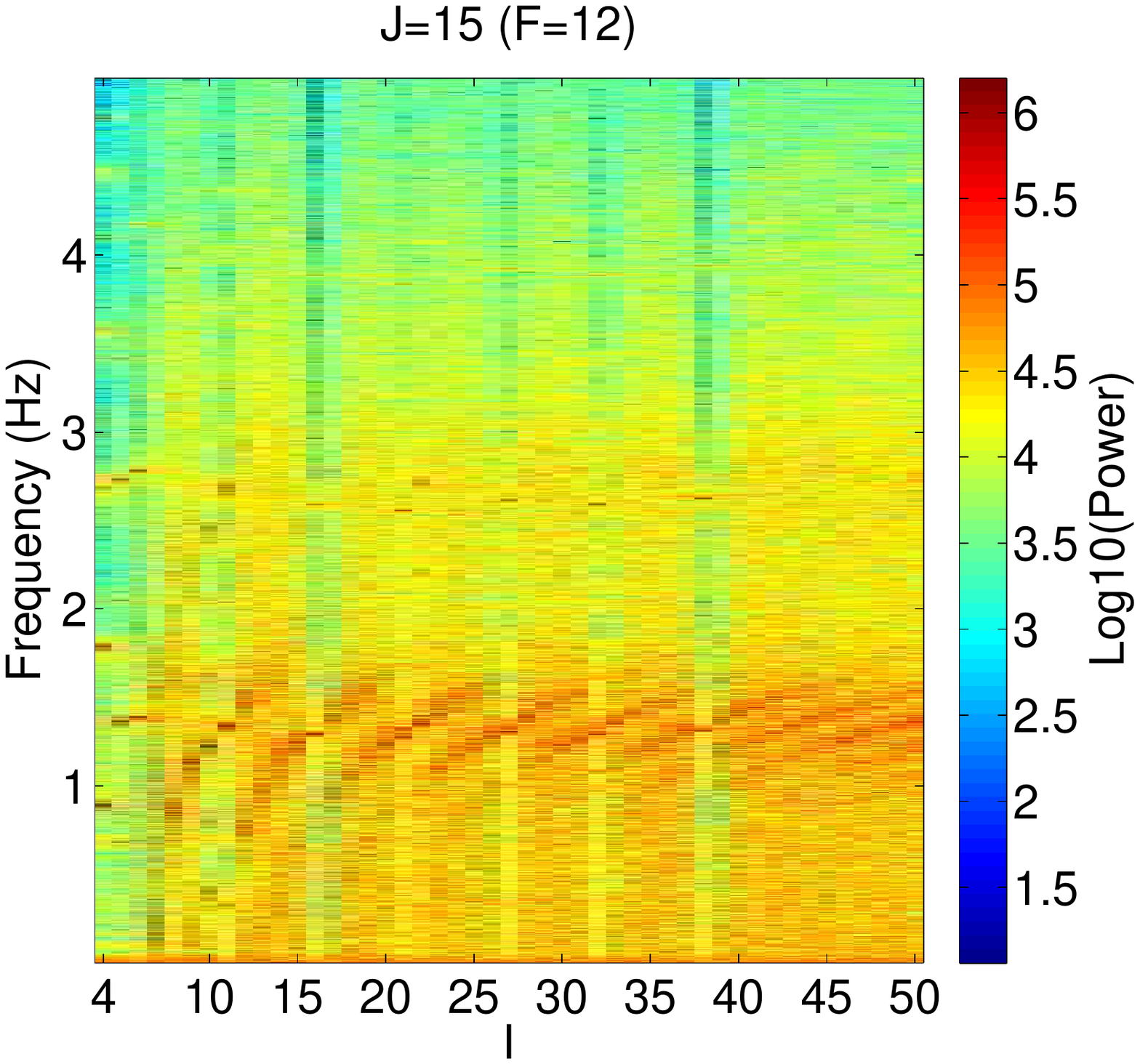}&
		\hspace{-.7cm}\includegraphics[width=.29\columnwidth,trim=1cm 5cm .5cm 6cm,clip]{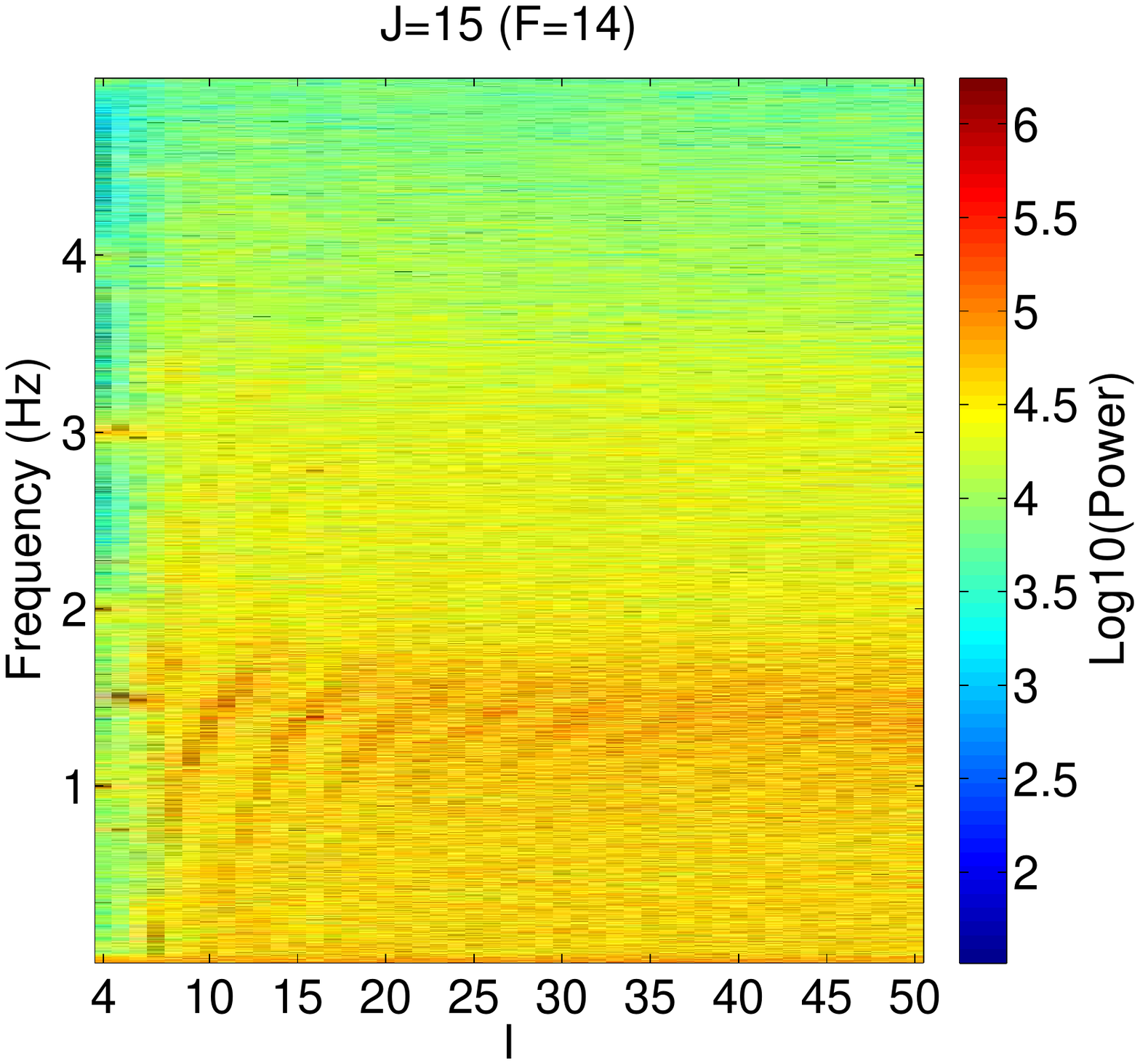}\\
		%%%%%%%%%%%%%%
		\hspace{-.7cm}\includegraphics[width=.29\columnwidth,trim=1cm 5cm .5cm 6cm,clip]{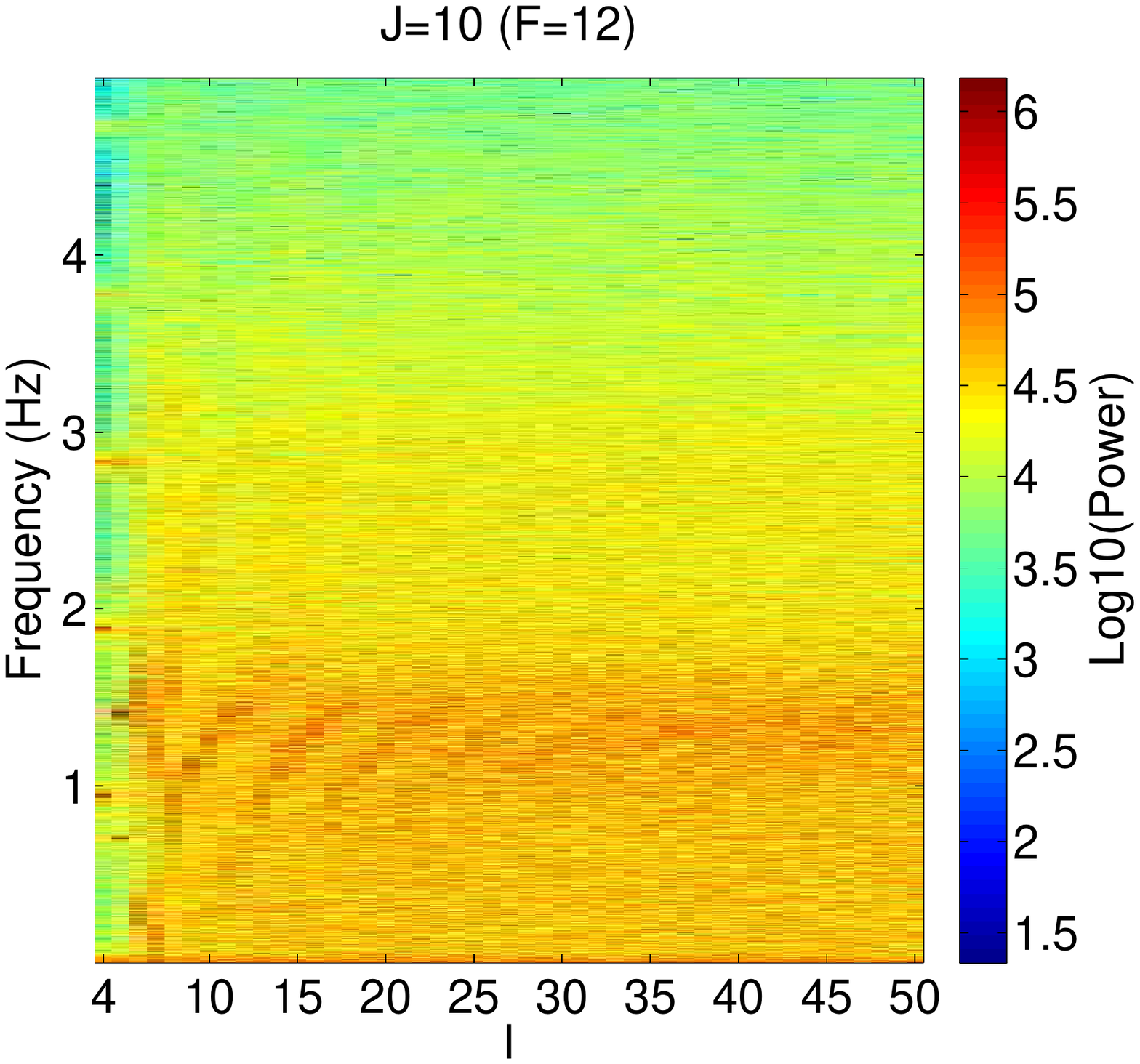}&
		\hspace{-.7cm}\includegraphics[width=.29\columnwidth,trim=1cm 5cm .5cm 6cm,clip]{specMaster_15_12_logtest.pdf}&
		\hspace{-.7cm}\includegraphics[width=.29\columnwidth,trim=1cm 5cm .5cm 6cm,clip]{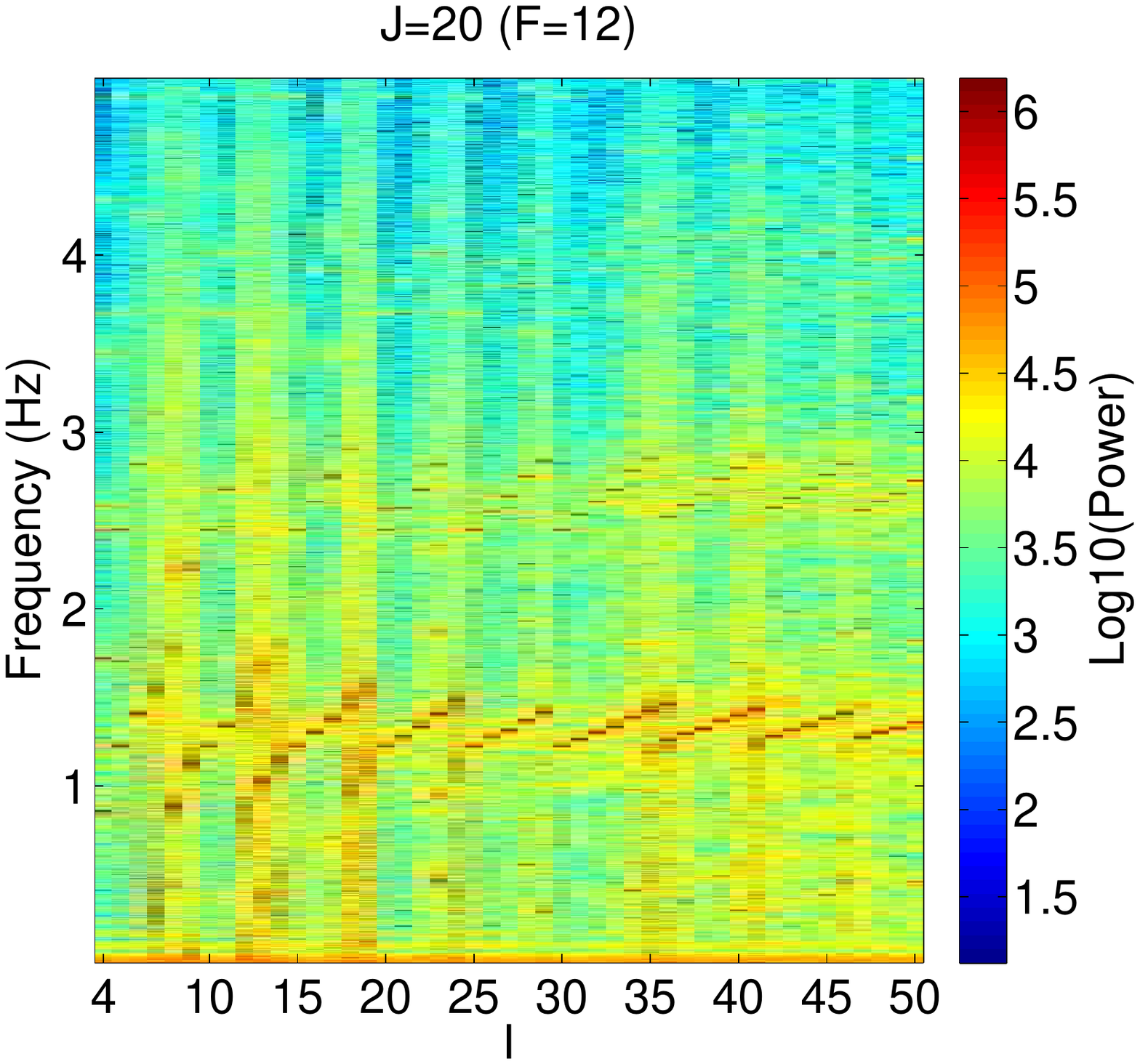}&
		\hspace{-.7cm}\includegraphics[width=.29\columnwidth,trim=1cm 5cm .5cm 6cm,clip]{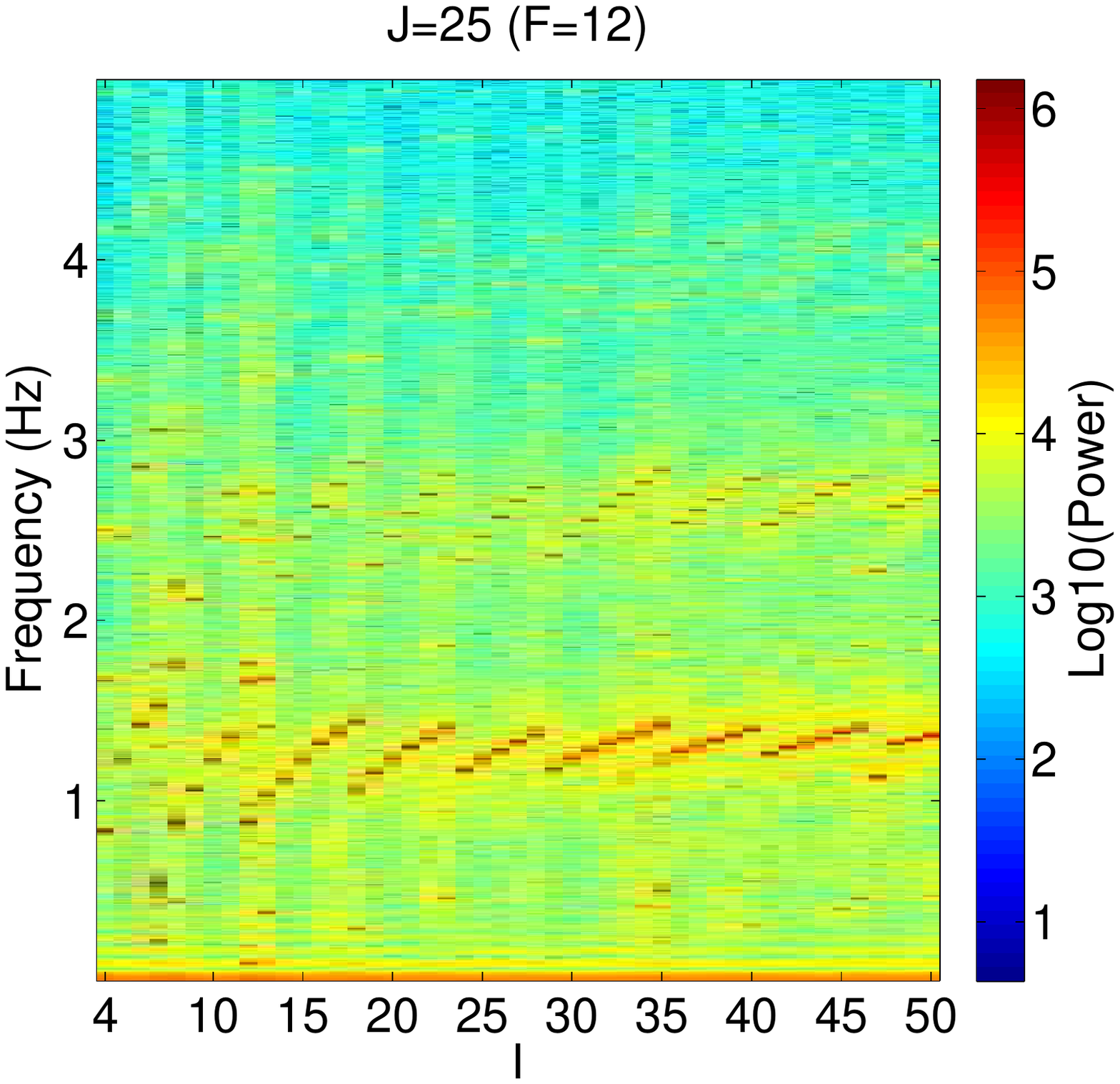}\\
	\end{array}$
	\caption{For these plots, the x-axis represents choices of $I$, the y-axis represents different frequencies, and the color represents the power spectrum of the trajectory of a slow oscillator at the corresponding parameter choice. (Top Row) $J=15$ while $F\in[8,10,12,14]$. (Bottom Row) $F=12$ while $J\in[10,15,20,25]$.}
	\label{fig4}
\end{figure}

\begin{figure}[!h]
	\centering
	$\begin{array}{ccccccc}
\hspace{-1cm}\multirow{9}{*}{\begin{overpic}[scale=.3,trim=4.5cm 8cm 4.5cm 6cm,clip]{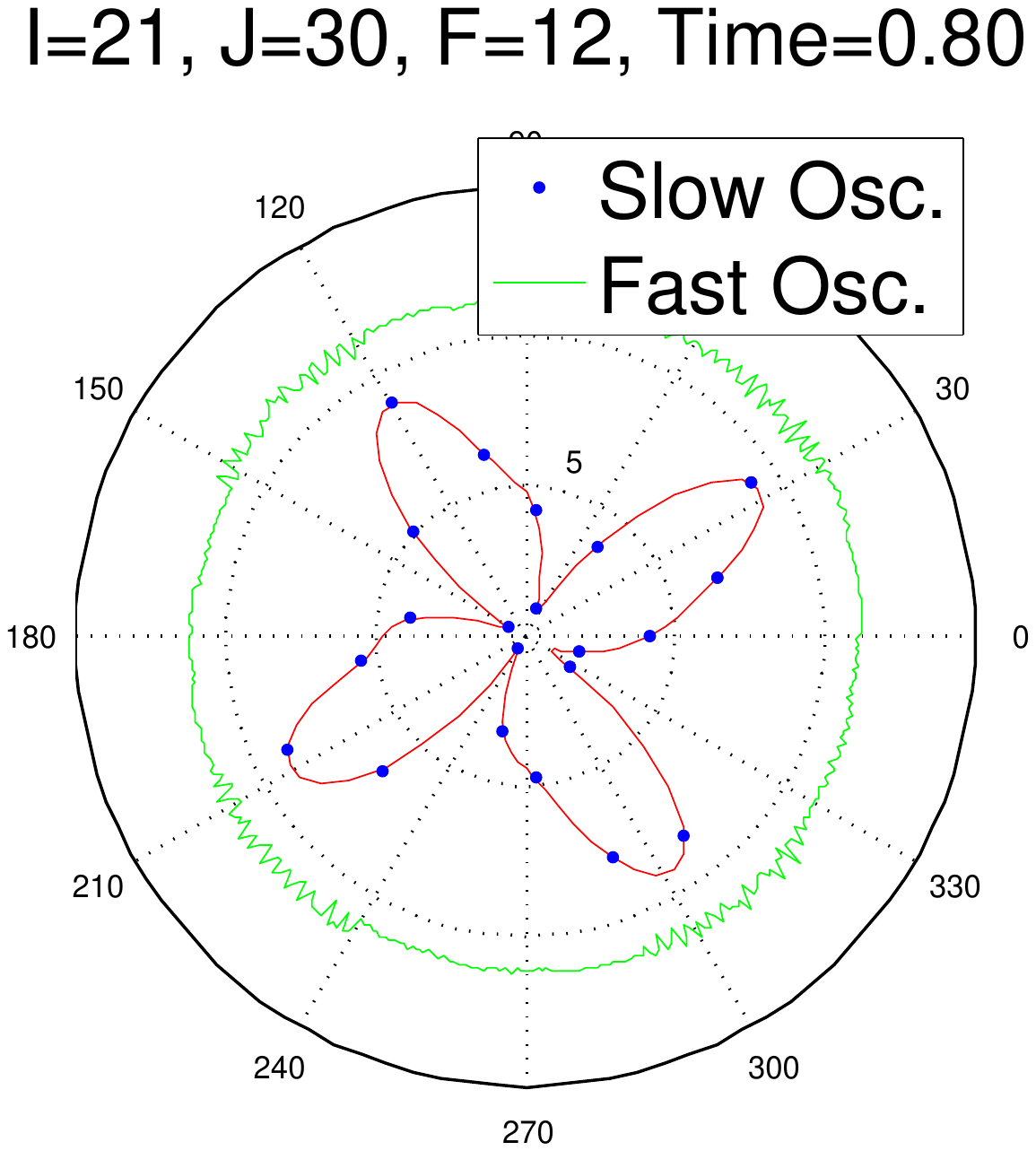}\put(1,32){\colorbox{white}{\fbox{\text{A}}}}\end{overpic}}&&\hspace{-.5cm}\multirow{9}{*}{\includegraphics[scale=.3,trim=4.5cm 8cm 4.5cm 6cm,clip]{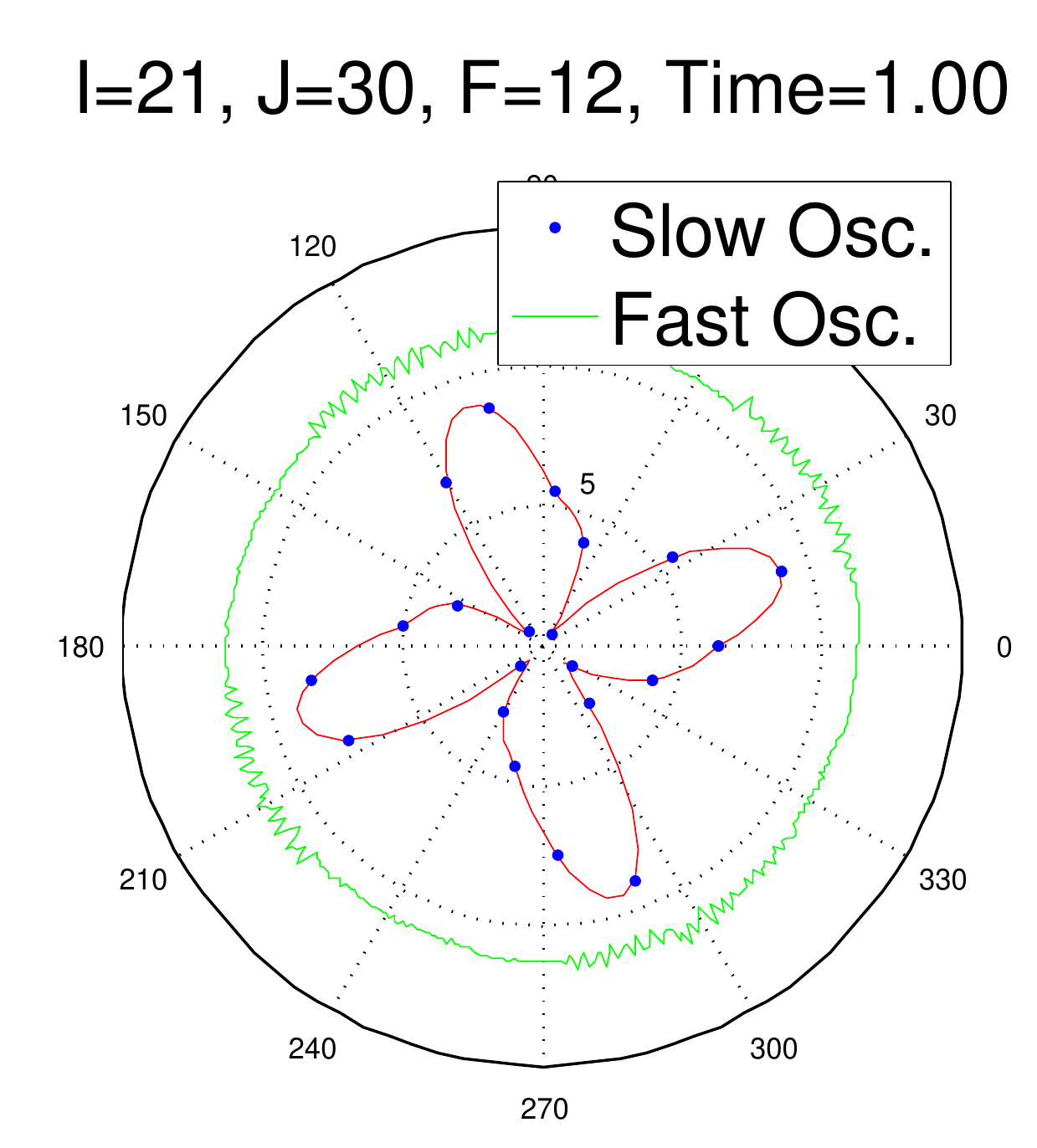}}&&\hspace{-.5cm}\multirow{9}{*}{\includegraphics[scale=.3,trim=4.5cm 8cm 4.5cm 6cm,clip]{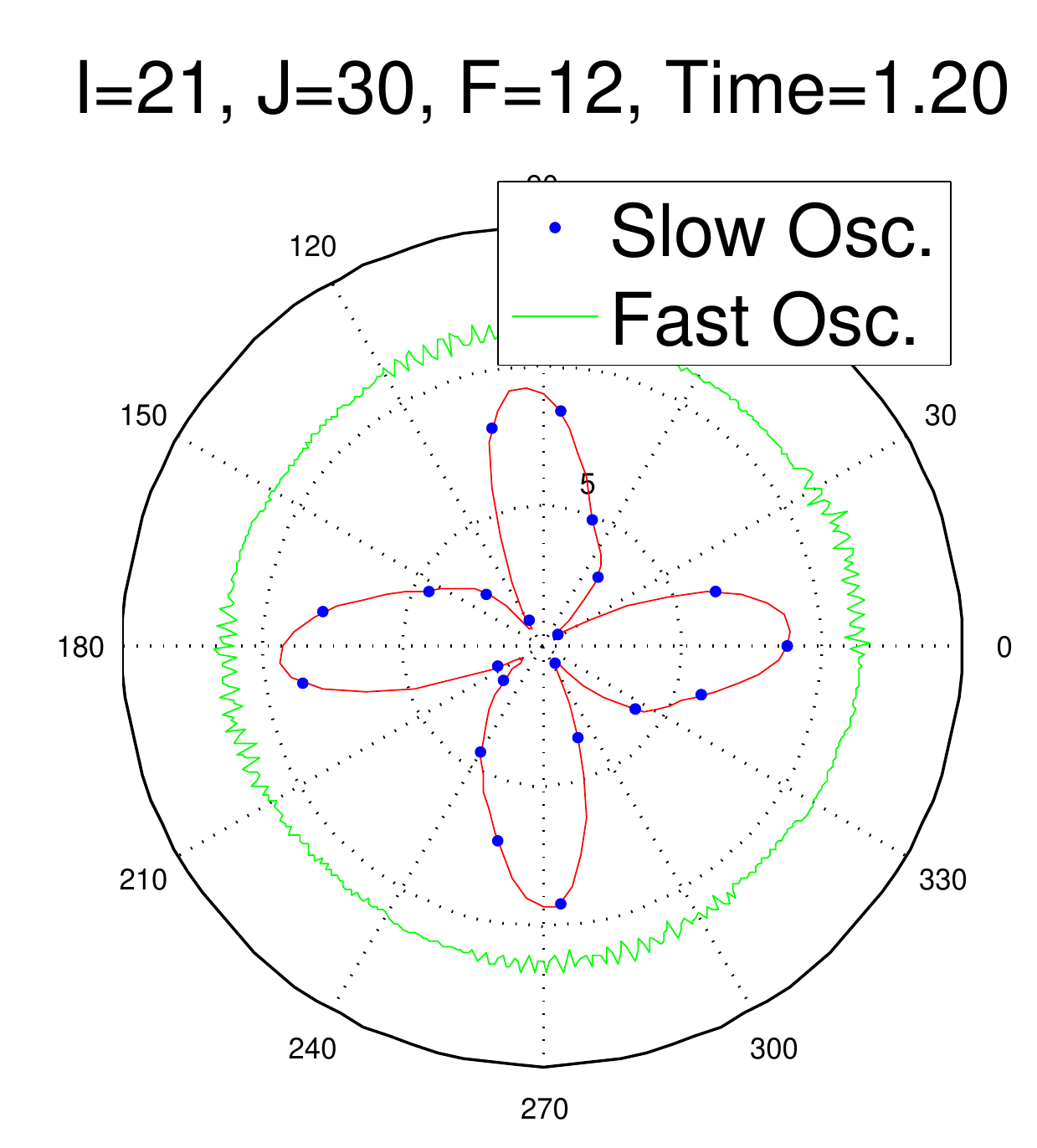}}&&\hspace{-.5cm}\multirow{9}{*}{\includegraphics[scale=.3,trim=4.5cm 8cm 4.5cm 6cm,clip]{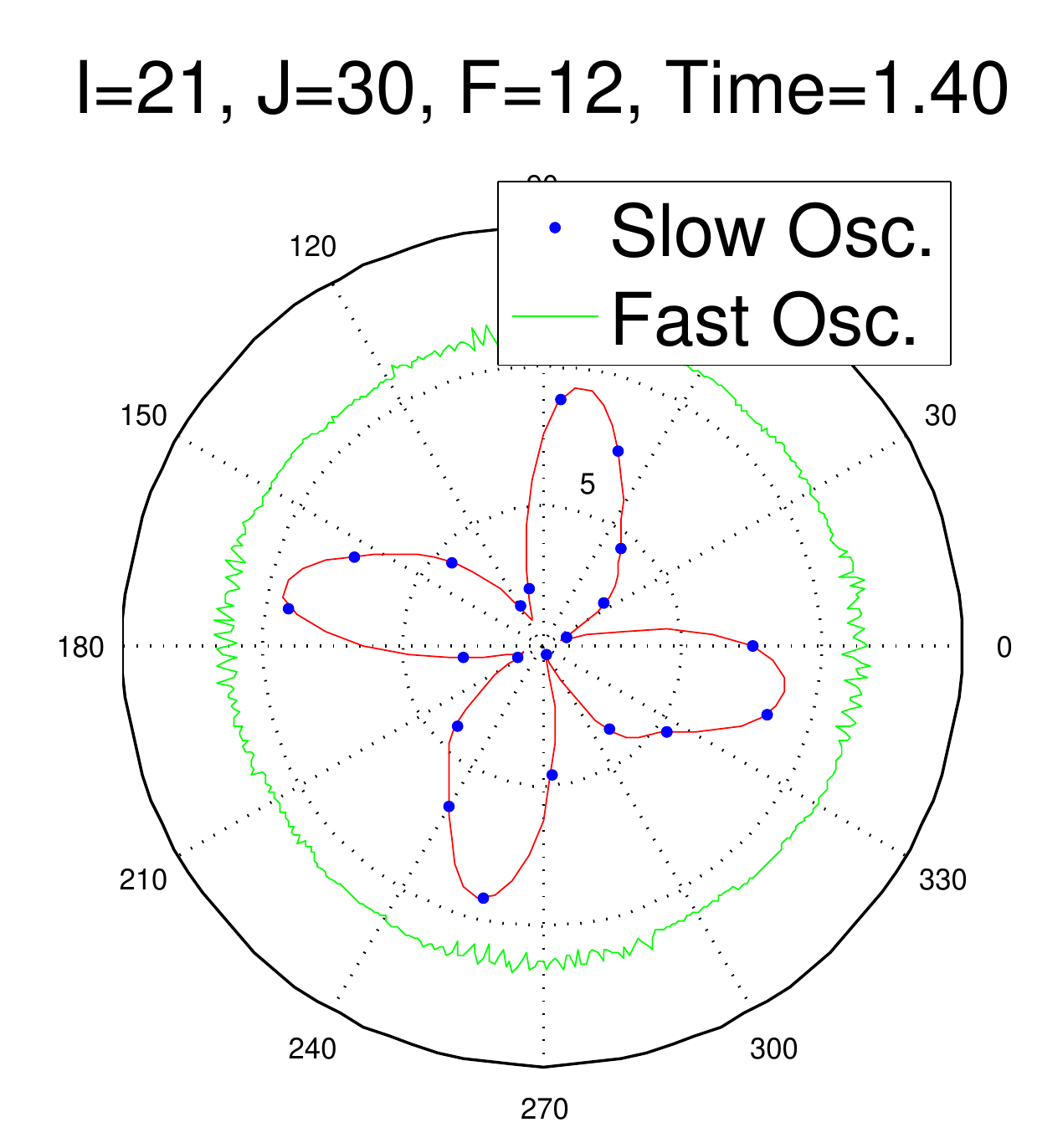}}\\
&&&&&&\\
&&&&&&\\
&&&&&&\\
&&&&&&\\
&\hspace{-.5cm}\rightarrow&&\hspace{-.5cm}\rightarrow&&\hspace{-.5cm}\rightarrow&\\
&&&&&&\\
&&&&&&\\
&&&&&&\\
&&&&&&\\ \hline
	\end{array}$
	$\begin{array}{ccc}
	\hspace{-1cm}\begin{overpic}[width=.25\columnwidth,trim=4.5cm 8cm 4.5cm 6cm,clip]{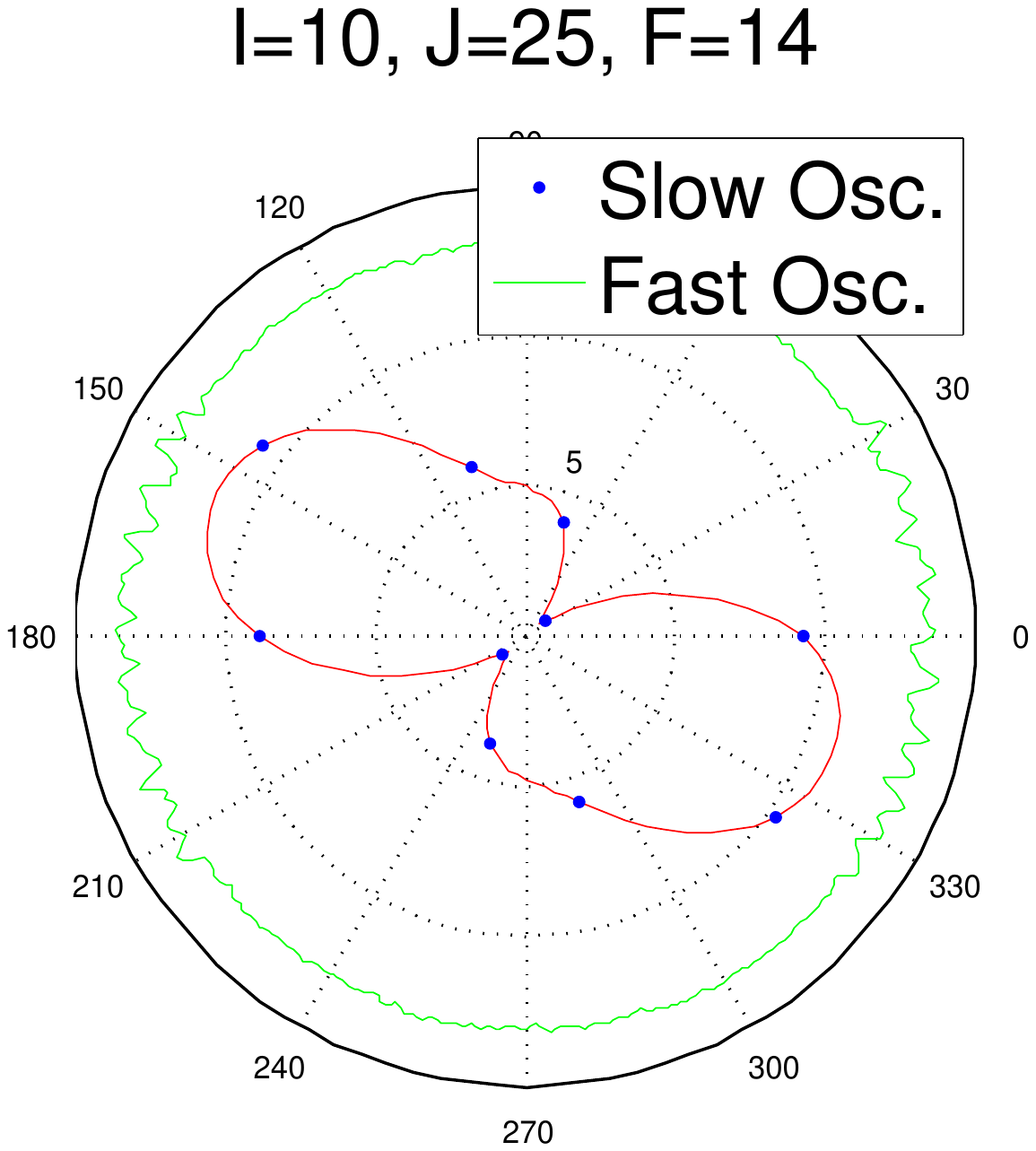}\put(1,35){\colorbox{white}{\fbox{\text{B}}}}\end{overpic}&\includegraphics[width=.25\columnwidth,trim=4.5cm 8cm 4.5cm 6cm,clip]{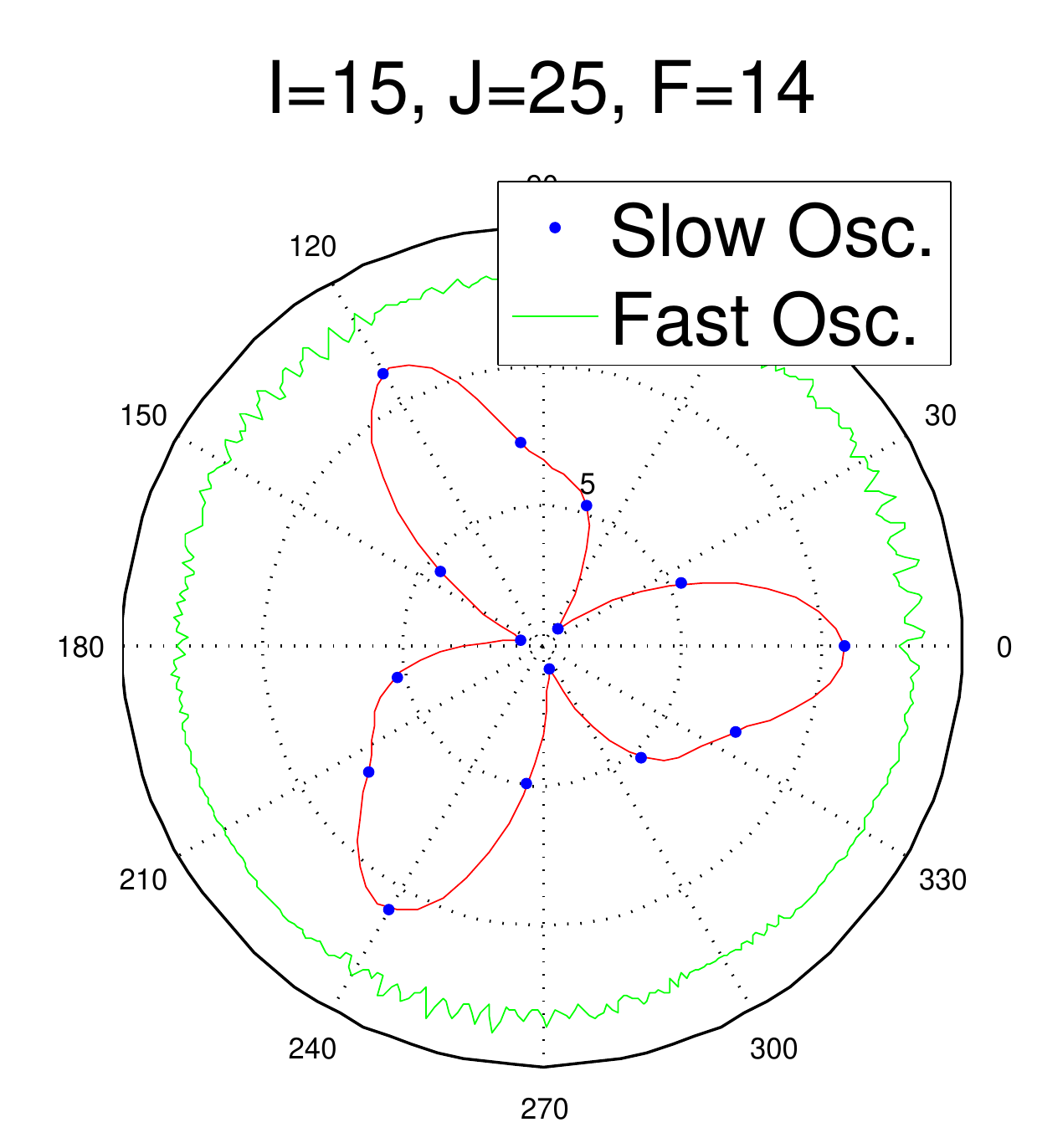}&\includegraphics[width=.25\columnwidth,trim=4.5cm 8cm 4.5cm 6cm,clip]{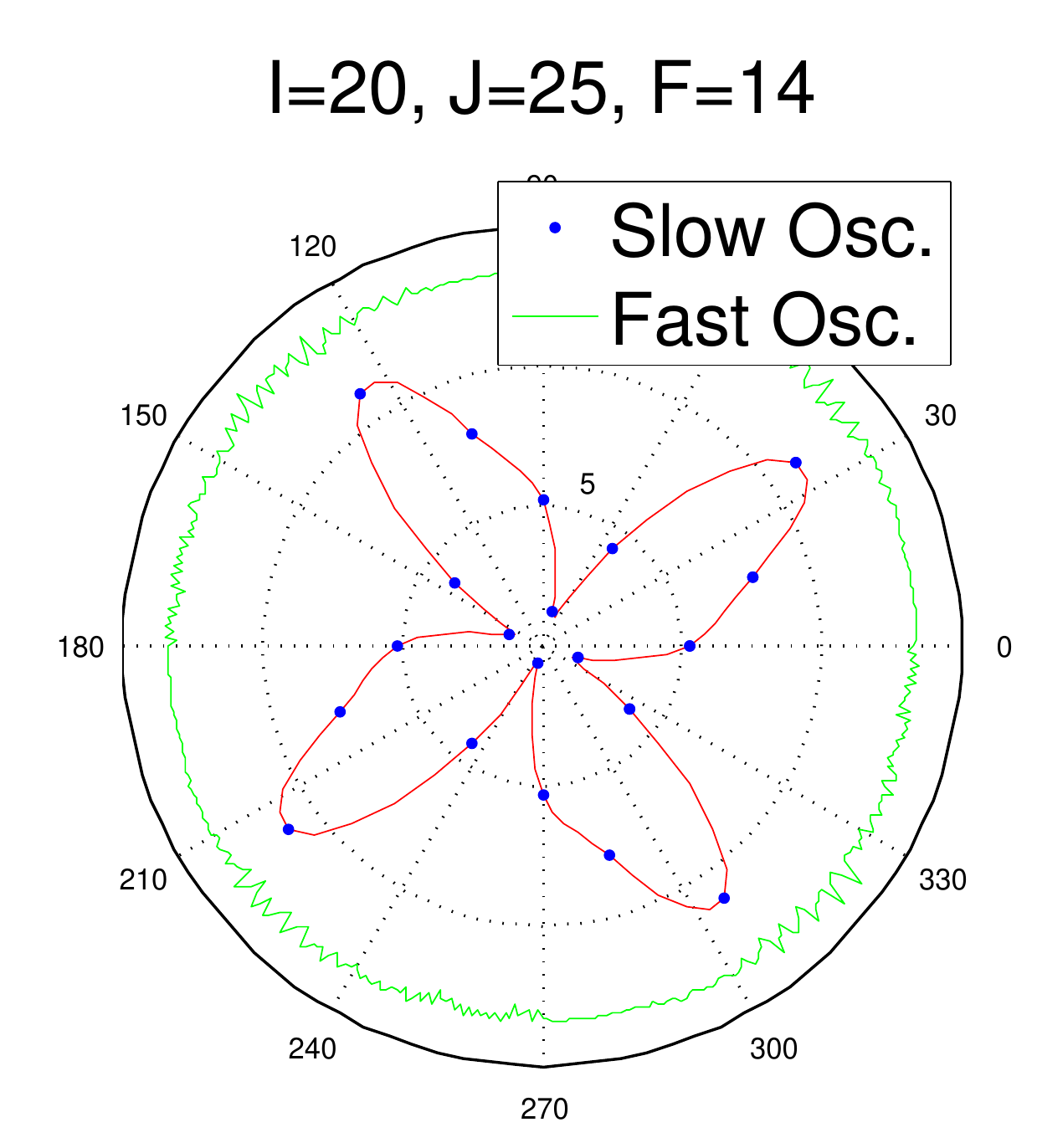}\\
	\hspace{-1cm}\includegraphics[width=.25\columnwidth,trim=4.5cm 8cm 4.5cm 6cm,clip]{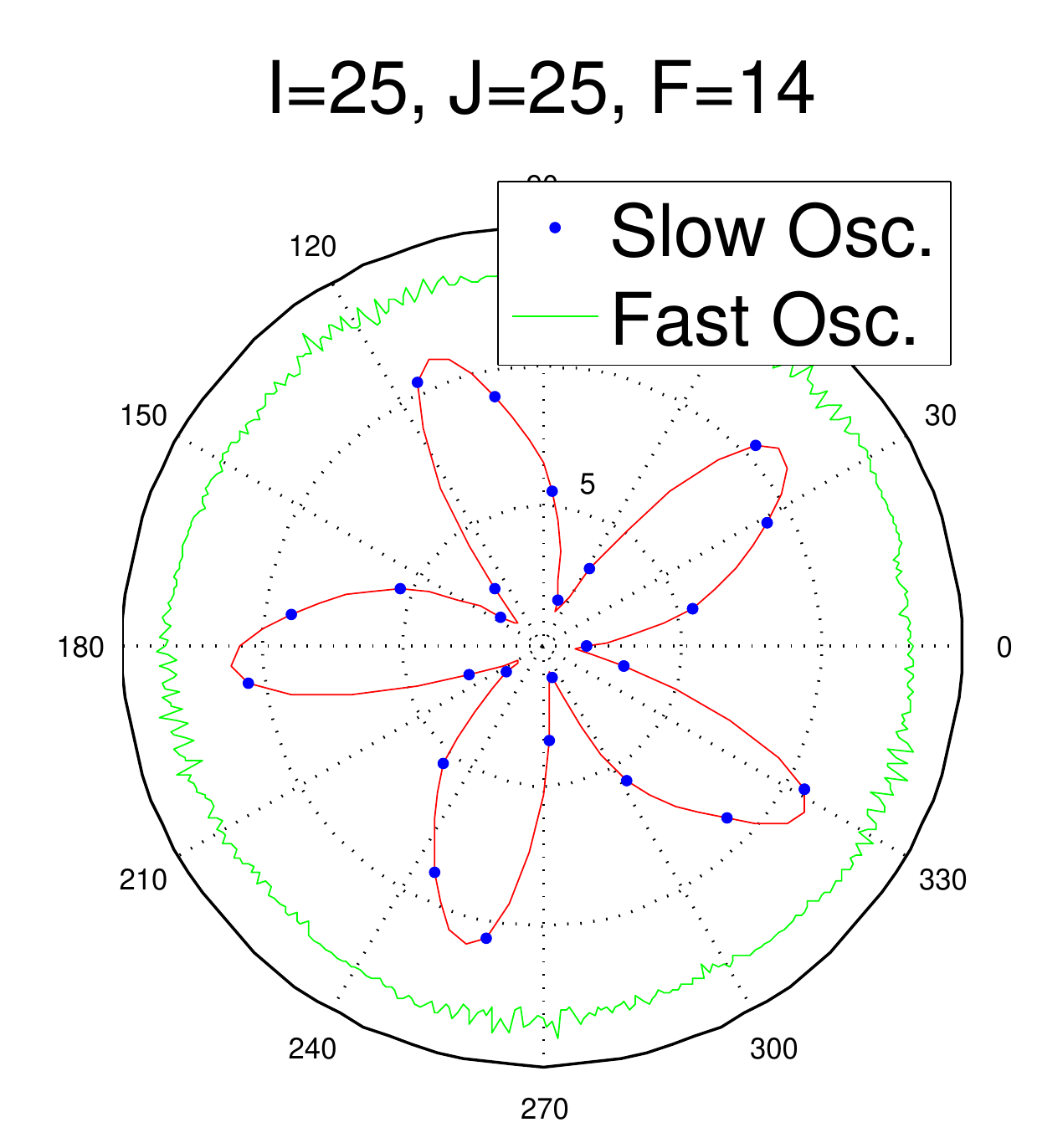}&\includegraphics[width=.25\columnwidth,trim=4.5cm 8cm 4.5cm 6cm,clip]{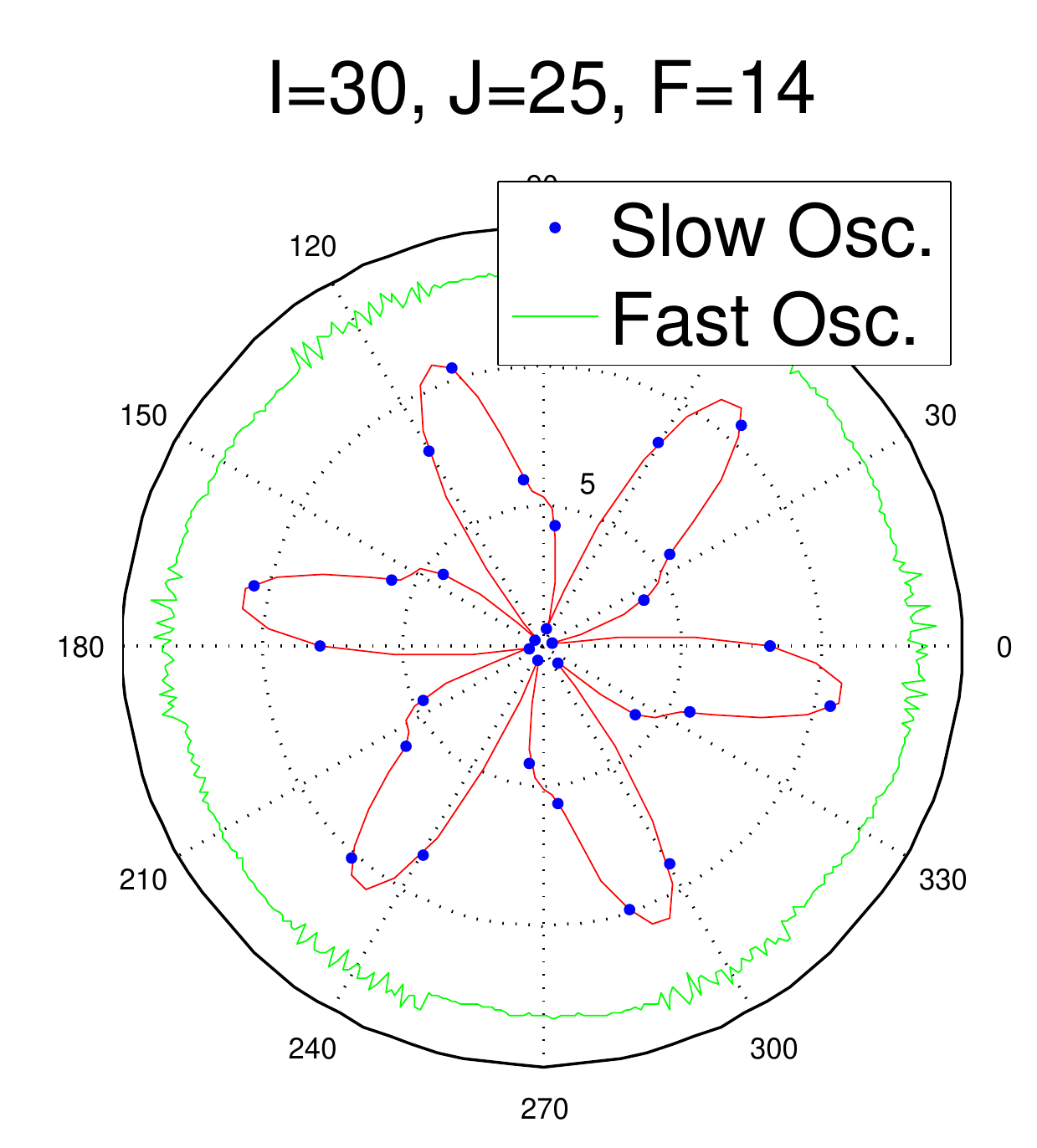}&\includegraphics[width=.25\columnwidth,trim=4.5cm 8cm 4.5cm 6cm,clip]{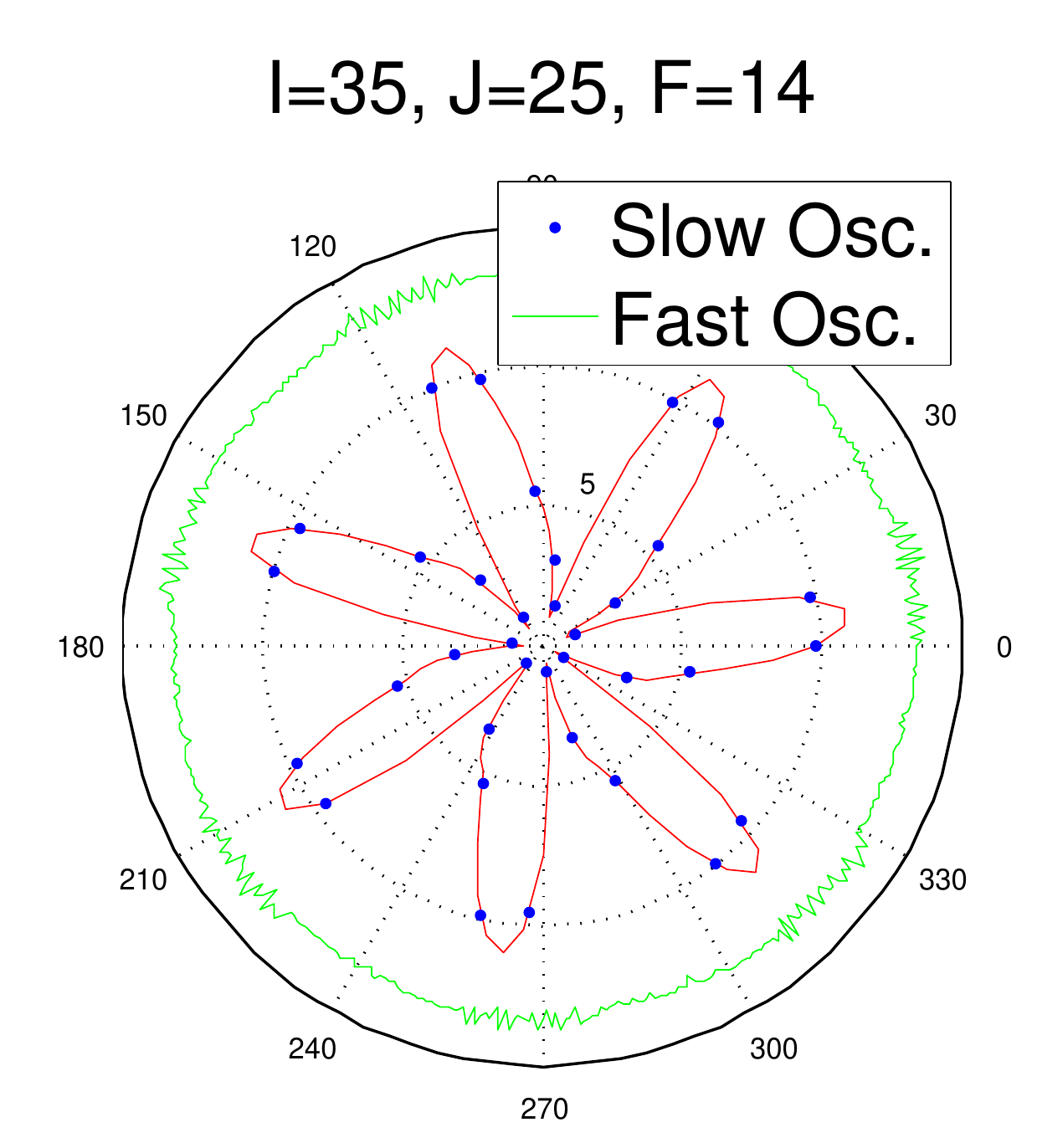}\\
	\hspace{-1cm}\includegraphics[width=.25\columnwidth,trim=4.5cm 8cm 4.5cm 6cm,clip]{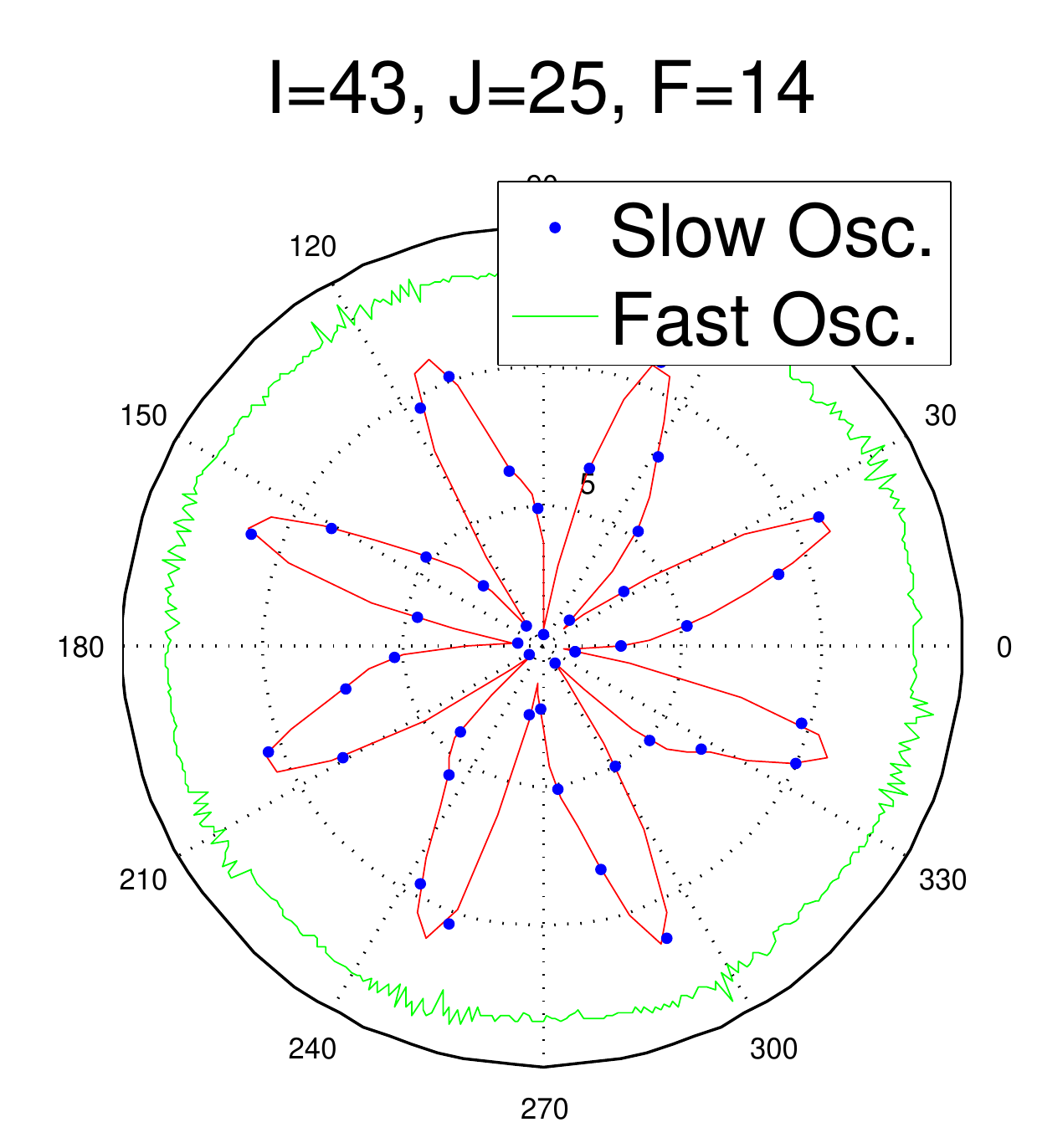}&\includegraphics[width=.25\columnwidth,trim=4.5cm 8cm 4.5cm 6cm,clip]{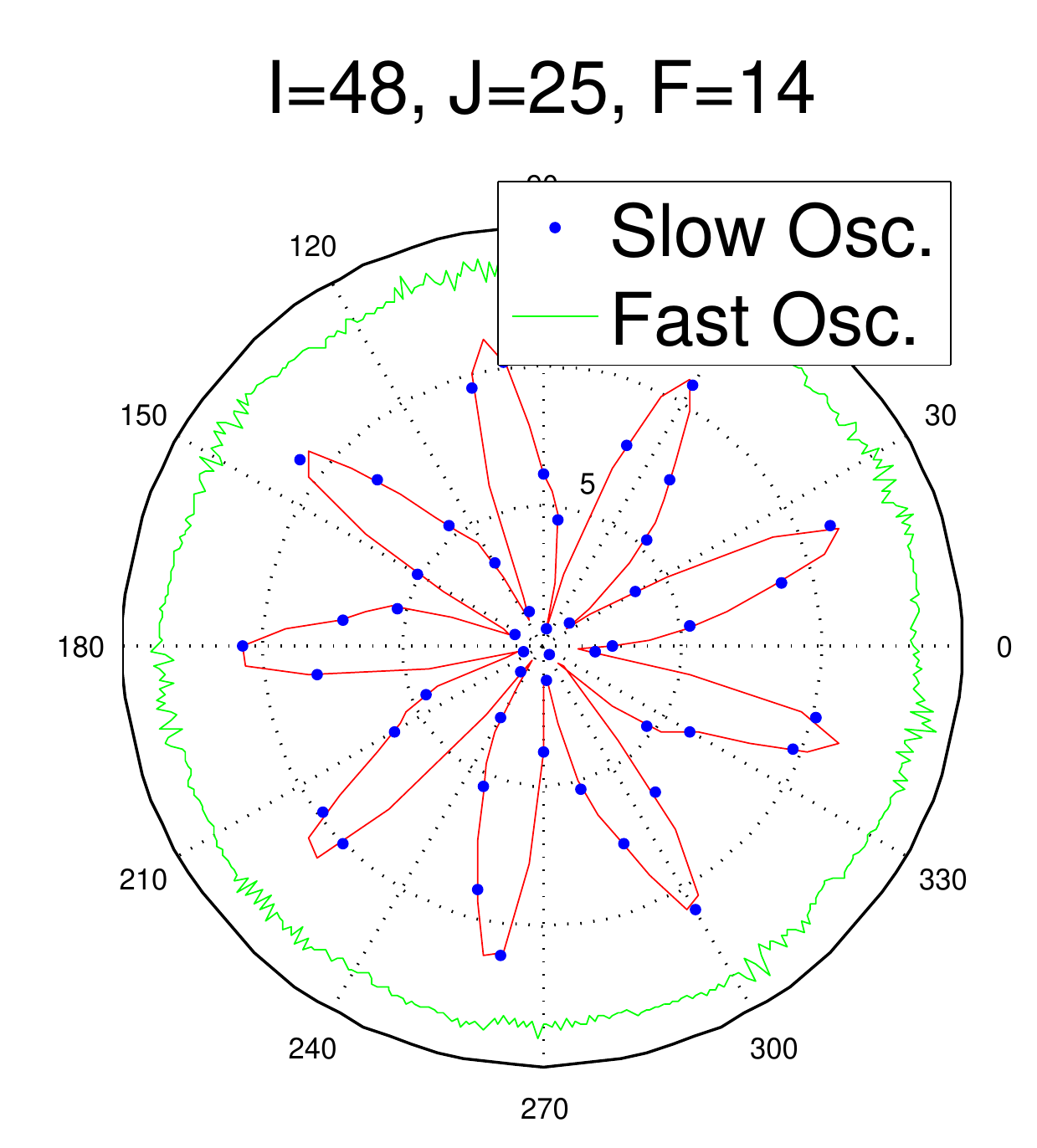}&\includegraphics[width=.25\columnwidth,trim=4.5cm 8cm 4.5cm 6cm,clip]{chaotic_30_5_14.pdf}\\
	\end{array}$
	\caption{(A) For $I=21$, $J=30$, and $F=12$, we plot the trajectories of the fast and slow oscillators. This parameter choice yields a stable attractor as indicated by four snapshots of the standing waves, which travel clockwise around the ring of slow modes. (B) We show different parameter choices yielding different numbers of standing waves (from 2-9). The plot in the bottom right corner represents a snapshot of a trajectory on a chaotic attractor and shows much more irregularity than the standing waves. Animations of these time series can be found \href{http://www.uvm.edu/storylab/share/papers/frank2014a/L96Stability.avi}{here} (full link in references \cite{mfVid00}).}
	\label{fig5}
	\clearpage
\end{figure}

\indent It is common in the literature referencing the Lorenz '96 model to see the parameters $I$, $J$, and $F$ chosen to ensure that the system exhibits sufficient amounts of chaos to make the prediction problem interesting. For example, it is well-known that $F>6$ will usually result in a weakly chaotic system for reasonable choices of $I$ and $J$ \cite{wilks00,karimi00}. Beyond this, $I=8$ and $J=4$ for a total of 40 oscillators is a popular choice, so much so that it is commonly known as the ``Lorenz 40-variable" model \cite{hungLi00}. Generalizing from these standards, we explore the parameter space for $I$, $J$, and $F$ systematically and characterize the resulting dynamical systems. \\
\indent We first measure the largest Lyapunov exponent for several choices of $I$, $J$, and $F$ in Fig. (\ref{fig3}) (top row). We observe that the lower portion of the plots (i.e. small $J$) exhibit strong, positive largest Lyapunov exponents (red \& yellow regions). As $J$ is increased, we observe the emergence of greatly reduced largest Lyapunov exponent (blue regions). This region of reduced chaotic activity returns to a region of increased largest Lyapunov exponent as we continue to increment $J$. Furthermore, we observe that the top and bottom borders of the blue regions oscillate with increasing $I$. The blue region of reduced chaos appears to occur at larger values of $J$ as $F$ is increased, while the range of the blue regions remain fairly constant in $J$.\\
%%%%%%%%%%%
%%%%%%%%%%%
\indent We observe the percentage of positive Lyapunov exponents in the middle row of Fig. (\ref{fig3}). Green vertical windows of increased percentage of positive Lyapunov exponents correspond to the peaks of the blue regions observed in the largest Lyapunov exponent plots. Interestingly, we find that as we continue to increment $J$ beyond these green vertical strips, the percentage of positive Lyapunov exponents sharply declines.\\
\indent The normalized Lyapunov dimension is shown in the bottom row of Fig. (\ref{fig3}). Here, we observe green and yellow vertical striations representing regions of reduced fractal dimensionality relative to the high fractal dimensionality red regions around them. These unstable dimension striations are in locations corresponding to the observed regions of reduced largest Lyapunov exponent, and the vertical striations of increased percentage of positive Lyapunov exponents. A periodicity in $I$ is again apparent here.\\
\indent We are surprised by these regions of reduced chaotic activity and endeavor to explore them using a frequency spectrum analysis. To this end, we examine frequency spectrum bifurcation diagrams representing slices through $I$-$J$ space with a fixed $F$ \cite{orrell00}. A subset of these slices are presented in Fig. (\ref{fig4}). We fix $J=15$ and increase the forcing $F$ moving from left-to-right along the top row of Fig. (\ref{fig4}). Along the bottom row of Fig. (\ref{fig4}), We fix $F=12$ while increasing the number of fast variables $J$ moving from left-to-right. \\
\indent Examining the top row of Fig. (\ref{fig4}) for $8\leq F\leq 12$, we find increased power at many frequencies for most choices of $I$, but, interestingly, we also observe periodic windows in the frequency spectrum bifurcation diagram where power is organized into just two different frequencies. Furthermore, these periodic windows of reduced spectral dispersion correspond to choices of $I$ that result in the stable behavior found in Fig. (\ref{fig3}). We observe that when $F\geq14$ there is power at many frequencies for $I\geq6$ and periodic windows do not exist. This observation corresponds to the rise of the blue region of reduced largest Lyapunov exponent as $F$ is increased (Fig. (\ref{fig3})). \\
%%%%%%%%%%%%%%
%%%%%%%%%%%%%%%%
%\noindent 
\indent We look at the frequency spectrum bifurcation diagram in the bottom row of Fig. (\ref{fig4}) by fixing $F=12$ and varying $J$.  Fig. (\ref{fig3}) suggests that we will observe reduced largest Lyapunov exponent for $10\leq I\leq 35$ for most choices of $I$, and this is reflected in Fig. (\ref{fig4}) where we observe power at many frequencies for $J=10$. We take steps through this region of reduced chaotic activity as we increase $J$, and again find periodic windows in the frequency spectrum bifurcation diagram, where large amounts of power are only found at a finite number of different frequencies. Again these windows of reduced spectral activity occur at $I$ values corresponding to peaks in the blue regions from Fig. (\ref{fig3}). Furthermore, we see evidence that increasing $J$ may have similar effects as reducing $F$. \\
\indent Through further analysis of the frequency bifurcation spectrum diagrams, we observe that in general frequencies between one and two have more power, suggesting that slow oscillators tend to exhibit these frequencies even for parameter choices resulting in chaotic dynamics. We find more interesting frequency behavior in the many windows of organized spectral activity, where the dominant and subdominant frequencies, namely the frequency with the most power and the frequency with the second most power, appear to oscillate as a function of $I$. For $J=20$ and $J=25$ in the bottom row of Fig. (\ref{fig4}), we see that the dominant and subdominant frequencies fluctuate every fifth or sixth increment as we increase $I$. Also, the fluctuations become less severe as $I$ approaches 50. \\
\indent The frequency spectrum bifurcation diagrams show us that several parameter choices constrain the slow oscillators to two distinct frequencies. This suggests that we should see a strong regularity in the time series for these parameter choices. In Fig. (\ref{fig5}), we provide  example snapshots of stable attractors, which resemble rose-plots in polar coordinates, and a chaotic attractor with a large positive largest Lyapunov exponent, which resembles an amoeba (bottom right). Each petal of the stable attractors is in fact a standing wave traveling around the slow oscillators over time as shown by Fig. (\ref{fig5})A. We see that the oscillations of the stable attractors show signs of being comprised of two frequencies, as individual slow oscillators seem to achieve both a relative local maximum and a global maximum. Furthermore, as we increase $I$ we see additional petals added to the stable attractor. If $I$ is chosen so that it falls between two windows of increased spectral organization, then we see the dynamics attempt to add an additional 
%%%%%%%%%%%%%%%%%%%%%%%%%%%%%%%%%%
\clearpage
\clearpage
\begin{figure}[!h]
	\centering
           \fbox{\includegraphics[scale=.5,trim=0cm 5cm 0cm 5cm,clip]{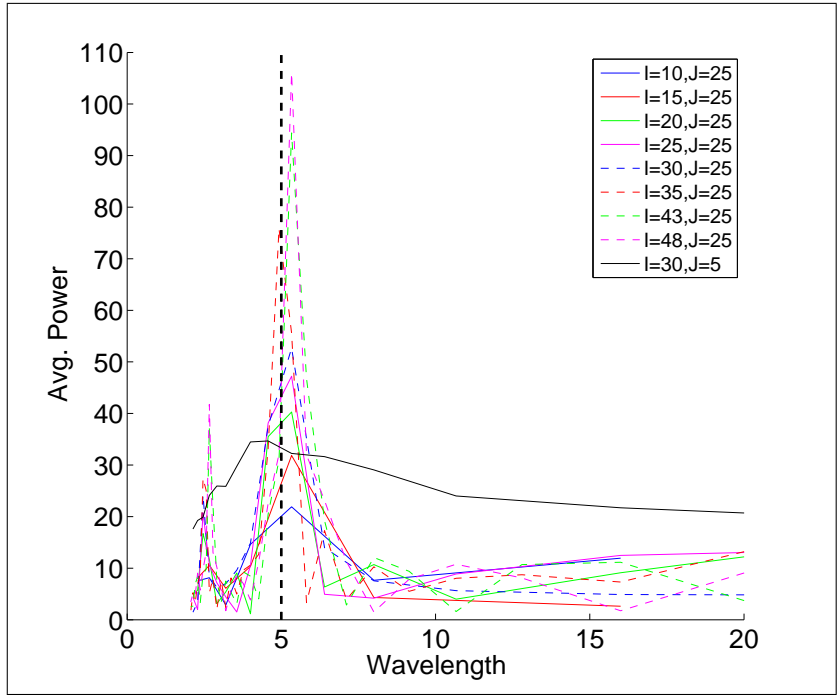}}
	\caption{We measure the wavelength for the attractors in each of the nine example plots in Fig. (\ref{fig5}) by examining the states of the slow oscillators in the frequency domain. The colored lines represent the stable attractors, and the black solid line represents the same analysis for the chaotic attractor.}
        \label{fig6}
\end{figure}
%%%%%%%%%%%%%%%%%%
\noindent petal, but this petal will dissipate over time in a repeating process that prevents the trajectory from stabilizing. We propose a simple function describing the stable behavior in the Appendix. \\
\indent Fig. (\ref{fig6}) quantifies the wavelengths preferred by the stable attractors and the chaotic attractor in Fig. (\ref{fig5}) by examining spatial fast fourier transforms. The colored lines represent the stable attractors, and we observe that these curves exhibit peaks around a wavelength of 5 indicating that the standing waves in the stable attractors usually involve about five slow oscillators. The black solid line represents the same analysis for the chaotic attractor in Fig. (\ref{fig5}), and we observe a smoother curve indicating that many different wavelengths are more equally preferred by flow traveling around the slow oscillators in the chaotic attractor.\\
\indent We have provided evidence that stability emerges amongst regions of chaos in parameter space for the Lorenz '96 system, and that there appears to be a relationship between the usual bifurcation parameter, $F$, and the parameters controlling the dimension of the system, $I$ \& $J$. Fig. (\ref{fig7}) shows a few bifurcation diagrams where $I$, the number of slow oscillators, is used as a bifurcation parameter. These bifurcation diagrams clearly display regions of stability and chaos as a function of $I$. Furthermore, we again observe evidence of the regularity in the trajectories of the slow oscillators for parameter choices leading to stability since the values of the local maxima of the slow oscillators in such regions are roughly constant across each bifurcation diagram.\\
\indent Figures 7A, 7B are example trajectories corresponding to the dashed lines in figures 7G, 7H, respectively. The dots in these time series indicate local maxima of the trajectories \cite{lorenz01}. Fig. (\ref{fig7})A demonstrates that values of the local maxima can fluctuate wildly, while Fig. (\ref{fig7})B shows a parameter choice for which local maxima tend towards only two different values. The middle row of Fig. (\ref{fig7}) (panels C-F) exhibits windows of both stable and chaotic dynamics as a function of the dimensional parameter $I$. We again observe windows of stability and chaos in panels G-J where $F$, a physical parameter, is tuned as the bifurcation parameter for several choices of $I$ and $J$. For a fixed $I$, increasing $J$ seems to condense the dynamics, constraining them to the envelope of values observed.\\
\indent Fig. (\ref{fig8}) allows us to relate the effects of varying the dimensional parameter $J$ to varying the physical coupling parameter $h$. We vary $I$ from 4 to 50 and vary $h$ from 0 to 1 while holding fixed $J=50$ (note that $h=1$ in all previous figures, consistent with the literature). We observe a pattern reminiscent of those observed in the top row of Fig. (\ref{fig3}), which suggests that the parameters $h$ and $J$ may have an analogous effect on the system.
%%%%%%%%%%%%%%%%%%%%%%%%%%%%%%%%%%%%%%%%%%%%%%%%%%%%%\end{multicols}
\begin{figure}[!t]
	\centering
$\begin{array}{cc}
		\hspace{-1cm}\begin{overpic}[width=.47\columnwidth,trim=1cm 3.5cm 2cm 7cm,clip]{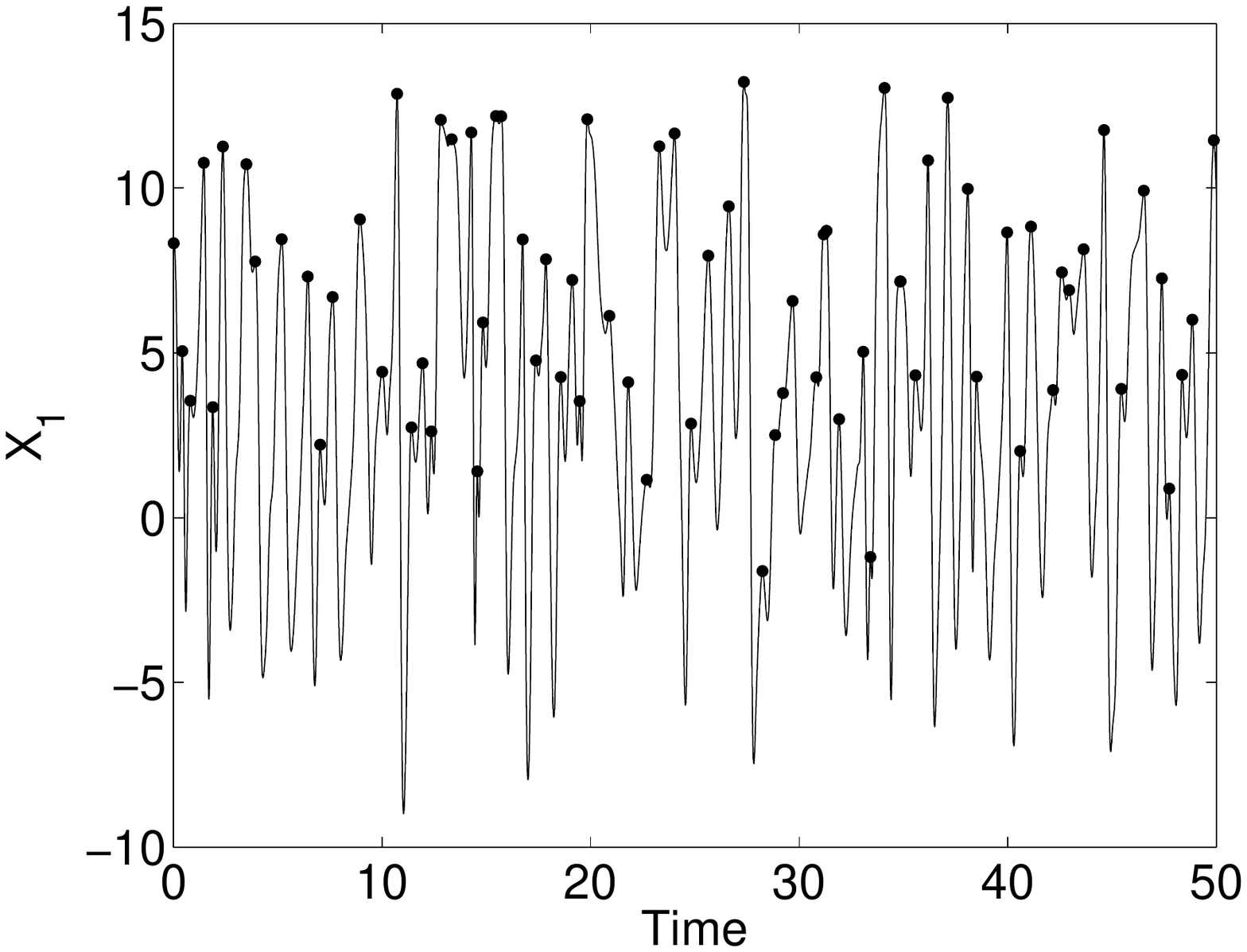}\put(15,68){\colorbox{white}{\fbox{\text{A}}}}\end{overpic}&\begin{overpic}[width=.47\columnwidth,trim=1cm 3.5cm 2cm 7cm,clip]{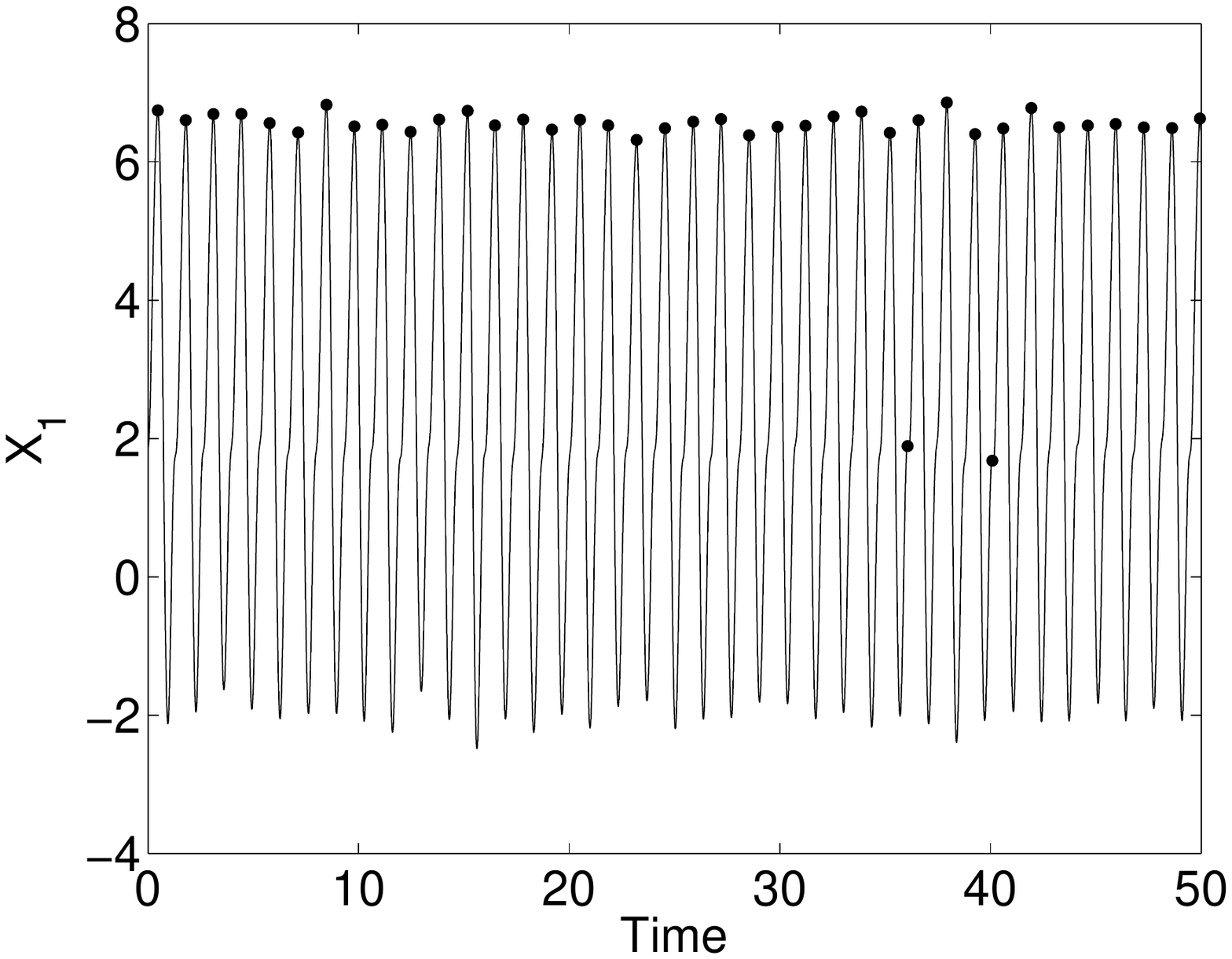}\put(15,68){\colorbox{white}{\fbox{\text{B}}}}\end{overpic}\\
	\end{array}$
	\vspace{.5cm}
	$\begin{array}{cccc}
		\begin{overpic}[scale=.21,trim=2.5cm -7cm -1cm 7cm]{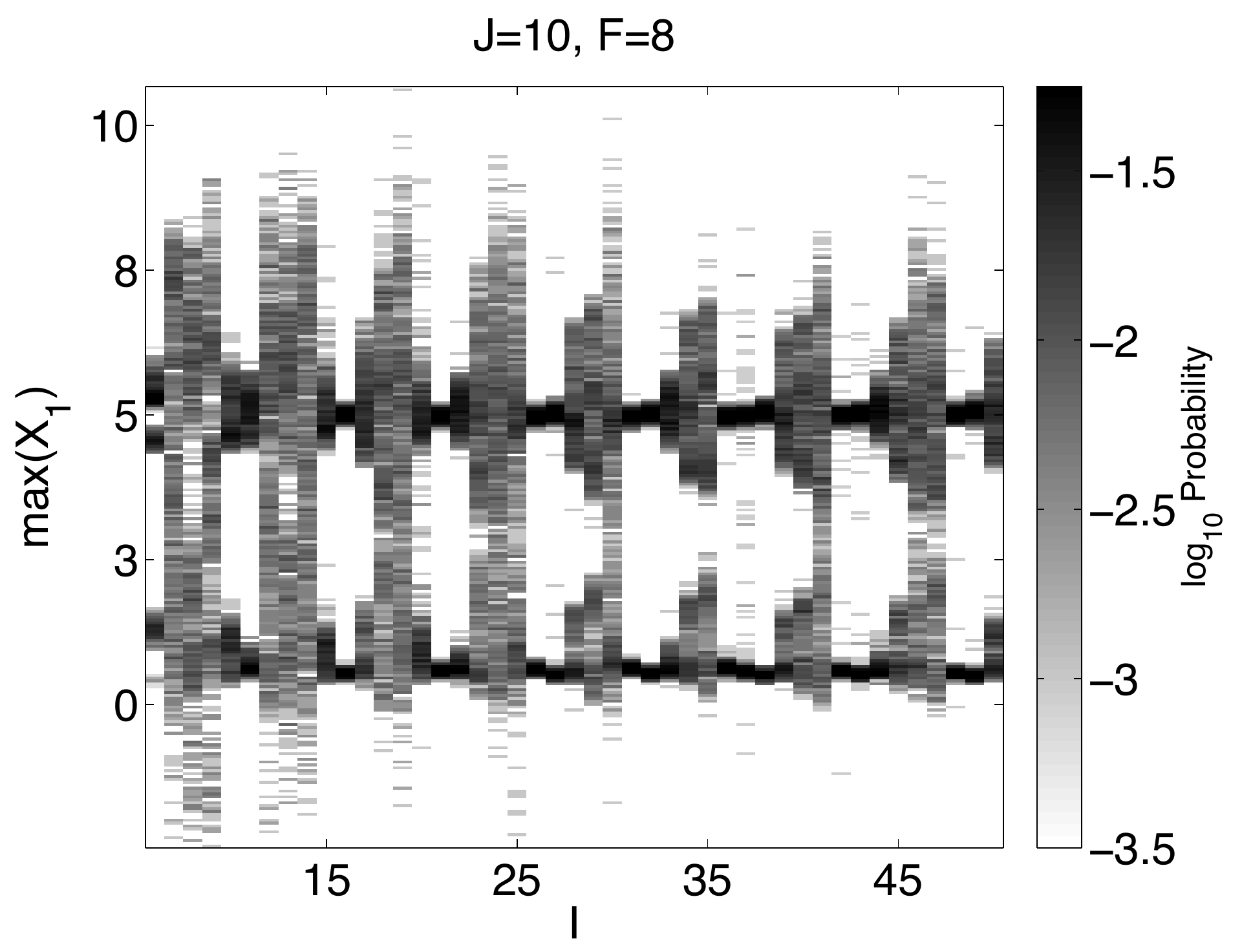}\put(1,42){\colorbox{white}{\fbox{\text{C}}}}\end{overpic}&		
		\begin{overpic}[scale=.21,trim=3cm -7cm -1cm 7cm]{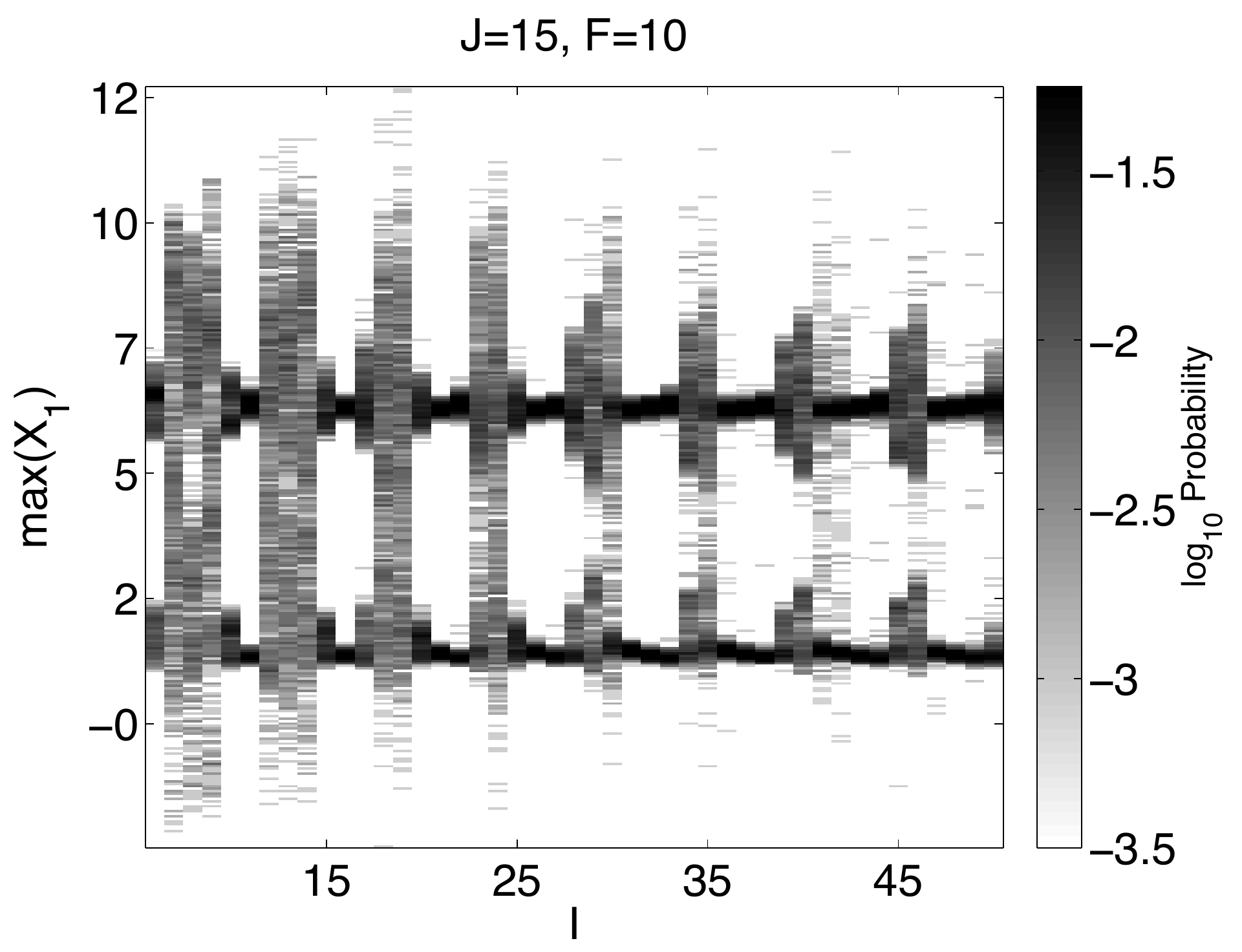}\put(1,42){\colorbox{white}{\fbox{\text{D}}}}\end{overpic}&
		\begin{overpic}[scale=.21,trim=3cm -7cm -1cm 7cm]{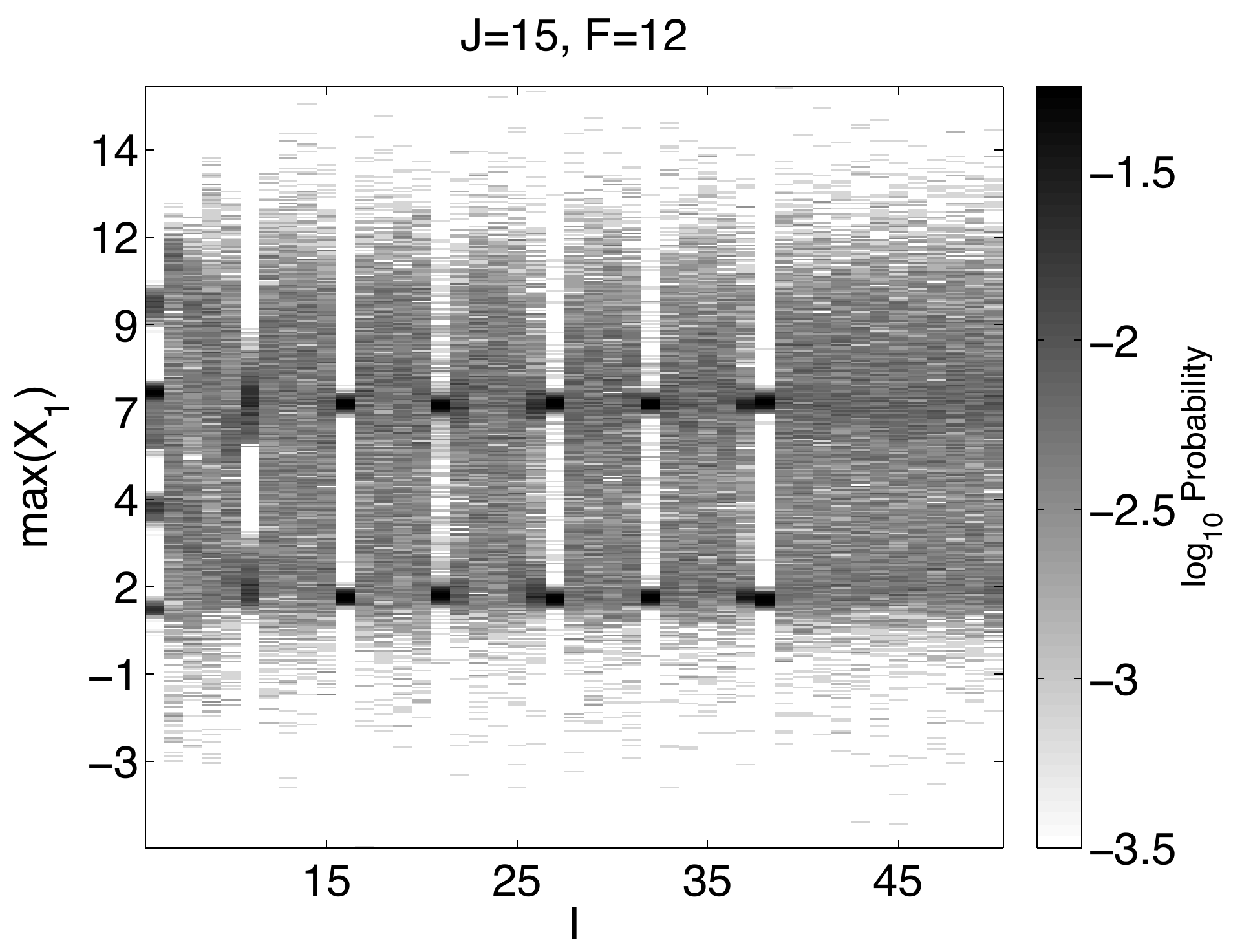}\put(1,42){\colorbox{white}{\fbox{\text{E}}}}\end{overpic}&
		\begin{overpic}[scale=.21,trim=3cm -7cm -1cm 7cm]{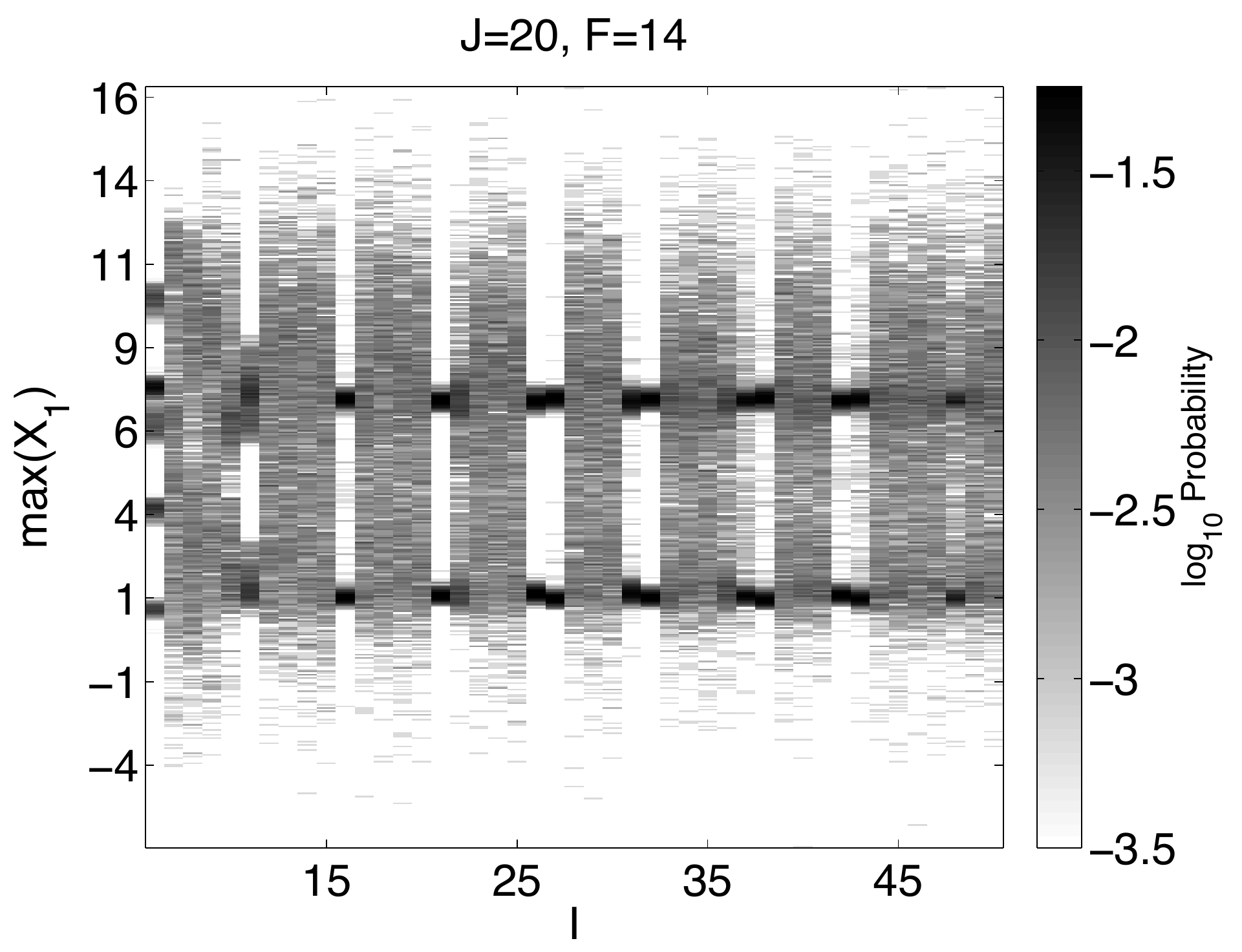}\put(1,42){\colorbox{white}{\fbox{\text{F}}}}\end{overpic}\\
	\begin{overpic}[scale=.26,trim=1.5cm 3cm -1cm 7cm]{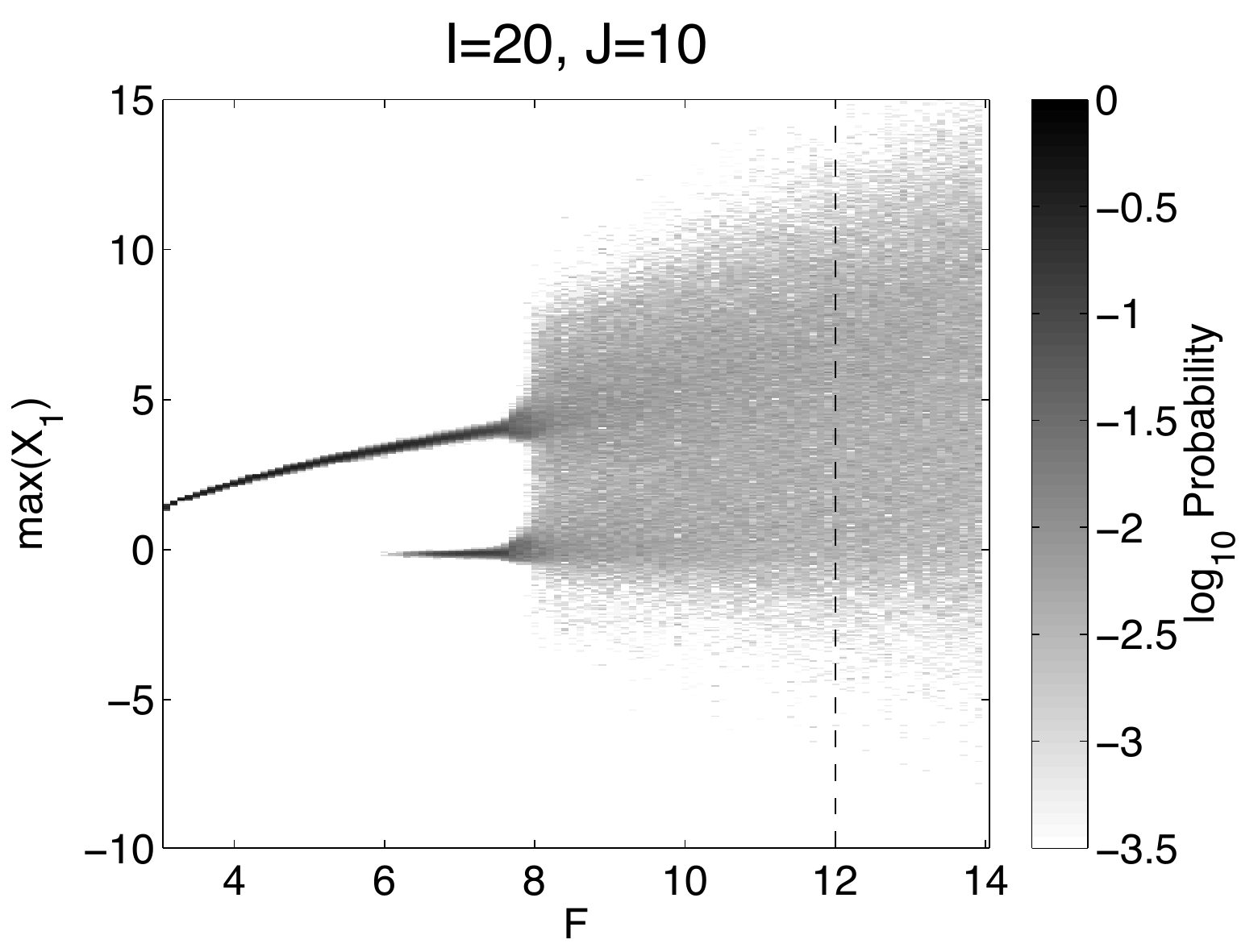}\put(2,13){\colorbox{white}{\fbox{\text{G}}}}\end{overpic}&		 
	\begin{overpic}[scale=.26,trim=2cm 3cm -1cm 7cm]{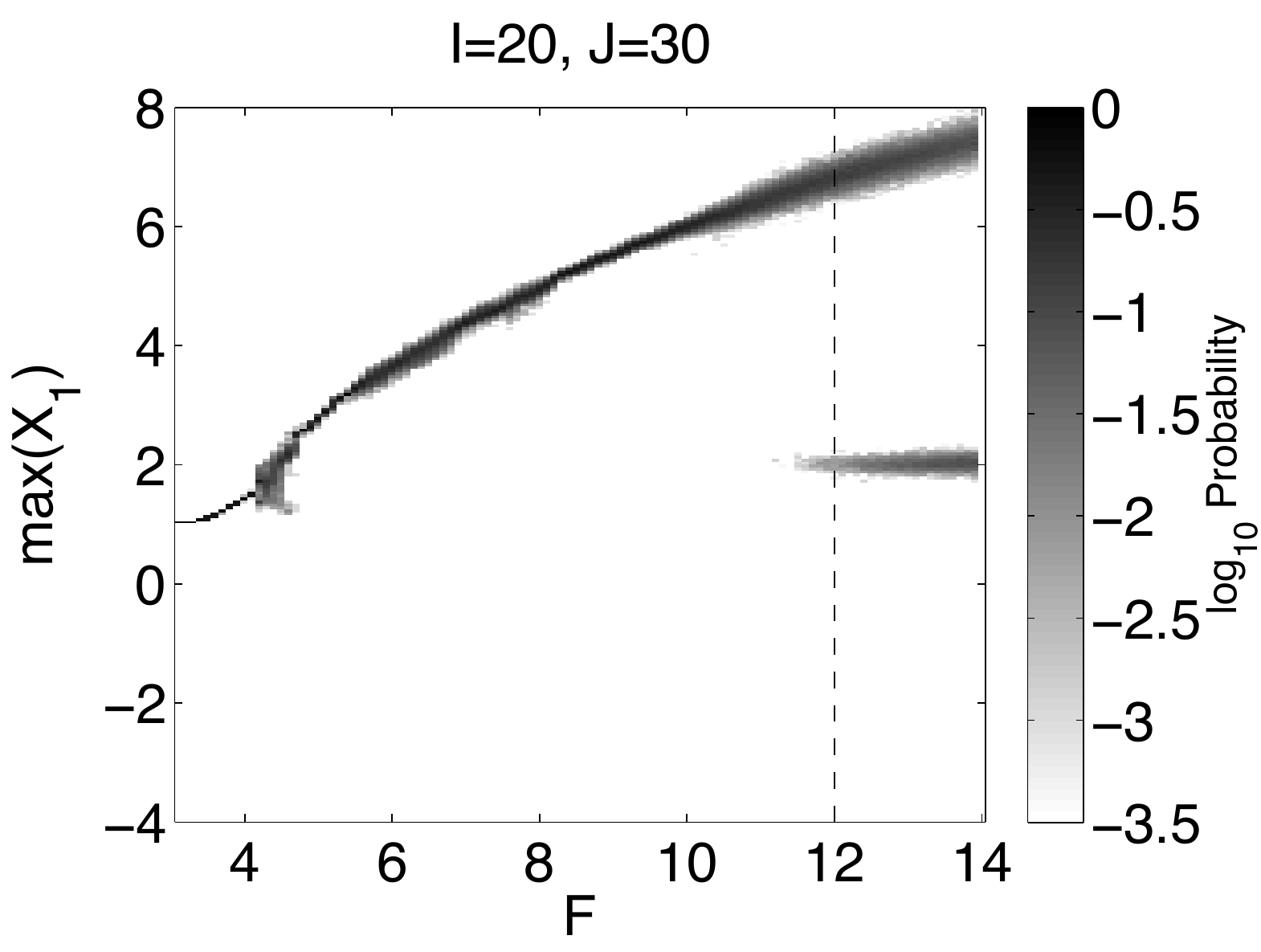}\put(1,13){\colorbox{white}{\fbox{\text{H}}}}\end{overpic}&
		\begin{overpic}[scale=.21,trim=3cm 4cm -1cm 7cm]{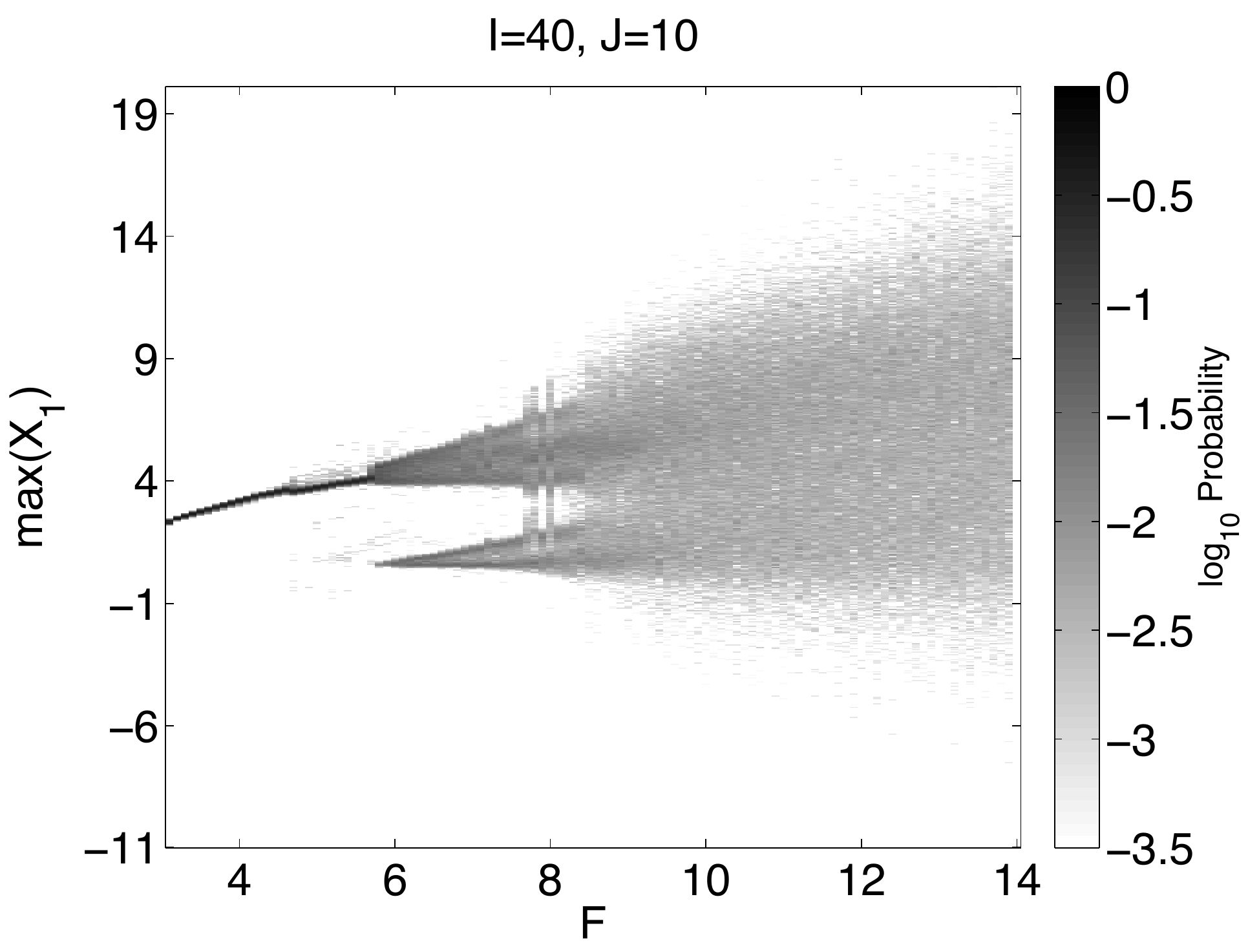}\put(1,13){\colorbox{white}{\fbox{\text{I}}}}\end{overpic}&
		\begin{overpic}[scale=.21,trim=3cm 4cm -1cm 7cm]{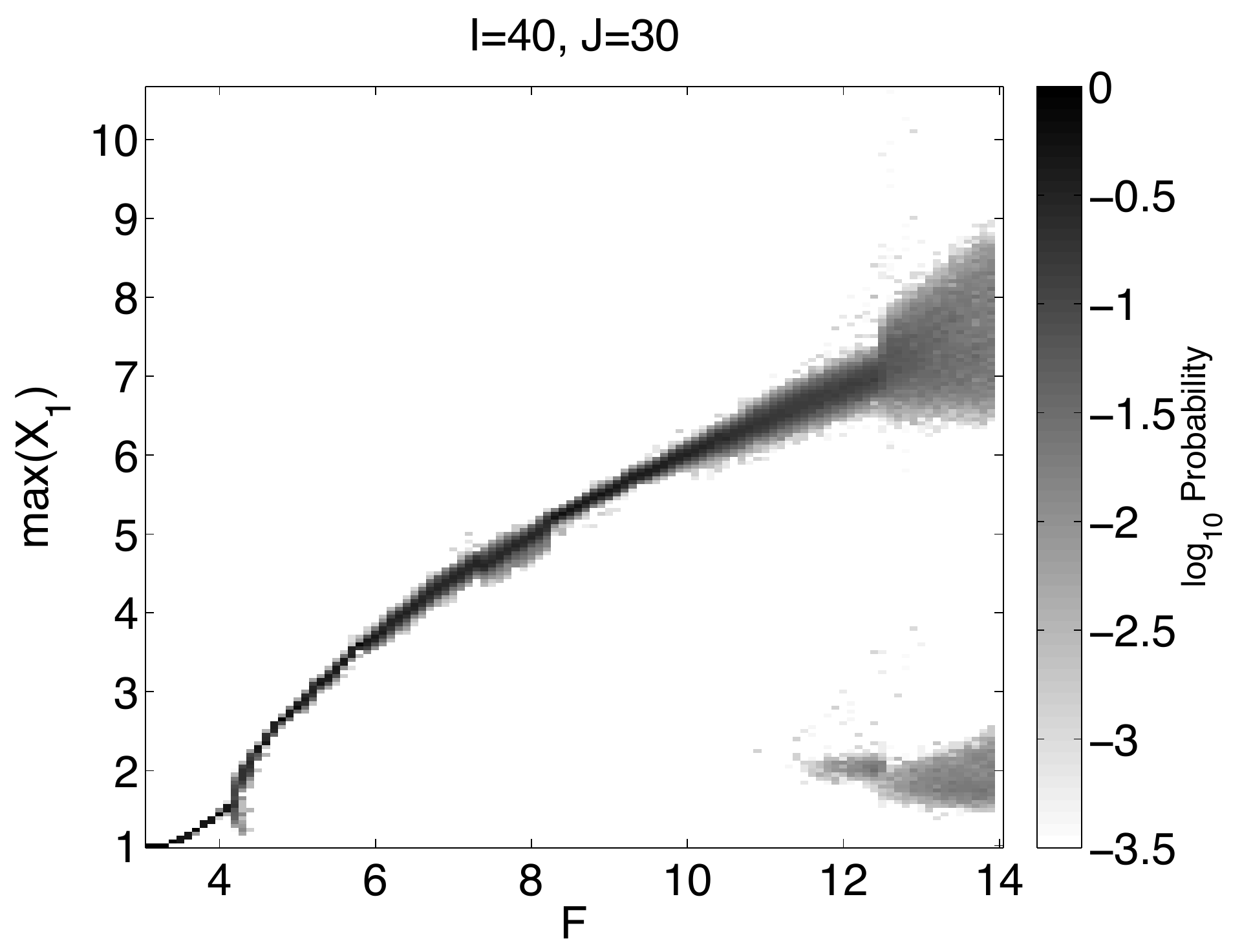}\put(1,13){\colorbox{white}{\fbox{\text{J}}}}\end{overpic}\\
	\end{array}$
	\caption{In the top row (A \& B), we show two example trajectories representative of $I=20$, $J=10$, $F=12$ and $I=20$, $J=30$, $F=12$, respectively. Black circles indicate local maxima of the trajectories. These time series are example trajectories taken from the bifurcation diagrams; panel A corresponds to the dashed line in panel G, and panel B corresponds to the dashed line in panel H. In the middle row (C-F), we provide bifurcation diagrams for several choices of $J$ and $F$ while $I$ is varied as the bifurcation parameter. The y-axis indicates the values of the $x_1$ local maxima. Note that ranges of the y-axes are different for each figure. The x-axis represents different choices of $F$ in the bottom row (G-J) for a few choices of $I$ and $J$. We observe both windows of stability and windows of chaos. }
	\label{fig7}
\end{figure}
\begin{figure}[!h]
	\centering
	\includegraphics[width=.45\columnwidth,trim=0cm .3cm 0cm .25cm,clip]{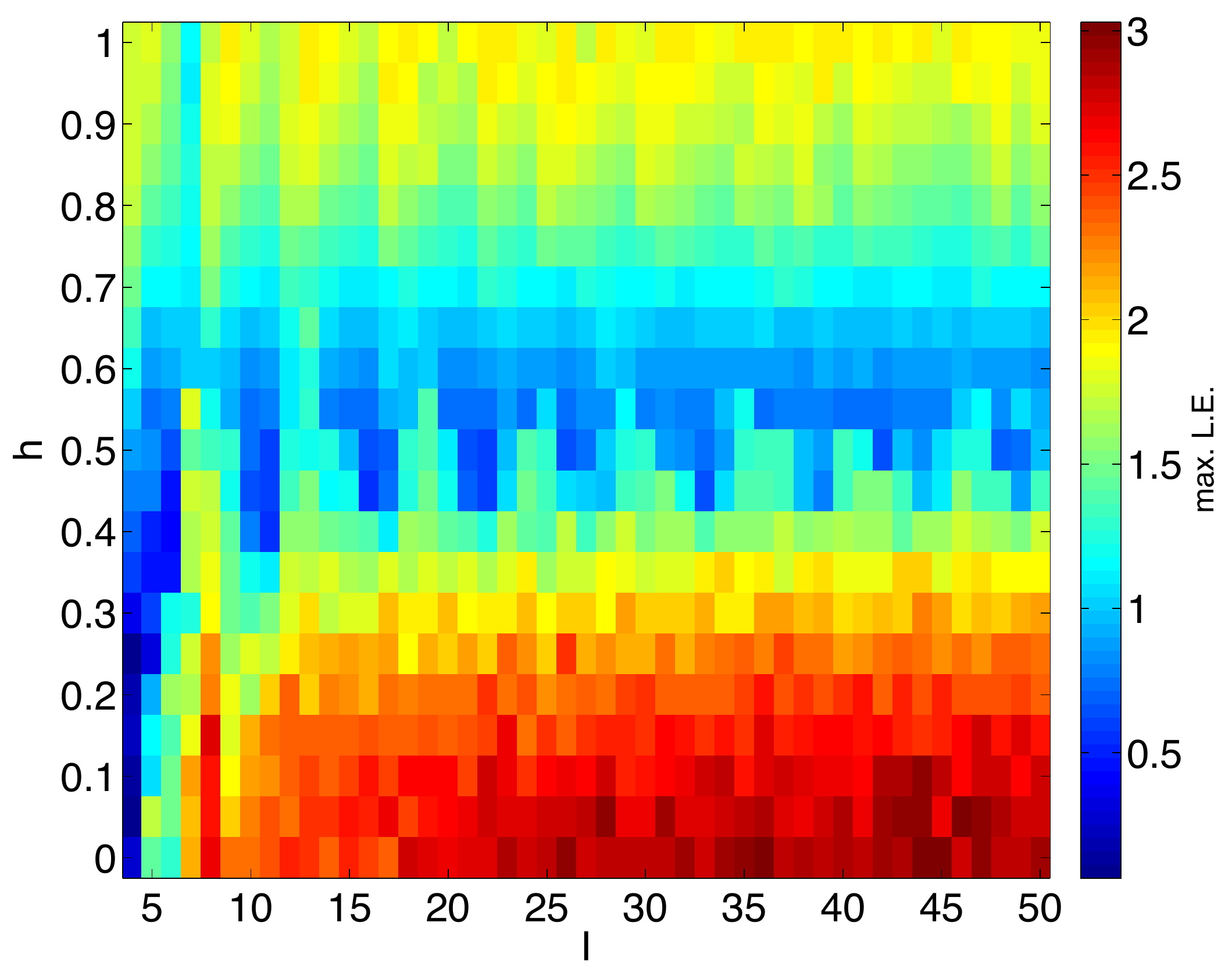}
	\caption{We examine the largest Lyapunov exponent as we vary $I$ on the x-axis and $h$, the coupling parameter, on the y-axis. $J$ is fixed to be 50. We find a pattern similar to the ones observed in Fig. (\ref{fig3}).}
	\label{fig8}
\end{figure}
\section{Discussion}
\begin{figure}[!h]
	\centering
	$\begin{array}{ccc}
	\hspace{-1cm}\includegraphics[width=.25\columnwidth,trim=4.5cm 8cm 4cm 6cm,clip]{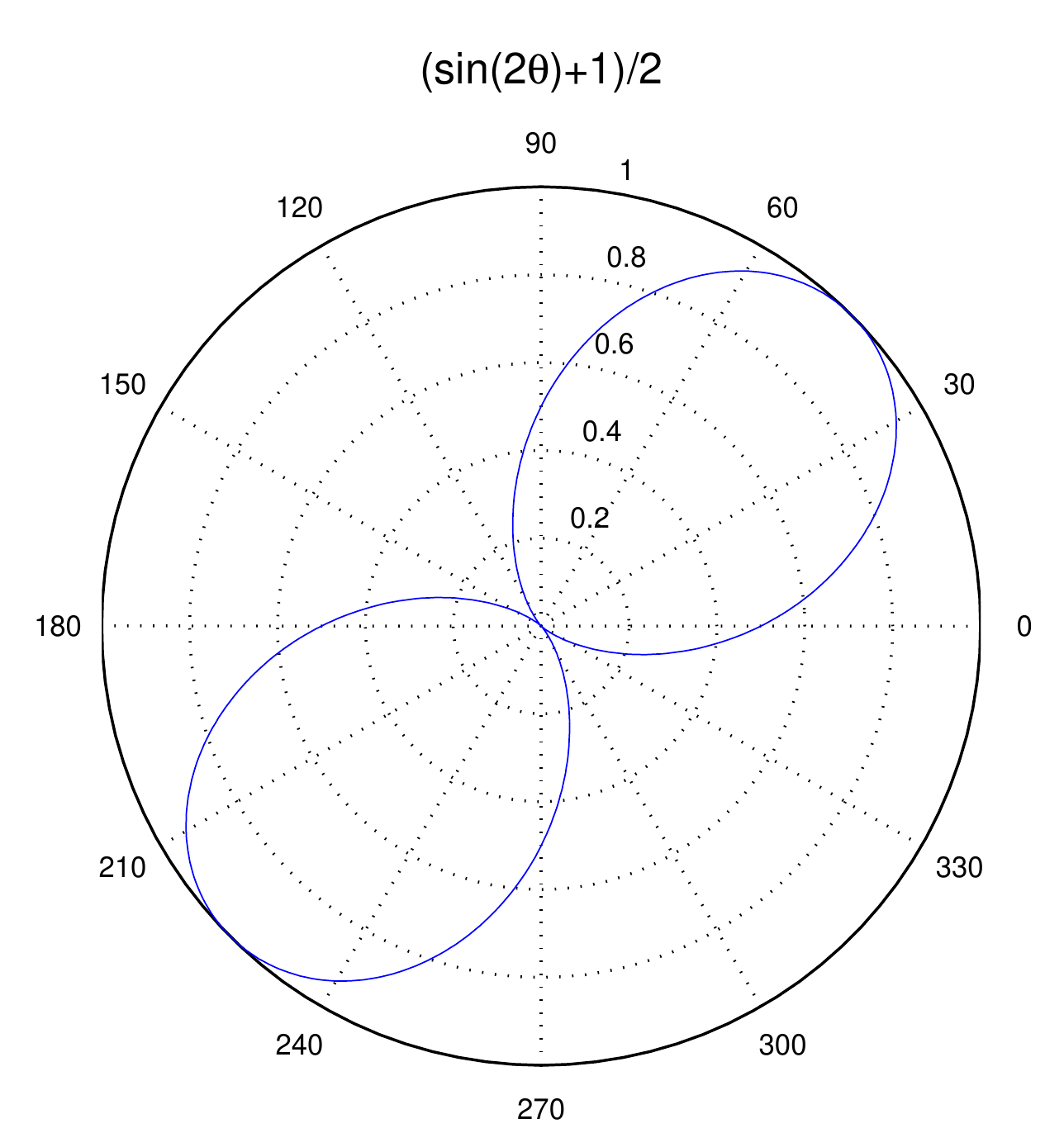}&\includegraphics[width=.25\columnwidth,trim=4.5cm 8cm 4cm 6cm,clip]{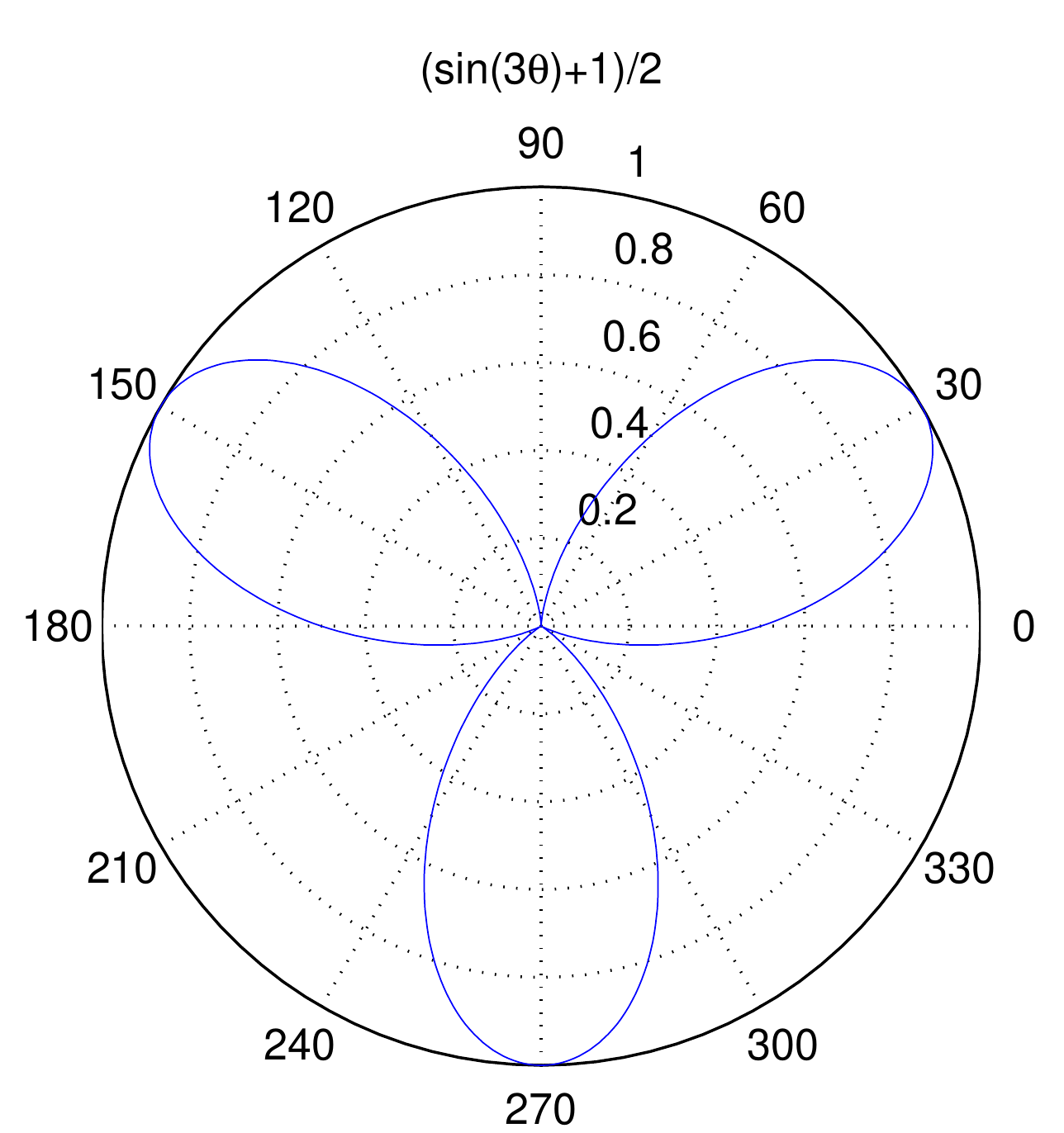}&\includegraphics[width=.25\columnwidth,trim=4.5cm 8cm 4cm 6cm,clip]{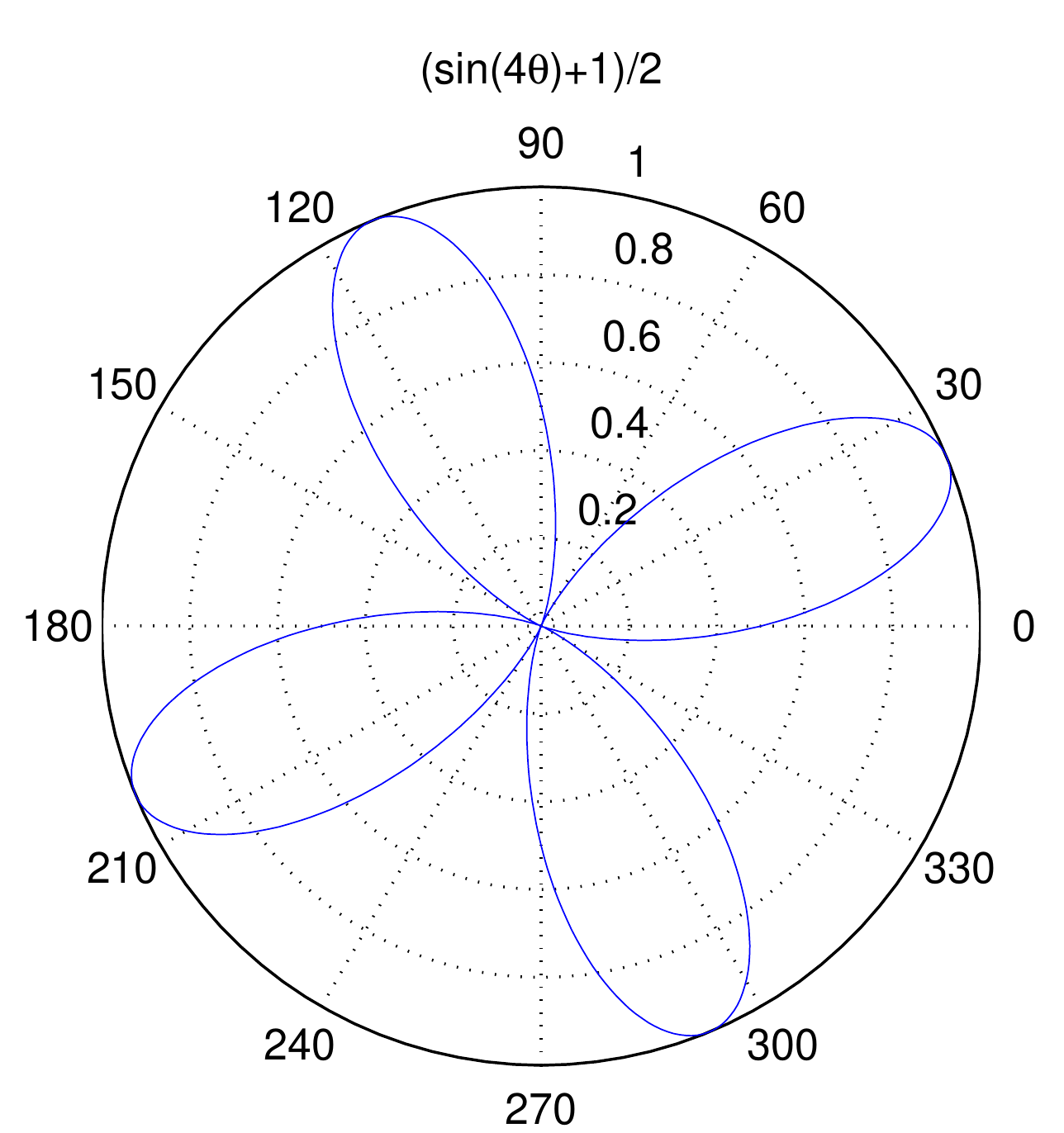}\\
	\hspace{-1cm}\includegraphics[width=.25\columnwidth,trim=4.5cm 8cm 4cm 6cm,clip]{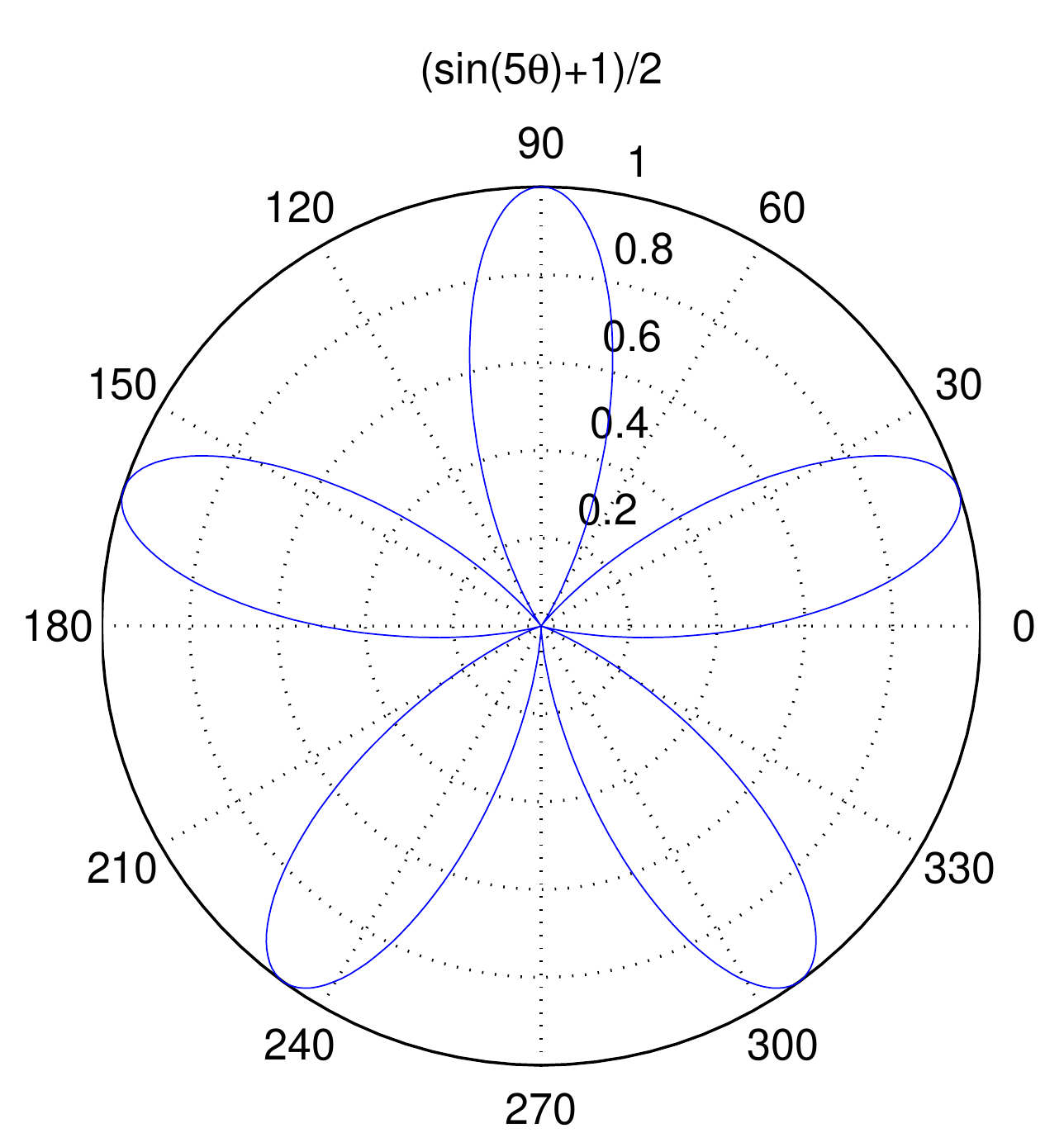}&\includegraphics[width=.25\columnwidth,trim=4.5cm 8cm 4cm 6cm,clip]{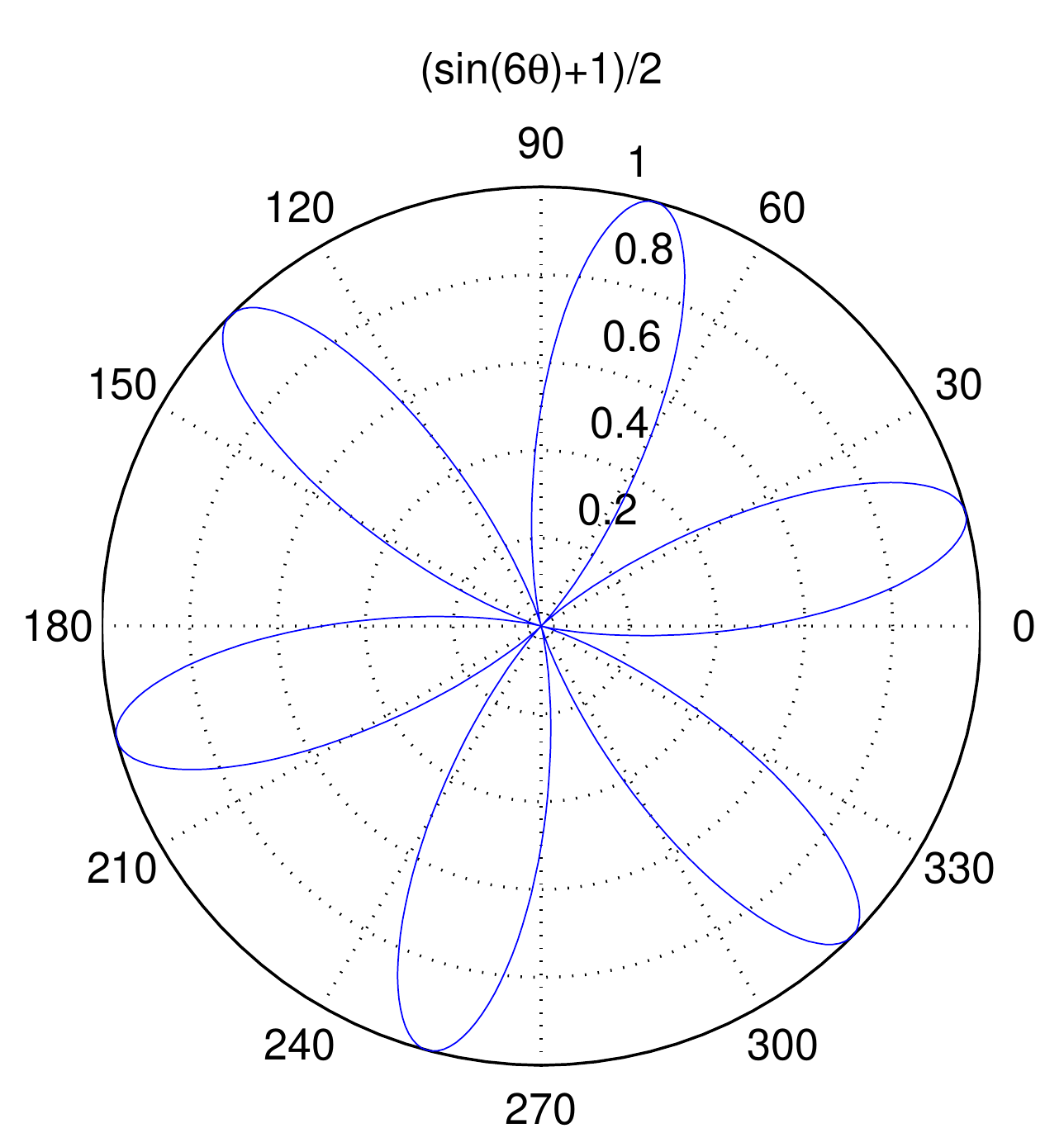}&\includegraphics[width=.25\columnwidth,trim=4.5cm 8cm 4cm 6cm,clip]{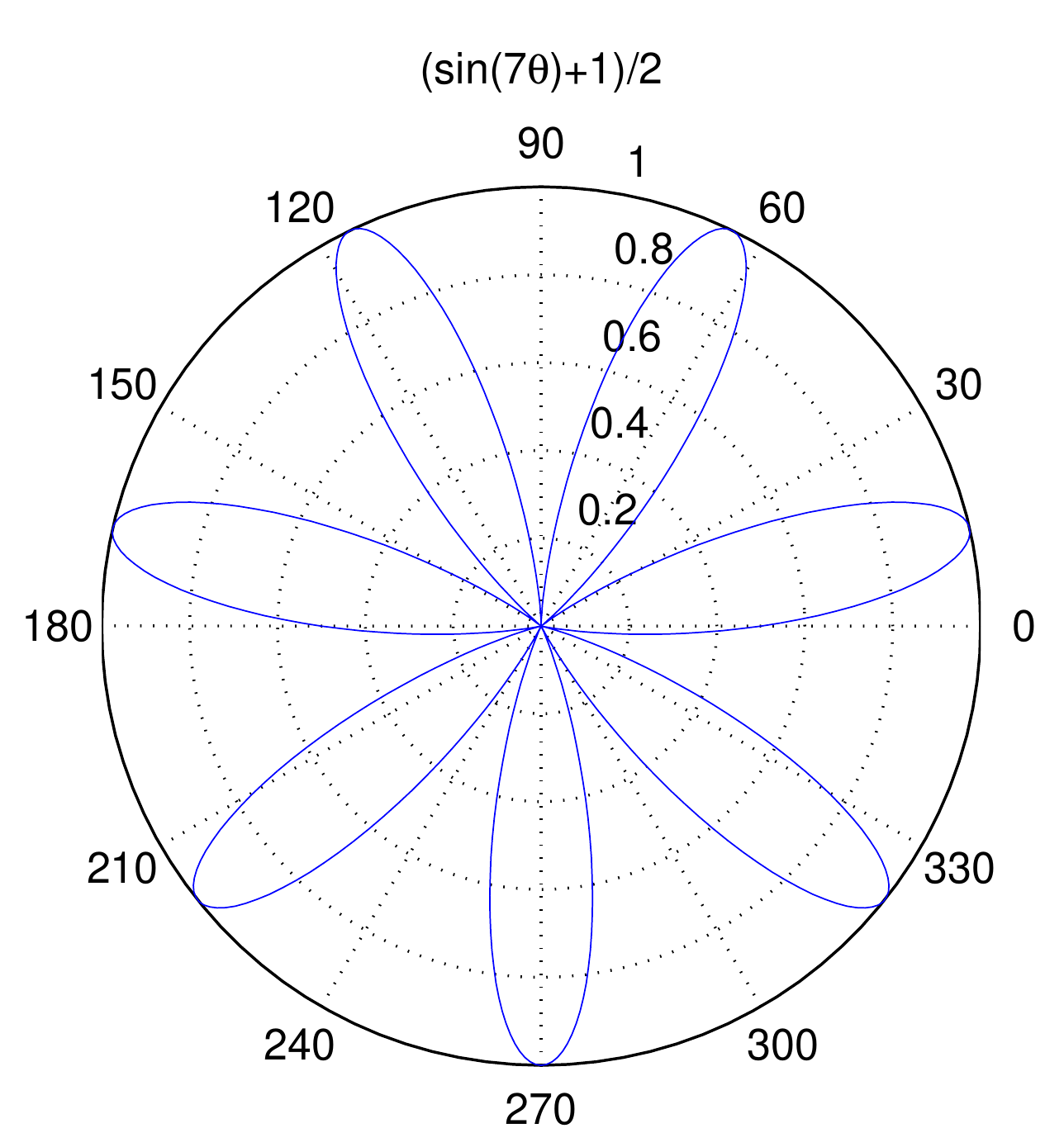}\\
	\hspace{-1cm}\includegraphics[width=.25\columnwidth,trim=4.5cm 8cm 4.5cm 6cm,clip]{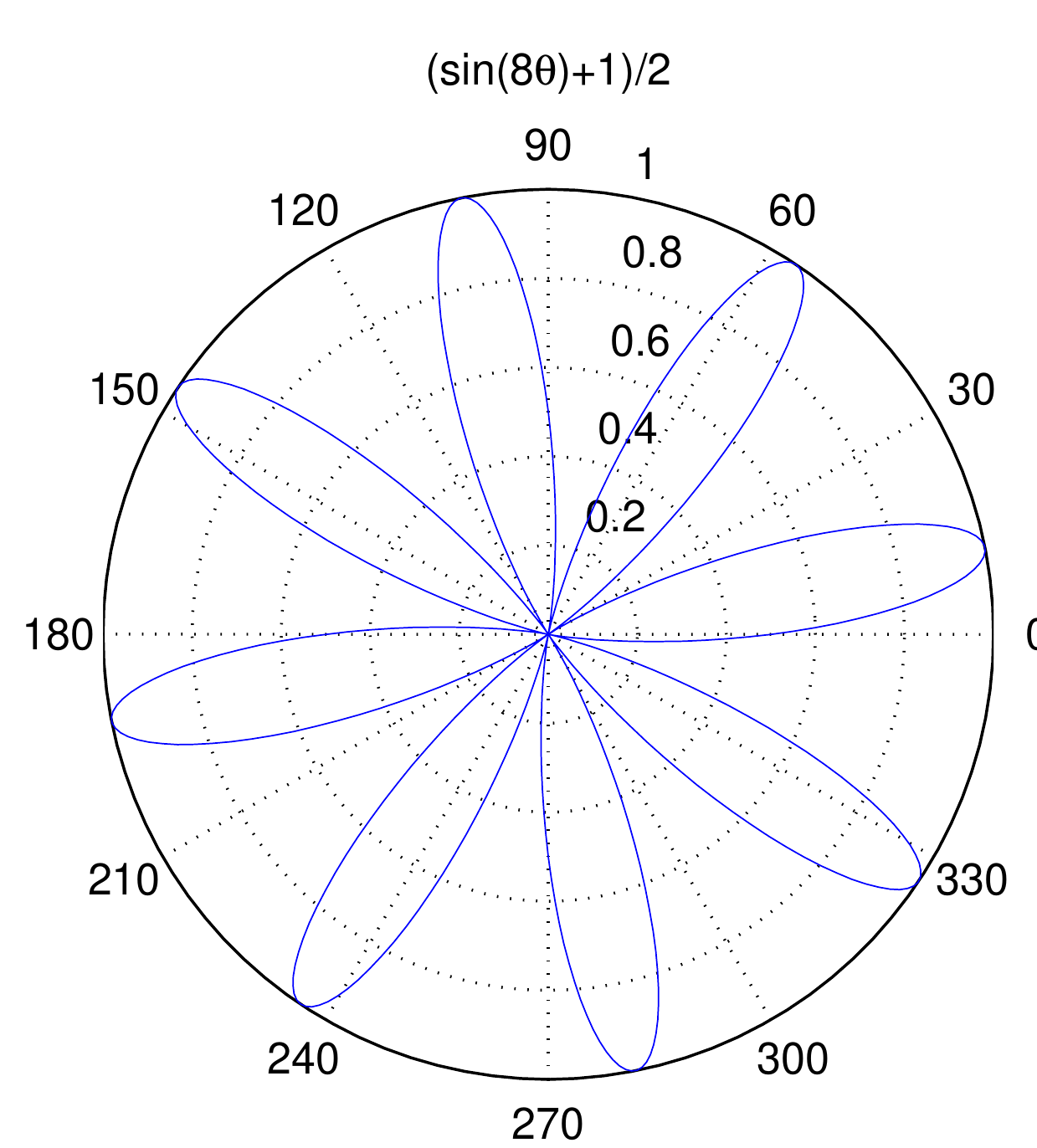}&\includegraphics[width=.25\columnwidth,trim=4.5cm 8cm 4.5cm 6cm,clip]{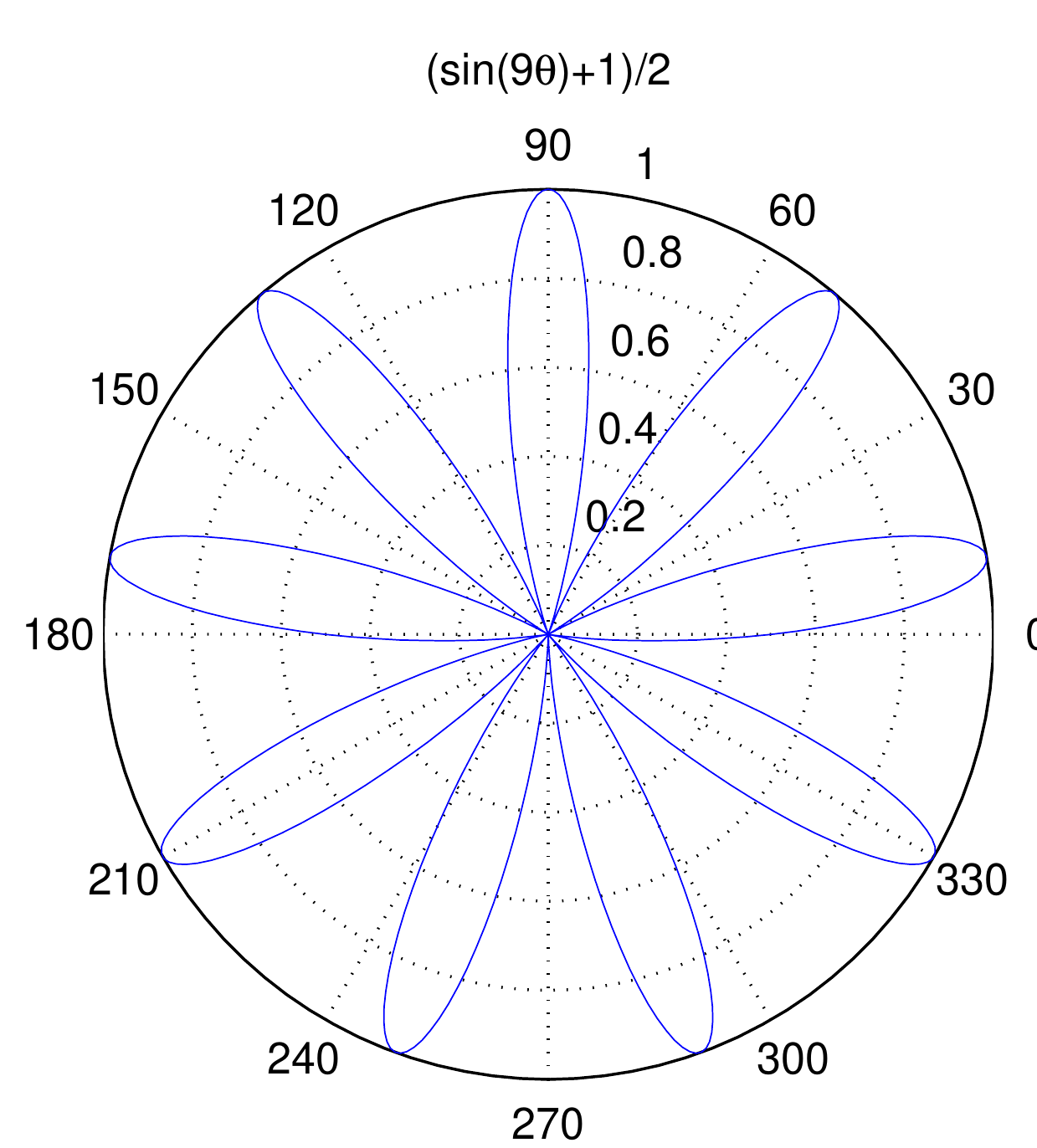}&
		\end{array}$
	\caption{Example trajectories from the model for stable behavior. We examine the results from the model for $t=0$ with $N\in\{2,3,4,5,6,7,8,9\}$. The resulting trajectories are comparable to the stable trajectories shown in Fig. (\ref{fig5}).}
	\label{figA}
\end{figure}

\indent The Lorenz '96 model is a popular choice for atmospheric scientists attempting to improve prediction techniques. This is largely due to the reduction in degrees of freedom offered by the Lorenz '96 system in comparison to more sophisticated models used to make real-world weather predictions. Despite this simplification, the Lorenz '96 model is known for being a computationally manageable model that exhibits tunable levels of chaos, making it an appropriate tool for testing prediction techniques. However, our inspection of parameter space reveals regions of unexpected structural stability. In matters of complexity, adding simple agents often leads to more complexity, but in the case of the Lorenz '96 model we see that there exists a bounded range of $J$ which organizes the dynamics and results in a systemic dampening of chaos. \\
\indent We attempt to explain the observed regions of stability by inspecting the equations for the Lorenz '96 system. Considering Eq. (\ref{eq2}), the sum of the fast oscillators coupled with a given slow oscillator has a dampening effect on the velocity of the slow oscillator, while we also find that the slow oscillator provides positive feedback to the fast oscillators to which it is coupled in Eq. (\ref{eq3}). Therefore, since each slow oscillator has many fast oscillators coupled to it, we expect any excitement of the slow oscillator to be quickly damped away by the fast oscillators. We find evidence of this in Fig. (\ref{fig5}), where peaks in the trajectories of the slow oscillators (points on the inner circle) correspond to increased activity in the fast oscillators coupled to it (the radially adjacent region in the outer circle). If one continues to increase $J$ beyond the observed regions of stability, then the increasingly chaotic dynamics observed in Fig. (\ref{fig3}) may be a result of increased apparent forcing. The magnitude of the sum of the fast oscillators for a given slow oscillator may be large enough to act as a driving force for the dynamics of the slow oscillator (see Eq. (\ref{eq2})). \\
\indent To test this theory, Fig. (\ref{fig8}) shows the largest Lyapunov exponents as we vary $I$ and $h$, the coupling parameter, while holding $J$ fixed at 50. Recalling Eq. (\ref{eq2}), we see that reducing $h$ dampens the sum of the fast oscillators coupled to each slow oscillator. We observe that Fig. (\ref{fig8}) exhibits a similar pattern to Fig. (\ref{fig3}), supporting the claim that reducing the sum of the fast oscillators leads to the stable behavior we observe.\\
\indent The frequency spectrum bifurcation diagrams in Fig. (\ref{fig4}) reveal that the parameter choices for reduced chaotic activity in Fig. (\ref{fig3}) yield surprisingly regular stable attractors with slow oscillators whose trajectories are comprised of only two frequencies. In fact, so long as the choices of $I$, $J$, and $F$ are such that the Lorenz '96 system is in one of the stable regions of parameter space, the trajectories of any slow oscillator exhibits approximately the same dynamics since the dominant and subdominant frequencies for stable attractors lie between 1-2, and 2.5-3, respectively, as seen in the frequency spectrum bifurcation diagrams in Fig. (\ref{fig4}). Indeed, the local maxima of the trajectories of the slow oscillators remain roughly constant across parameter choices leading to stability as shown in Fig. (\ref{fig6}). For a given choice of $F$ and $J$, as $I$ is increased from one stable region in parameter space to the next, we observe the addition of a petal, or a wave, to the attractor. When $I$ lies in between regions of stability in parameter space, we observe attractors that periodically try to grow an additional petal that will eventually dissipate over time. These interesting attractor behaviors appear to occur periodically as a function of $I$. \\
\indent Finally, we note that interactive versions of many of the figures in this manuscript can be found in an online \href{http://www.uvm.edu/~storylab/share/papers/frank2014a/thesis.html}{appendix} (full url \cite{thesis00}).\\
%%%%%%%%%%%%%%%%%%%%%%%%%%%%%%%%%%%%%%%%%%%%%%
\nonumsection{Appendix}
\appendix{}
\indent We attempt to further understand the stable behavior observed in the Lorenz Õ96 model by proposing a parameterization of the petals observed in Fig. (\ref{fig5}). We model the normalized magnitude of a standing wave among the slow modes (as observed in Fig. (\ref{fig5})) with $N(\approx I/5)$ waves at time $t$ using 
\begin{equation}
	r(\theta,t)=\frac{\sin\big(N(\theta+2\pi f\cdot t)\big)+1}{2}.
\end{equation}
where $f$ is the frequency of a representative slow mode. The frequency of the slow mode can be obtained by looking at the dominant frequency from the spectra illustrated in Figure 4 (a function of $I$ and $F$). Scaling $f$ by $2\pi$ and by the number of waves ($N$) yield the desired angular velocity for the standing waves resulting from the model. Example waves resulting from the model at $t=0$ are presented in Fig. \ref{figA}.
%%%%%%%%%%%%%%%%%%%%%%%%%%%%%%%%%%%%%%%%%%%%%%%

\end{document}